\documentclass{vldb}
\usepackage{verbatim}
\usepackage{graphicx}
\usepackage{multirow}

\usepackage[labelfont=bf,textfont=bf]{caption}
\usepackage[labelfont=bf,textfont=bf,singlelinecheck=off,justification=raggedright]{subcaption}
\newcommand{\setHeight}{2cm}
\usepackage[normalem]{ulem}
\usepackage{longtable}
\usepackage{array}
\usepackage{url}
\usepackage{balance}
\usepackage{color}
\usepackage[table]{xcolor}
\usepackage{hhline}
\usepackage{amsmath}

\usepackage{amsthm}
\usepackage{bm}
\newtheorem{defn}{Definition}
\graphicspath{{.}}

\newif\ifJournal
\newif\ifFuture
\newif\ifGrey
\newif\ifColor

\Greytrue

\vldbTitle{The Lernaean Hydra of Data Series Similarity Search: An Experimental Evaluation of the State of the Art}
\vldbAuthors{Karima Echihabi, Kostas Zoumpatianos, Themis Palpanas, and Houda Benbrahim}
\vldbDOI{https://doi.org/10.14778/3282495.3282498}
\vldbVolume{12}
\vldbNumber{2}
\vldbYear{2018}

\begin{document}

\title{The Lernaean Hydra of Data Series Similarity Search:\\ An Experimental Evaluation of the State of the Art	
}

\numberofauthors{4}

\author{
%
%
\alignauthor
Karima Echihabi \\
       \affaddr{
       IRDA, Rabat IT Center, \\
       ENSIAS, Mohammed V Univ.}\\ 
       \affaddr{{karima.echihabi@gmail.com}}\\
\alignauthor Kostas Zoumpatianos \\
\affaddr{Harvard Univ.}\\ 
\affaddr{{kostas@seas.harvard.edu}}
\alignauthor
Themis Palpanas \\
      \affaddr{Paris Descartes Univ.}\\ 
      \affaddr{{themis@mi.parisdescartes.fr}}\\
\and
\alignauthor Houda Benbrahim \\
       \affaddr{
       	IRDA, Rabat IT Center, \\
       	ENSIAS, Mohammed V Univ.}\\ 
       \affaddr{{houda.benbrahim@um5.ac.ma}}
}

\maketitle

\begin{abstract}
Increasingly large data series collections are becoming commonplace across many different domains and applications.
A key operation in the analysis of data series collections is similarity search, 
which has attracted lots of attention and effort over the past two decades.
Even though several relevant approaches have been proposed in the literature, none of the existing studies provides a detailed evaluation against the available alternatives.
The lack of 
comparative results is further exacerbated by the non-standard use of terminology,  
which has led to confusion and misconceptions.
In this paper, we provide definitions for the different flavors of similarity search that have been studied in the past, and present the first systematic experimental evaluation of the efficiency of data series similarity search techniques.
Based on the experimental results, we describe the strengths and weaknesses of each approach and give recommendations for the best approach to use under typical use cases.
Finally, by identifying the shortcomings of each method, our findings lay the ground for solid further developments in the field.
\end{abstract}

\section{Introduction}
\label{sec:introduction}

\noindent{\bf Data Series.}
A data series is an ordered sequence of data points\footnote{When the sequence is ordered on time, it is called a \emph{time series}.
However, the order can be defined by angle (e.g., in radial profiles), mass (e.g., in mass spectroscopy), position (e.g., in genome sequences), and others~\cite{conf/sofsem/Palpanas2016}. The terms \emph{data series}, \emph{time series} and \emph{sequence} are used interchangeably.}.
Data series are one of the most common types of data, covering virtually every scientific and social domain, such as astrophysics, neuroscience, seismology, environmental monitoring, biology, health care, energy, finance, criminology, social studies, video and audio recordings, and many others~\cite{KashinoSM99,Shasha99,humanbehaviorpatterns,volker,DBLP:conf/edbt/MirylenkaCPPM16,HuijseEPPZ14,percomJournal,windturbines,spikesorting,VALMOD,journal/jte/Williams2003,conf/compstats/Hebrail2000}.
As more devices, applications, and users are connected with IoT technologies, an increasingly large number of data series is generated, leading to multi-TB collections~\cite{DBLP:journal/sigmod/Palpanas15}.
%
%
We note that, when these collections are analyzed, the common denominator of most analysis algorithms and machine learning methods, e.g., outlier detection~\cite{journal/csur/Chandola2009,journal/vldb/Dallachiesa2014}, frequent pattern mining~\cite{conf/kdd/Mueen2012}, clustering~\cite{conf/kdd/Keogh1998,conf/sdm/Rodrigues2006,conf/icdm/Keogh2011, journal/pattrecog/Warren2005}, and classification~\cite{journal/jmlr/Chen2009}, is that they are based on similarity search.
That is, they require to compute distances among data series, and this operation is repeated many times.

\noindent{\bf Data Series Similarity Search.}
Similarity search is the operation of finding the set of data series in a collection, which is close to a given query series according to some definition of distance (or similarity).
A key observation is that similarity search needs to process a sequence (or subsequence) of values as a single object, rather than as individual, independent points, which is what makes the management and analysis of data sequences a hard problem with a considerable cost.
Therefore, improving the performance of similarity search can improve the scalability of data analysis algorithms for massive data series collections.

Nevertheless, despite the significance of data series similarity search, and the abundance of relevant methods that have been proposed in the past two decades~\cite{conf/fodo/Agrawal1993, conf/vldb/Ciaccia1997,conf/kdd/shieh1998, conf/icde/Rafiei99, conf/kdd/ColeSZ05, conf/icdm/Camerra2010, conf/kdd/Karras2011, conf/kdd/Mueen2012,journal/edbt/Schafer2012,conf/vldb/Wang2013, journal/kais/Camerra2014,journal/vldb/Zoumpatianos2016, code/Mueen2017,dpisax,ulisse,journal/vldb/linardi18,conf/bigdata/peng18}, no study has ever attempted to compare these methods under the same conditions. 
We also point out that we focus on the \emph{efficiency} of similarity search methods, whereas previous works studied the \emph{accuracy} of dimensionality reduction techniques and similarity measures, focusing on classification~\cite{journal/dmkd/Keogh2003, conf/vldb/Ding2008, DBLP:journals/datamine/BagnallLBLK17}. 

In this experimental and analysis paper, we thoroughly assess different data series similarity search methods, in order to lay a solid ground for future research developments in the field.
In particular, we focus on the problem of \emph{exact whole matching similarity search in collections with a very large number of data series}, i.e., similarity search that produces exact (not approximate) results, by calculating distances on the whole (not a sub-) sequence.
This problem represents a common use case across many domains~\cite{url:adhd,url:sds,conf/compstats/Hebrail2000,SENTINEL-2}. 
This work is the most extensive experimental comparison of the efficiency of 
similarity search methods ever conducted. 


\noindent{\bf Contributions.}
We make the following contributions:

1. We present a thorough discussion of the data series similarity search problem, formally defining its different variations that have been studied in the literature under diverse and conflicting names. Thus, establishing a common language that will facilitate further work in this area.

2. We include a brief survey of data series similarity search approaches, bringing together studies presented in different communities that have been treated in isolation from each other.
These approaches range from smart serial scan methods to the use of indexing, and are based on a variety of classic and specialized data summarization techniques.

3. We make sure that all approaches are evaluated under the same conditions, so as to guard against implementation bias.
To this effect, we used implementations in C/C++ for all approaches, and reimplemented in C the ones that were only available in other programming languages.
Moreover, we conducted a careful inspection of the code bases, and applied to all of them the same set of optimizations (e.g., with respect to memory management, Euclidean distance calculation, etc.), leading to considerably faster performance.

4. We conduct the first comprehensive experimental evaluation of the efficiency of data series similarity search approaches, using several synthetic and 4 real datasets from diverse domains.
In addition, we report the first large scale experiments with carefully crafted query workloads that include queries of varying difficulty, which can effectively stress-test all the approaches.
Our results reveal characteristics that have not been reported in the literature, and lead to a deep understanding of the different approaches and their performance.
Based on those, we provide recommendations for the best approach to use under typical use cases, and identify promising future research directions.

5. 
We make available online all source codes, datasets, and query workloads used in our study~\cite{url/DSSeval}. 
This will render our work reproducible and further help the community to agree on and establish a much needed data series similarity search benchmark~\cite{journal/dmkd/Keogh2003, conf/kdd/Zoumpatianos2015, johannesjoural2018}.



\section{Definitions and Terminology}
\label{sec:problem}


Similarity search represents a common problem in various areas of computer science.
However, in the particular case of data series, there exist several different flavors that have been studied in the literature, often times using overloaded and conflicting terms.
This has contributed to an overall confusion, which hinders further advances in the field.

In this section, we discuss the different flavors of data series similarity search, and provide corresponding definitions, which set a common language for the problems in this area.

\noindent{\bf On Sequences.}
A \textit{\textbf{data series}} $S(p_1,p_2,...,p_n)$ is an ordered sequence of points, $p_i$, $1 \leq i \leq n$.
The number of points, $|S|=n$, is the length of the series.
We denote the $i$-th point in $S$ by $S[i]$; then $S[i:j]$ denotes the \textit{\textbf{subsequence}} $S(p_i,p_{i+1},...,p_{j-1},p_j)$, where $1 \leq i \leq j \leq n$.
We use $\mathbb{S}$ to represent all the series in a collection (dataset).

In the above definition, if each point in the series represents the value of a single variable (e.g., temperature) then each point is a scalar, and we talk about a \textit{\textbf{univariate series}}. Otherwise, if each point represents the values of multiple variables (e.g., temperature, humidity, pressure, etc.) then each point is a vector, and we talk about a \textit{\textbf{multivariate series}}.
The values of a data series may also encode measurement errors, or imprecisions, in which case we talk about uncertain data series~\cite{DBLP:conf/ssdbm/AssfalgKKR09,DBLP:conf/edbt/YehWYC09,DBLP:conf/kdd/SarangiM10,DBLP:journals/pvldb/DallachiesaNMP12,journal/vldb/Dallachiesa2014}.

Especially in the context of similarity search, a data series of length $n$ can also be represented as a single point in an $n$-dimensional space. 
Then the values and length of $S$ are referred to as \emph{dimensions} and \emph{dimensionality}, respectively.

%



\noindent{\bf On Distance Measures.}
A data series \textit{\textbf{distance}} is a function that measures the (dis)similarity of two data series.
The distance between a query series, $S_Q$, and a candidate series, $S_C$, is denoted by $d(S_Q,S_C)$.

Even though several distance measures have been proposed in the literature~\cite{berndt1994using,das1997finding,DBLP:conf/edbt/AssfalgKKKPR06,DBLP:conf/icde/ChenNOT07,journal/dmkd/Wang2013,DBLP:conf/ssdbm/MirylenkaDP17}, the Euclidean distance is the one that is the most widely used, as well as one of the most effective for large data series collections~\cite{conf/vldb/Ding2008}.
We note that an additional advantage of Euclidean distance is that in the case of Z-normalized series (mean=$0$, stddev=$1$), which are very often used in practice~\cite{conf/kdd/Zoumpatianos2015}, it can be exploited to compute Pearson correlation~\cite{conf/icde/Rafiei99}.

In addition to the distance used to compare data series in the high-dimensional space, some similarity search methods also rely on \textit{lower-bounding}~\cite{journal/kais/Camerra2014, journal/vldb/Zoumpatianos2016, journal/edbt/Schafer2012, conf/vldb/Wang2013, dpisax, ulisse, conf/vldb/Ciaccia1997, conf/kdd/Karras2011} and \textit{upper-bounding} distances~\cite{conf/vldb/Wang2013, conf/kdd/Karras2011}. 
A \textit{\textbf{lower-bounding distance}} is a distance defined in the reduced dimensional space satisfying the lower-bounding property, i.e., the distance between two series in the reduced space is guaranteed to be smaller than or equal to the distance between the series in the original space~\cite{conf/sigmod/Faloutsos1994}. 
Inversely, an \textit{\textbf{upper-bounding distance}} ensures that distances in the reduced space are larger than the distances in the original space~\cite{conf/vldb/Wang2013, conf/kdd/Karras2011}.

\noindent{\bf On Similarity Search Queries.}
We now define the different forms of data series similarity search queries.
We assume a data series collection, $\mathbb{S}$, a query series, $S_Q$, and a distance function $d(\cdot,\cdot)$.

A \textit{\textbf{k-Nearest-Neighbor (k-NN) query}} identifies the $k$ series in the collection with the smallest distances to the query series.

\begin{defn} \label{def:knnquery}
Given an integer $k$, a \textit{\textbf{k-NN query}} retrieves the set of series $\mathbb{A} = \{ \{S_{C_1},...,S_{C_k}\} \subseteq \mathbb{S} | \forall \ S_C \in \mathbb{A} \ and \ \forall \ S_{C'} \notin \mathbb{A}, \ d(S_Q,S_C) \leq d(S_Q,S_{C'})\}$.
\end{defn}

An \textit{\textbf{r-range query}} identifies all the series in the collection within range $r$ form the query series.

\begin{defn} \label{def:rquery}
Given a distance $r$, an \textit{\textbf{r-range query}} retrieves the set of series $\mathbb{A} = \{S_C \in \mathbb{S} | d(S_Q,S_C) \leq r\}$.
\end{defn}

We additionally identify the following two categories of k-NN and range queries.
In \textit{\textbf{whole matching (WM)}} queries, we compute the similarity between an entire query series and an entire candidate series.
All the series involved in the similarity search have to have the same length.
In \textit{\textbf{subsequence matching (SM)}} queries, we compute the similarity between an entire query series and all subsequences of a candidate series.
In this case, candidate series can have different lengths, but should be longer than the query series. 

\begin{defn} \label{def:wholematch}
	A \textit{\textbf{whole matching query}} finds the candidate data series $S \in \mathbb{S}$ that matches $S_Q$, where $|S|=|S_Q|$. 
\end{defn}

\begin{defn} \label{def:submatch}
	A \textit{\textbf{subsequence matching query}} finds the subsequence $S[i:j]$ of a candidate data series $S \in \mathbb{S}$ that matches $S_Q$, where $|S[i:j]| = |S_Q| < |S|$.
\end{defn}



In practice, we encounter situations that cover the entire spectrum: WM queries on large collections of short series~\cite{SENTINEL-2,url:sds}, SM queries on large collections of short series~\cite{url:adhd}, and SM queries on collections of long series~\cite{url/data/seismic}.

Note that SM queries can be converted to WM: create a new collection that comprises all overlapping subsequences (each long series in the candidate set is chopped into overlapping subsequences of the length of the query), and perform a WM query against these subsequences~\cite{ulisse,journal/vldb/linardi18}.

\noindent{\bf On Similarity Search Methods.}
When a similarity search algorithm (k-NN or range) produces answers that are (by definition) always correct and complete: we call such an algorithm \textit{\textbf{exact}}.
Nevertheless, we can also develop algorithms without such strong guarantees: we call such algorithms \textit{\textbf{approximate}}.
As we discuss below, there exist different flavors of approximate similarity search algorithms.


An {\textit{\bf$\bm{\epsilon}$-approximate}} algorithm guarantees that its distance results have a relative error no more than $\epsilon$, i.e., the approximate distance is at most $(1+\epsilon)$ times the exact one. 

\begin{defn} \label{def:epsmatch}
Given a query $S_Q$, and $\epsilon \geq 0$, an \textit{\textbf{$\bm{\epsilon}$-approximate}} algorithm guarantees that all results, $S_C$, are at a distance $d(S_Q,S_C) \leq (1+\epsilon)d(S_Q,[\text{k-th NN of }S_Q])$ in the case of a $k$-NN query, and distance $d(S_Q,S_C) \leq (1+\epsilon)r$ in the case of an r-range query.
\end{defn}

A {\bf $\bm{\delta}$-$\bm{\epsilon}$-approximate} algorithm, guarantees that its distance results will have a relative error no more than $\epsilon$ (i.e., the approximate distance is at most $(1+\epsilon)$ times the exact distance), with a probability of at least $\delta$.
	

\begin{defn} \label{def:probmatch}
Given a query $S_Q$, $\epsilon \geq 0$, and $\delta \in [0,1]$, a \textit{\textbf{$\bm{\delta}$-$\bm{\epsilon}$-approximate}} algorithm produces results, $S_C$, for which $Pr[d(S_Q,S_C)$ $\leq (1+\epsilon)d(S_Q,[\text{k-th NN of }S_Q])] \geq \delta$ in the case of a $k$-NN query, and $Pr[d(S_Q,S_C) \leq (1+\epsilon)r] \geq \delta$) in the case of an r-range query.
\end{defn}


An \textit{\textbf{ng-approximate}} (no-guarantees approximate) algorithm does not provide any guarantees (deterministic, or probabilistic) on the error bounds of its distance results.

\begin{defn} \label{def:appmatch}
Given a query $S_Q$, an \textit{\textbf{ng-approximate}} algorithm produces results, $S_C$, that are at a distance $d(S_Q,S_C) \leq (1+\theta)d(S_Q,[\text{k-th NN of }S_Q])$ in the case of a $k$-NN query, and distance $d(S_Q,S_C) \leq (1+\theta)r$ in the case of an r-range query, for an arbitrary value $\theta \in \mathbb{R}_{>0}$.
\end{defn}

In the data series literature, \textit{ng-approximate} algorithms have been referred to as \emph{approximate}, or \emph{heuristic} search~\cite{journal/kais/Camerra2014, journal/vldb/Zoumpatianos2016, journal/edbt/Schafer2012, conf/vldb/Wang2013, dpisax, ulisse}.
Unless otherwise specified, for the rest of this paper we will refer to \textit{ng-approximate} algorithms simply as approximate. Approximate matching in the data series literature
consists of pruning the search space, by traversing one path of an index structure representing the data, visiting at most one leaf, to get a baseline best-so-far (bsf) match.

Observe that when $\delta = 1$, a $\delta$-$\epsilon$-approximate method becomes $\epsilon$-approximate, and when $\epsilon=0$, an $\epsilon$-approximate method becomes exact~\cite{conf/icde/Ciaccia2000}.
It it also possible that the same approach implements both approximate and exact algorithms~\cite{conf/kdd/shieh1998,conf/vldb/Wang2013,journal/kais/Camerra2014,journal/vldb/Zoumpatianos2016,journal/edbt/Schafer2012}. 
Methods that provide exact answers with probabilistic guarantees are considered $\delta$-0-approximate. 
These methods guarantee distance results to be exact with probability at least $\delta$ ($0 \leq \delta \leq 1$ and $\epsilon$ = 0).
(We note that in the case of $k$-NN queries, Def.~\ref{def:epsmatch} corresponds to the \emph{approximately correct NN}~\cite{conf/icde/Ciaccia2000} and \emph{$(1+\epsilon)$-approximate NN}~\cite{journal/acm/Arya1998}, while  Def.~\ref{def:probmatch} corresponds to the \emph{probably approximately correct NN}~\cite{conf/icde/Ciaccia2000}.)

\noindent{\bf Scope.}
In this paper, we focus on \emph{univariate} series with \emph{no uncertainty},
and we examine \emph{exact} methods for \emph{whole matching} in collections with a \emph{very large number of series}, using \emph{$k$-NN queries} and the \emph{Euclidean distance}.
This is a very popular problem that lies at the core of several other algorithms, and is important for many applications in various domains in the real world~\cite{journal/pattrecog/Warren2005,conf/kdd/Zoumpatianos2015,conf/sofsem/Palpanas2016}, ranging from fMRI clustering~\cite{golay1998new} to mining earthquake~\cite{kakizawa1998discrimination}, energy consumption~\cite{kovsmelj1990cross}, and retail data~\cite{DBLP:conf/kdd/KumarPW02}.
Note also that some of the insights gained by this study could carry over to other settings, such as, $r$-range queries, dynamic time warping distance, or approximate search.

\section{Similarity Search Primer}
\label{sec:approaches}


Similarity search methods can be classified into sequential, and indexing methods.
Sequential methods proceed in one step to answer a similarity search query. Each candidate is read sequentially from the raw data file and compared to the query. Particular optimizations can be applied to limit the number of these comparisons~\cite{conf/kdd/Mueen2012}. 
Some sequential methods work with the raw data in its original high-dimensional representation~\cite{conf/kdd/Mueen2012}, while others perform transformations on the raw data before comparing them to the query~\cite{code/Mueen2017}.

On the other hand, answering a similarity query using an index involves two steps: a filtering step where the pre-built index is used to prune candidates and a refinement step where the surviving candidates are compared to the query in the original high dimensional space~\cite{conf/sigmod/Guttman1984,conf/vldb/weber1998, conf/cikm/hakan2000, journal/kais/Camerra2014,journal/vldb/Zoumpatianos2016,journal/edbt/Schafer2012,conf/vldb/Wang2013,conf/icmd/Beckmann1990,dpisax,ulisse, journal/vldb/linardi18}. Some indexing methods first summarize the original data and then index these summarizations~\cite{conf/icmd/Beckmann1990,journal/edbt/Schafer2012, conf/vldb/weber1998, conf/cikm/hakan2000}, while others interwine data reduction and indexing~\cite{journal/kais/Camerra2014,journal/vldb/Zoumpatianos2016,conf/vldb/Wang2013}. 
Some methods index high dimensional data directly~\cite{conf/vldb/Ciaccia1997}. 
We note that all indexing methods depend on lower-bounding, since it allows indexes to prune the search space with the guarantee of no false dismissals~\cite{conf/sigmod/Faloutsos1994} (the DSTree index~\cite{conf/vldb/Wang2013} also supports an upper-bounding distance, but does not use it for similarity search).
Metric indexes (such as the M-tree~\cite{conf/vldb/Ciaccia1997}) additionally require the distance measure triangle inequality to hold.
Though, there exist (non-metric) indexes for data series that are based on distance measures that are not metrics~\cite{journal/kis/Keogh2005}. 

There also exist hybrid approaches that fall in-between indexing and sequential methods.
In particular, multi-step approaches, where data are transformed and re-organized in levels.
Pruning then occurs at multiple intermediate filtering steps as levels are sequentially read one at a time.{ 
	
Stepwise is such a method~\cite{conf/kdd/Karras2011}, relying on Euclidean distance, and lower- and upper-bounding distances.

\subsection{Summarization Techniques}

We now briefly outline the summarization techniques used by the methods that we examine in this study.

The {\it Discrete Haar Wavelet Transform} (DHWT)~\cite{conf/icde/Chan1999} uses the Haar wavelet decomposition to transform each data series $S$ into a multi-level hierarchical structure.
Resulting summarizations are composed of the first $l$ coefficients. 

The {\it Discrete Fourier Transform} (DFT)~\cite{conf/fodo/Agrawal1993,conf/sigmod/Faloutsos1994,conf/sigmod/Rafiei1997,journal/corr/Rafiei1998} decomposes $S$ into frequency coefficients.
A subset of $l$ coefficients constitutes the summary of $S$. 
In our experiments, we use the Fast Fourier Transform (FFT) algorithm,
which is optimal for whole matching scenarios (the MFT algorithm~\cite{conf/icdsp/Albrecht1997} is faster than FFT for computing DFT on sliding windows, thus beneficial for subsequence matching queries).

The {\it Piecewise Aggregate Approximation} (PAA)~\cite{journal/kais/Keogh2001} and {\it Adaptive Piecewise Constant Approximation} (APCA)~\cite{journal/acds/Chakrabarti2002} methods are segmentation techniques that divide $S$ into $l$ (equi-length and varying-length, respectively) segments. Each segment represents the mean value of the corresponding points. 
The {\it Extended Adaptive Piecewise Approximation} (EAPCA)~\cite{conf/vldb/Wang2013} technique extends APCA by using more information to represent each segment. 
In addition to the mean, 
it also stores the standard deviation of the segment.
With the {\it Symbolic Aggregate Approximation} (SAX)~\cite{conf/dmkd/LinKLC03}, $S$ is first transformed using PAA into $l$ real values, 
and then a discretization technique is applied to map PAA values 
to discrete set of symbols (alphabet) that can be succinctly represented in binary form.
A SAX representation consists of $l$ such symbols.
An $i$SAX (indexable SAX)~\cite{conf/kdd/Shieh2008} representation can have an arbitrary alphabet size for each segment.

Similarly to SAX, the {\it Symbolic Fourier Approximation} (SFA) \cite{journal/edbt/Schafer2012} is also a symbolic approach. 
However, instead of PAA, it first transforms $S$ 
into $l$ DFT coefficients using FFT (or MFT for subsequence matching), then extends the discretization principle of SAX to support both equi-depth and equi-width binning, and to allow each dimension to have its own breakpoints. 
An SFA 
summary consists of $l$ symbols.

Using the VA+file method~\cite{conf/cikm/hakan2000}, $S$ of length $n$ is first transformed using the Karhunen\textendash Lo\`{e}ve transform (KLT) into $n$ real values,
which are then quantized to discrete symbols.
As we will detail later, we modified the VA+file to use DFT instead of KLT, for efficiency reasons. 

Figure~\ref{fig:summarizations} presents a high-level overview of the summarization techniques presented above.

\begin{figure}[tb]
	\captionsetup{justification=centering}
	\includegraphics[scale =0.70]{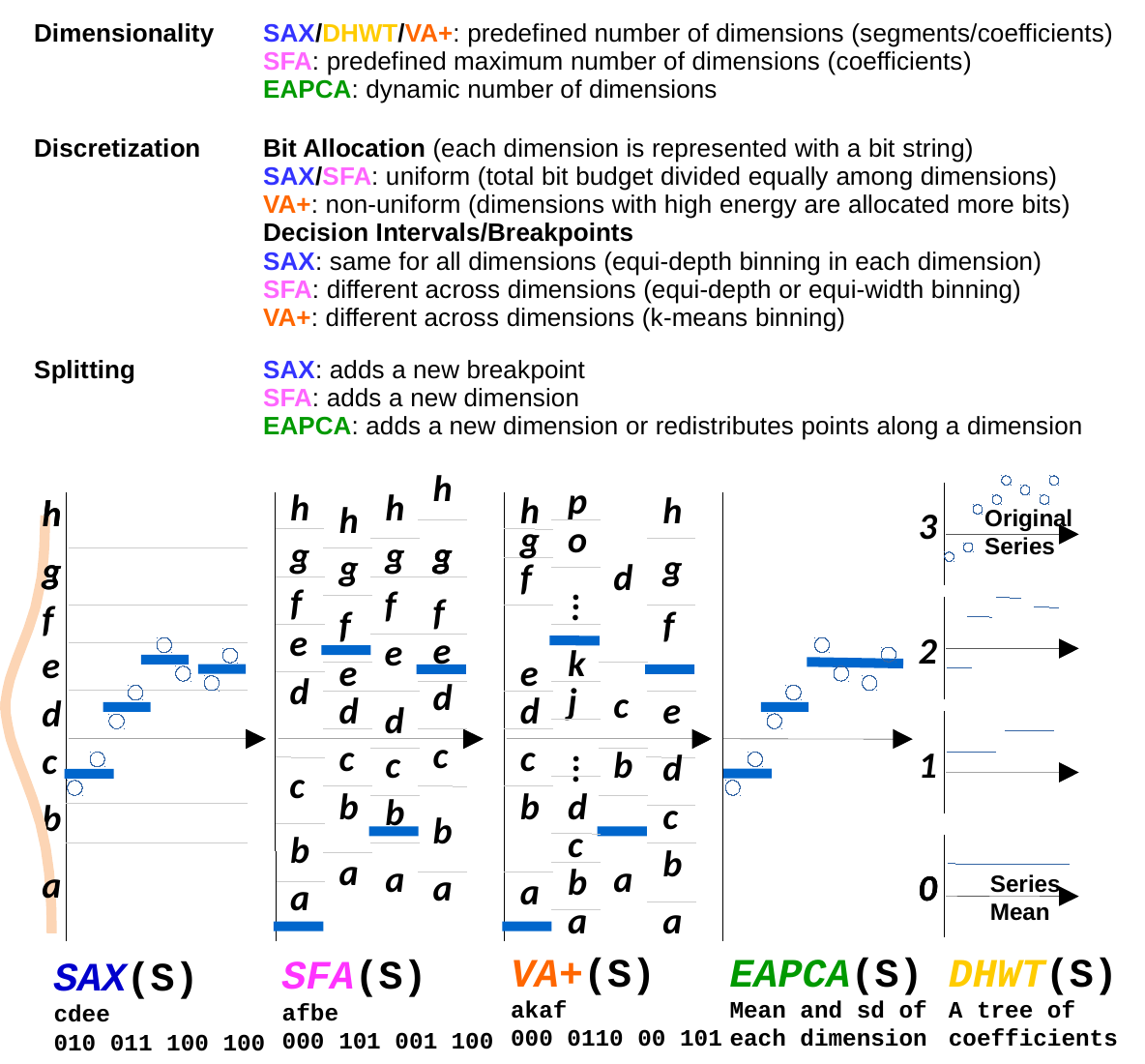}
	\caption{Summarizations}
	\label{fig:summarizations}
\end{figure}

\begin{table*}[tb]
	\caption{Similarity search methods}
	{\small
		\centering
		\hspace*{1cm}
		\begin{tabular*}{\linewidth}{|*{12}{c|}} 
			\cline{3-12} 
			\multicolumn{1}{c}{}& & \multicolumn{4}{c|}{Matching Accuracy}  & \multicolumn{2}{c|}{Matching Type} & \multicolumn{2}{|c|}{Representation} & \multicolumn{2}{c|}{Implementation}\\    		
			\cline{3-12} 
			\multicolumn{1}{c}{}& & exact & ng-appr. & $\epsilon$-appr. &$\delta$-$\epsilon$-appr. & Whole & Subseq. & Raw & Reduced & Original  & New \\    		
			\cline{1-12}			 		 
			\multicolumn{1}{|c|}{\multirow{6}{*}{\rotatebox[origin=c]{90}{Indexes}}}
			& \multicolumn{1}{|c|}{ADS+} &\cite{journal/vldb/Zoumpatianos2016} &\cite{journal/vldb/Zoumpatianos2016} &  &  &  \checkmark &  &  & iSAX &  C &\\	
			\cline{2-12}			 		 
			& \multicolumn{1}{|c|}{DSTree} &\cite{conf/vldb/Wang2013} & \cite{conf/vldb/Wang2013} &  &  &  \checkmark &   & & EAPCA & Java & C\\	
			\cline{2-12}			 		 
			& \multicolumn{1}{|c|}{iSAX2+} & \cite{journal/kais/Camerra2014} & \cite{journal/kais/Camerra2014} &  &  & \checkmark & &  &  iSAX &  C\# & C\\	
			\cline{2-12}			 	 
			& \multicolumn{1}{|c|}{M-tree} & \cite{conf/vldb/Ciaccia1997}
			&  & \cite{conf/icde/Ciaccia2000} & \cite{conf/icde/Ciaccia2000} & \checkmark&  & \checkmark & & C++ &  \\	
			\cline{2-12}			 		 
			& \multicolumn{1}{|c|}{R*-tree} & \cite{conf/icmd/Beckmann1990} &  &  &  & \checkmark&  &  & PAA & C++  &\\	
			\cline{2-12}			 		 
			& \multicolumn{1}{|c|}{SFA trie} &\cite{journal/edbt/Schafer2012} & \cite{journal/edbt/Schafer2012} &  &  & \checkmark & \checkmark & & SFA & Java & C\\	
			\cline{2-12}			 		 
			& \multicolumn{1}{|c|}{VA+file} & {\cite{conf/cikm/hakan2000}} &  &  &  & {\checkmark} &  & & {DFT} & {MATLAB} & {C}\\	
			\cline{2-12}			 		 
			\cline{1-12}			 		 
			\multicolumn{1}{|c|}{\multirow{3}{*}{\rotatebox[origin=c]{90}{ Other }}}
			& \multicolumn{1}{|c|}{UCR Suite} & \cite{conf/kdd/Mueen2012} &  &  &  &  & \checkmark &  \checkmark &  & C &\\	
			\cline{2-12}			 		 
			& \multicolumn{1}{|c|}{MASS} &\cite{journal/dmkd/Yeh2017} &  &   &  &  & \checkmark &  & DFT &  C &\\	
			\cline{2-12}			 		 
			& \multicolumn{1}{|c|}{Stepwise} &\cite{conf/kdd/Karras2011} &  &  &  & \checkmark&  &   & DHWT & C & \\	
			\cline{1-12}			 		 
		\end{tabular*}
	} 
	\label{tab:multiprogram}
\end{table*}

\subsection{Similarity Search Methods}

In this study, we focus on algorithms that can produce exact results, and evaluate the ten methods outlined below (in chronological order).
The properties of these algorithms are also summarized in Table~\ref{tab:multiprogram}.



We also point out that there exist several techniques dedicated to approximate similarity search~\cite{conf/vldb/Gionis1999,conf/kdd/ColeSZ05,conf/vldb/sun14,journal/tpami/ge14, journal/corr/malkov16, conf/cvpr/yandex16}. 
	A thorough evaluation of all approximate methods deserves a study on its own, and we defer it to future work.

\noindent{\bf R*-tree.}
The R*-tree~\cite{conf/icmd/Beckmann1990} is a height-balanced spatial access method 
that partitions the data space into a hierarchy of nested overlapping rectangles.
Each leaf can contain either the raw data objects or pointers to those, along with the enclosing rectangle. 
Each intermediate node contains the minimum bounding rectangle that encompasses the rectangles of its children. 
Given a query $S_Q$, the R*-tree query answering algorithm visits all nodes whose rectangle intersects $S_Q$, starting from the root. Once a leaf is reached, all its data entries are returned. 
We tried multiple implementations of the R*-tree, and opted for the fastest~\cite{code/Marios2014}. 
We modified this code by adding support for PAA summaries.

\noindent{\bf M-tree.}
The M-tree~\cite{conf/vldb/Ciaccia1997} is a multidimensional, metric-space access method 
that uses hyper-spheres to divide the data entries according to their relative distances.
The leaves store data objects, and the internal nodes store routing objects; both store distances from each object to its parent. 
During query answering, the M-tree uses these distances to prune the search space. 
The triangle inequality that holds for metric distance functions guarantees correctness.
Apart from exact queries, it also supports $\epsilon$-approximate and $\delta$-$\epsilon$-approximate 
queries. 
We experimented with four different code bases:
two implementations that support bulk-loading~\cite{conf/ads/Ciaccia1998,journal/vldb/Dallachiesa2014}, the disk-aware mvptree~\cite{conf/sigmod/Bozkaya1997},
and a memory-resident implementation~\cite{journal/vldb/Dallachiesa2014}.
We report the results with the latter, because (despite our laborious efforts) it was the only one that scaled to datasets larger than 1GB. 
We modified it to use the same sampling technique as the original implementation~\cite{conf/ads/Ciaccia1998}, which chooses the number of initial samples based on the leaf size, minimum utilization, and dataset size. 

 
\noindent{\bf VA+file.}
The VA+file~\cite{conf/cikm/hakan2000} is an improvement of the VA-file method~\cite{conf/vldb/weber1998}. While both methods create a filter file containing quantization-based approximations of the high dimensional data, and share the same exact search algorithm, the VA+file does not assume that neighboring points (dimensions) in the sequence are uncorrelated. It thus improves the accuracy of the approximations by 
allocating bits per dimension in a non-uniform fashion, and partitioning each dimension using a k-means (instead of an equi-depth approach). We improved the efficiency of the original VA+file significantly by implementing it in C and modifying it to use DFT instead of KLT, since DFT is a very good approximation for KLT~\cite{conf/cikm/hakan2000} and is much more efficient~\cite{journal/acta/maccone2007}. 

\noindent{\bf Stepwise.}
The Stepwise method~\cite{conf/kdd/Karras2011} differentiates itself from indexing methods by storing DHWT summarizations vertically across multiple levels. 
This process happens in a pre-processing step.
When a query $S_Q$ arrives, the algorithm converts it to DWHT, and computes the distance between $S_Q$ and the DHWT of each candidate data series 
one level at a time,
using lower and upper bounding distances it filters out non-promising candidates.
When leaves are reached, the final refinement step consists of calculating the Euclidean distance between the raw representations of $S_Q$ and the candidate series.
We modified the original implementation to load the pre-computed sums in memory and answer one query at a time (instead of the batch query answering of the original implementation). 
We also slightly improved memory management to address swapping issues that occurred with the out-of-memory datasets. 



\noindent{\bf SFA trie.}
The SFA approach~\cite{journal/edbt/Schafer2012} first summarizes the series using SFA of length 1 and builds a trie with a fanout equal to the alphabet size on top of them. 
As leaves reach their capacity and split, the length of the SFA word for each series in the leaf is increased by one, and the series are redistributed among the new nodes. 
The maximum resolution is the number of DFT coefficients given as a parameter. 
SFA implements lower-bounding to prune the search space, as well as a bulk-loading algorithm.
We re-implemented SFA in C, optimized its memory management, and improved the sampling and buffering schemes. 
This resulted in a significantly faster implementation than the original one in Java.

\noindent{\bf UCR Suite.}
The UCR Suite~\cite{conf/kdd/Mueen2012} is an optimized sequential scan algorithm 
for exact subsequence matching. 
We adapted the original algorithm to support exact whole matching. 

\noindent{\bf DSTree.}
The DSTree~\cite{conf/vldb/Wang2013} approach uses the EAPCA summarization technique,
which allows, during node splitting, the resolution of a summarization to increase along two dimensions: vertically and horizontally.
(Instead, 
SAX-based indexes allow horizontal splitting by adding a breakpoint to the y-axis, and SFA allows vertical splitting by adding a new DFT coefficient.) 
In addition to a lower bounding distance, the DSTree also supports an upper bounding distance. 
It uses both distances to 
determine the optimal splitting policy for each node. 
We reimplemented the DSTree algorithm in C and we optimized its buffering and memory management, 
improving the performance of the algorithm by a factor of 4, compared to the original implementation (in Java).

\ifJournal
\begin{figure}[t]
	\captionsetup{justification=centering}
	\captionsetup[subfigure]{justification=centering}
	\begin{subfigure}{0.49\columnwidth}
	\centering
	\includegraphics[width=\columnwidth]{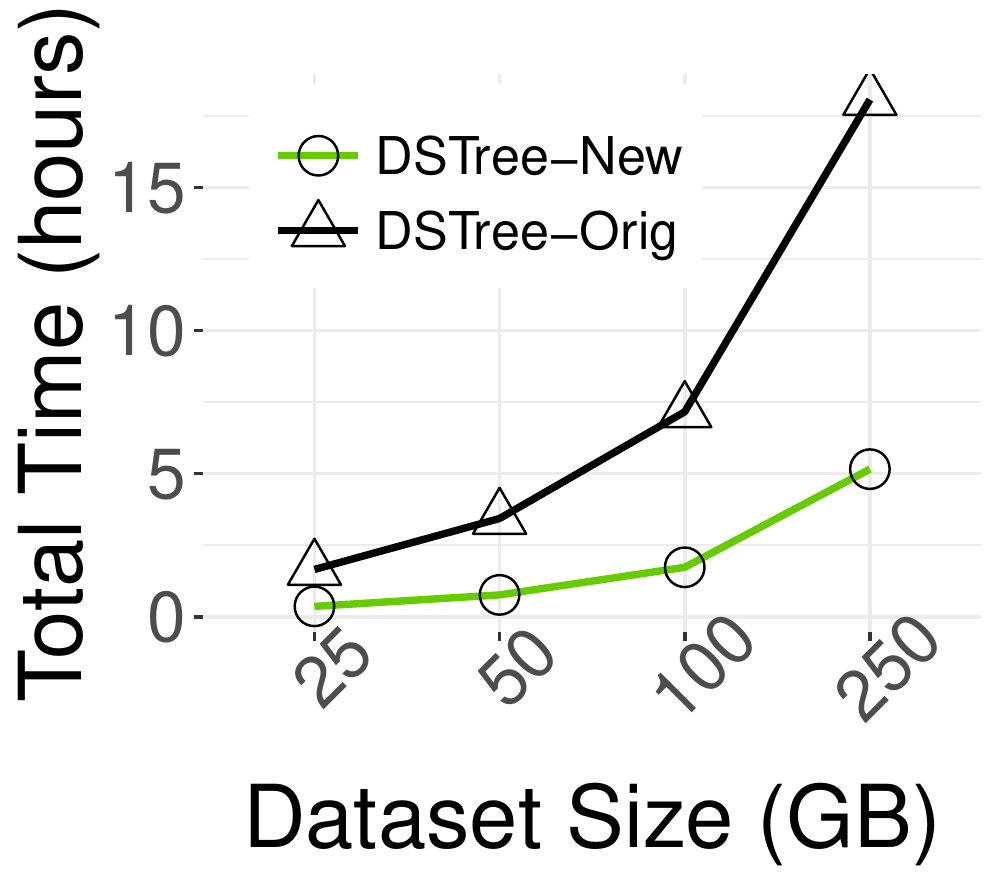}
	\caption{Total Time (Indexing and Answering 100 Exact Queries)}
	\label{fig:dstree:orig:new:combined}
	\end{subfigure}u
	\begin{subfigure}{0.49\columnwidth}
		\centering
		\includegraphics[width=\columnwidth]{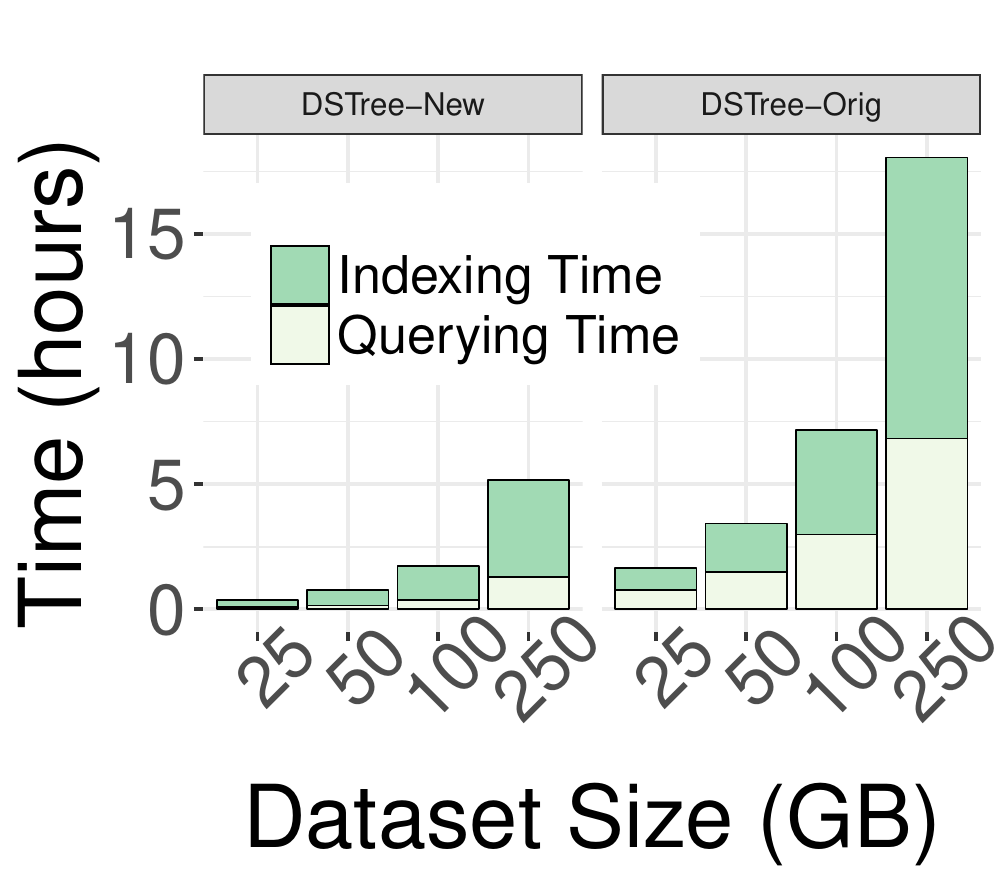}
		\caption{Detailed Times for Indexing and Answering 100 Exact Queries}
		\label{fig:dstree:orig:new:detailed}
	\end{subfigure}
\caption{DSTree Implementation Optimization}
\label{fig:dstree:orig:new}
}
\end{figure}
\fi

\noindent{\bf iSAX2+.}
The iSAX family of indexes has undergone several improvements. 
The iSAX 2.0 index~\cite{conf/icdm/Camerra2010} improved the splitting policy and added bulk-loading support to the original iSAX index~\cite{conf/kdd/shieh1998}.
iSAX2+~\cite{journal/kais/Camerra2014} further optimized bulk-loading. 
In the literature, competing approaches have either compared to iSAX, or iSAX 2.0. 
This is the first time that iSAX2+ is compared to other exact data series indexes. 
The index supports ng-approximate and exact query answering.
We reimplemented the original iSAX2+ algorithm from scratch using C, and optimized its memory management, leading to significant performance improvements.



\noindent{\bf ADS+.}
ADS+~\cite{journal/vldb/Zoumpatianos2016} is the first query adaptive data series index. 
It first builds an index tree structure using only the iSAX summarizations of the raw data, and then adaptively constructs the leaves and incorporates the raw data during query answering. 
For exact query answering, the SIMS algorithm is proposed. It first performs a fast ng-approximate search in the tree in order to acquire an initial best-so-far (bsf) distance, then prunes the search space by using the bsf and the lower bounds between the query and all iSAX summaries. Using that, it performs a skip-sequential search on the raw data that were not pruned.
In all our experiments involving ADS+ we use the SIMS algorithm for exact similarity search.
ADS-FULL is a non-adaptive version of ADS, that builds a full index using a double pass on the data.


\noindent{\bf MASS.}
MASS~\cite{journal/dmkd/Yeh2017} is an exact subsequence matching algorithm, which computes 
the distance between a query, $S_Q$, and every subsequence in the series, using 
the dot product of the DFT transforms of the series and the reverse of $S_Q$.
We adapted it to perform exact whole matching queries. 

\section{Experimental Evaluation}
\label{sec:experiments}

In order to provide an unbiased evaluation, 
we re-implemented 
in C all methods whose original language was other than C/C++. 
Our new implementations are more efficient (in space and time) than the original ones on all datasets we tested. 
All methods use single precision values, 
and the methods based on fixed summarizations use 16 segments/coefficients.
The same set of known optimizations for data series processing are applied to all methods.
All results, source codes, datasets and plots are available in~\cite{url/DSSeval}.



\subsection{Environment}
\label{subsec:environment}
All methods were compiled with GCC 6.2.0 under Ubuntu Linux 16.04.2 with level 2 optimization. Experiments were run on two different machines.
The first machine, called \emph{HDD}, is a server with two Intel Xeon E5-2650 v4 2.2GHz CPUs,
75GB\footnote{We used GRUB to limit the amount of RAM, so that all methods are forced to use the disk. Note that GRUB prevents the operating system from using the rest of the RAM as a file cache, which is what we wanted for our experiments.} of RAM, 
and 10.8TB (6 x 1.8TB) 10K RPM SAS hard drives 
in RAID0. 
The throughput of the RAID0 array is 1290 MB/sec.
The second machine, called \emph{SSD}, is a server with two Intel Xeon E5-2650 v4 2.2Ghz CPUs, 
75GB of RAM, 
and 3.2TB (2 x 1.6TB) SATA2 SSD in RAID0.
The throughput of the RAID0 array is 330 MB/sec.
All our algorithms are single-core implementations.

\subsection{Experimental Setup}
\label{subsec:framework}

\noindent{\textbf{Scope.}}
This work concentrates on exact whole-matching (WM) 1-NN queries.
Extending our experimental framework to cover $r$-range queries, subsequence matching and approximate query answering is part of our future work.

\noindent{\textbf{Algorithms.}}
This experimental study covers the ten methods described in Section~\ref{sec:approaches}, which all have native-support for Euclidean distance.
Our baseline is the Euclidean distance version of the UCR Suite~\cite{conf/kdd/Mueen2012}.
This is a set of techniques for performing very fast similarity computation scans.
These optimizations include: a) avoiding the computation of square root on Euclidean distance, b) early abandoning of Euclidean distance calculations, and c) reordering early abandoning on normalized data\footnote{Early abandoning of Z-normalization is not used since all datasets were normalized in advance.}.
We used these optimizations on all the methods that we examined.

\noindent{\textbf{Datasets.}}
Experiments were conducted using both synthetic and real datasets.
Synthetic data series were generated as random-walks (i.e., cumulative sums) of steps that follow a Gaussian distribution (0,1).
This type of data has been extensively used in the past~\cite{conf/sigmod/Faloutsos1994,journal/kais/Camerra2014,conf/kdd/Zoumpatianos2015}, and it is claimed to model the distribution of stock market prices~\cite{conf/sigmod/Faloutsos1994}.

Our four real datasets come from the domains of seismology, astronomy, neuroscience and image processing. 
The seismic dataset, \emph{Seismic}, was obtained from the IRIS Seismic Data Access archive~\cite{url/data/seismic}. It contains seismic instrument recording from thousands of stations worldwide and consists of 100 million data series of size 256. The astronomy dataset, \emph{Astro}, represents celestial objects and was obtained from~\cite{journal/aa/soldi2014}.
The dataset consists of 100 million data series of size 256. The neuroscience dataset, \emph{SALD}, obtained from~\cite{url/data/eeg} represents MRI data, including 200 million data series of size 128. The image processing dataset, \emph{Deep1B}, retrieved from~\cite{url/data/deep1b}, contains 267 million Deep1B vectors of size 96 extracted from the last layers of a convolutional neural network.
All of our real datasets are of size 100 GB. 
In the rest of the paper, the size of each dataset is given in GB instead of the number of data series. 
Overall, in our experiments, we use datasets of sizes between 25-1000GB.

\noindent{\textbf{Queries.}}
All our query workloads, unless otherwise stated, include 100 query series.
For synthetic datasets, we use two types of workloads: $Synth$-$Rand$ queries are produced using the same random-walk generator (with a different seed\footnote{All seeds can be found in~\cite{url/DSSeval}.}), while $Synth$-$Ctrl$ queries are created by extracting data series from the input data set and adding progressively larger amounts of noise, in order to control the difficulty of each query (more difficult queries tend to be less similar to their nearest neighbor~\cite{johannesjoural2018}).
For the real datasets, query workloads are also generated by adding progressively larger amounts of noise to data series extracted from the raw data, and we name them with the suffix -$Ctrl$.
For the Deep1B dataset, we additionally include a real workload that came with the original dataset; we refer to it as $Deep$-$Orig$.

\noindent{\textbf{Scenarios.}}
The experimental framework consists of three scenarios: parametrization, evaluation and comparison.
In parametrization (\S\ref{ssec:parametrization}), the optimal parameters for each method are identified.
In evaluation (\S\ref{ssec:evaluation}), the scalability and search efficiency for each method is evaluated under varying dataset sizes and data series lengths.
Finally, in comparison (\S\ref{ssec:comparison}), methods are compared together according to the following criteria:
a) scalability and search efficiency on more complex query workloads and more varied and larger datasets,
b) memory and disk footprint,
c) pruning ratio, and
d) tightness of the lower bound.

\noindent{\textbf{Measures}}
The measures we use are the following.

\noindent1. For scalability and search efficiency, we use two measures: \emph{wall clock time} and the \emph{number of random disk accesses}. \emph{Wall clock time} is used to measure input, output and total elapsed times. Then CPU time is calculated as the difference between the total time and I/O time. The \emph{number of random disk accesses} is measured for indexes. One random disk access corresponds to one leaf access for all indexes, except for the skip-sequential access method ADS+, for which one random disk access corresponds to one skip. As will be evident in the results, our measure of random disk accesses provides a good insight into the actual performance of indexes, even though we do not account for details such as caching, the number of disk pages occupied by a leaf and the numbers of leaves in contiguous disk blocks.

\noindent2. For footprint, the measures used are: \emph{total number of nodes}, \emph{number of leaf nodes}, \emph{memory size}, \emph{disk size}, \emph{leaf nodes fill factor} and \emph{leaf depth}.

\noindent3. We also consider the pruning ratio $P$, which has been widely used in the data series literature \cite{journal/kais/Keogh2001,journal/edbt/Schafer2012,conf/vldb/Wang2013,conf/vldb/Ding2008, conf/kdd/Karras2011} as an implementation-independent measure to compare the effectiveness of an index. It is defined as follows: 
\[P \ = \ 1-\frac{\# \ of \ Raw \ Data \ Series \ Examined}{\# \ of \ Data \ Series \ In \ Dataset} \]
Pruning ratio is a good indicator of the number of sequential I/Os incurred. However, since relevant data series are usually spread out on disk, it should be considered along with the number of random disk accesses (seeks) performed.

\noindent4. The \emph{tightness of the lower bound}, $TLB$ has been used in the literature as an implementation independent measure in various different forms~\cite{conf/kdd/shieh1998,journal/edbt/Schafer2012,journal/dmkd/Wang2013}.
         In this work we use the following version of the $TLB$ measure that better captures the performance of indexes:
    	   	\[TLB \ = \ \frac{Lower \ Bounding \ Distance (Q\prime, N)}{ Average \ True \ Distance(Q, N)}  \]
    	   Where $Q$ is the query, $Q\prime$ is the representation of $Q$ using the segmentation of a given leaf node $N$, and the average true distance between the query $Q$ and node $N$ is the average Euclidean distance between $Q$ and all data series in $N$. We report the average over all leaf nodes for all 100 queries.

\ifJournal

\noindent5. The \emph{accuracy of the approximate search} is measured by $\epsilon_{eff}$ defined as follows.
Given a query data series $S_Q$, an exact match $S_C$ and an approximate match $S_{C_{approx}}$, the \emph{effective error, $\epsilon_{\text{eff}}$} of $S_{C_{approx}}$ is:
\[\epsilon_{\text{eff}} = \frac {d(S_Q,S_{C_{approx}}) - d(S_Q, S_C)} {d(S_Q,S_C)} \]

Note that $\epsilon_{eff}$ and $TLB$ are different: $TLB$ measures how close the lower bounding distance $d_{lb}$ is to the real distance $d_{exact}$, whereas $\epsilon_{eff}$ measures how close the approximate distance $d_{approx}$ is to $d_{exact}$. 
The following inequality holds: 
$d_{lb} \ \leq  \ d_{exact} \leq d_{approx}$.
\fi

\ifFuture
{\color{blue}{\bf \\
		SCOPE: 	These items are only relevant for approximate and probabilistic evaluations
	}
	{\it
		\item Index exact answering methods are compared to approximate methods in two different ways:
		\begin{enumerate}
			\item The relative error of the approximate result to the exact result is measured. In the case of $k$-NN queries, the rank of the approximate result is also reported in order to evaluate the accuracy of the method used.
			\item When the exact method is actually a dual method, i.e., it answers both exact and approximate queries, the relative error of the the result returned by the approximate methods is compared to the result returned by the approximate answering of dual methods.
			\item The optimal parameters are evaluated for probabilistic methods (LSH and STATstream).
		\end{enumerate}
	}}
\fi

\begin{figure*}[tb]
	\captionsetup{justification=centering}
 \begin{subfigure}{\textwidth}
	\centering
  	\hspace{0.5cm}
  	\includegraphics[width=0.5\textwidth]{{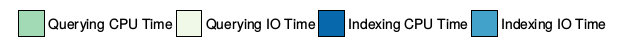}}
  \end{subfigure}	\captionsetup[subfigure]{justification=centering}
\hspace*{\fill} 
\hspace*{\fill} 

\begin{subfigure}{0.16\textwidth}
	\centering
	\includegraphics[width=\textwidth] {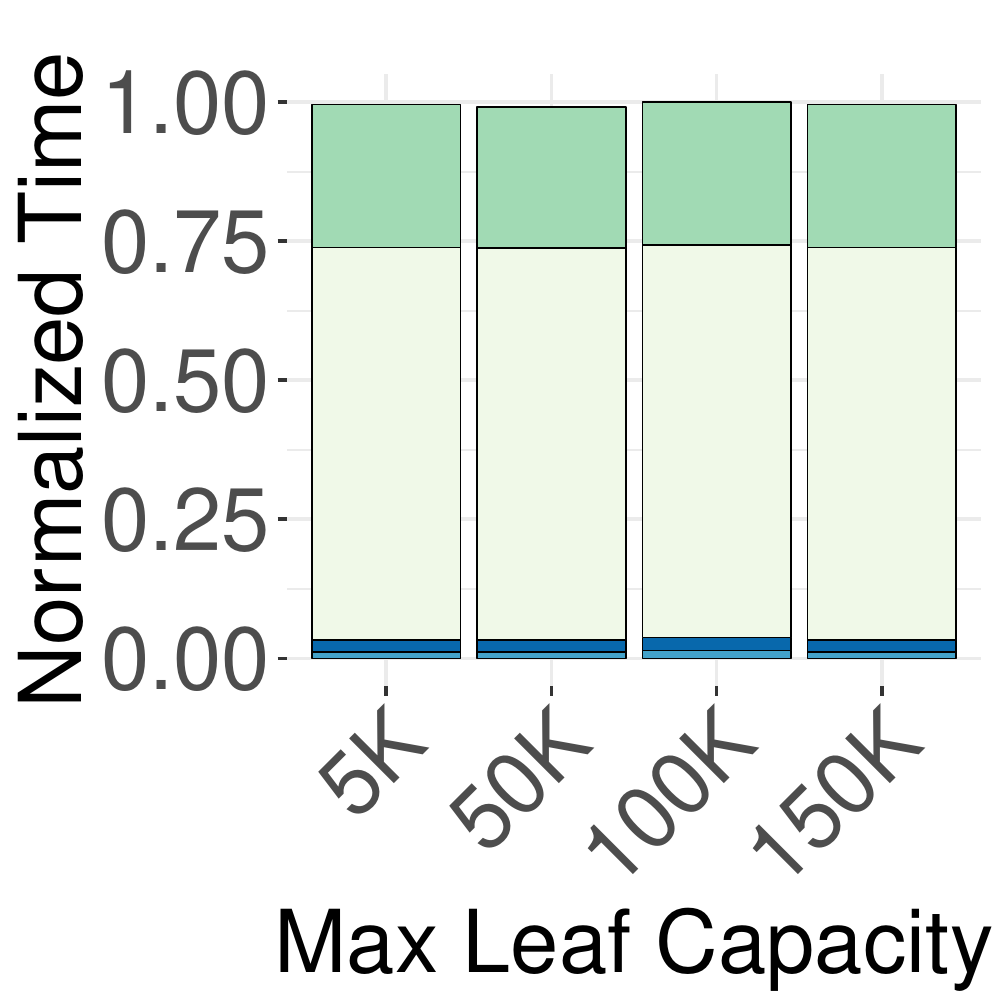}
	\caption{ADS+\\
		Dataset = 100GB}
	\label{fig:exact:leafsize:time:idxproc:ADS+}
\end{subfigure}
\begin{subfigure}{0.16\textwidth}
	\centering
	\includegraphics[width=\textwidth]{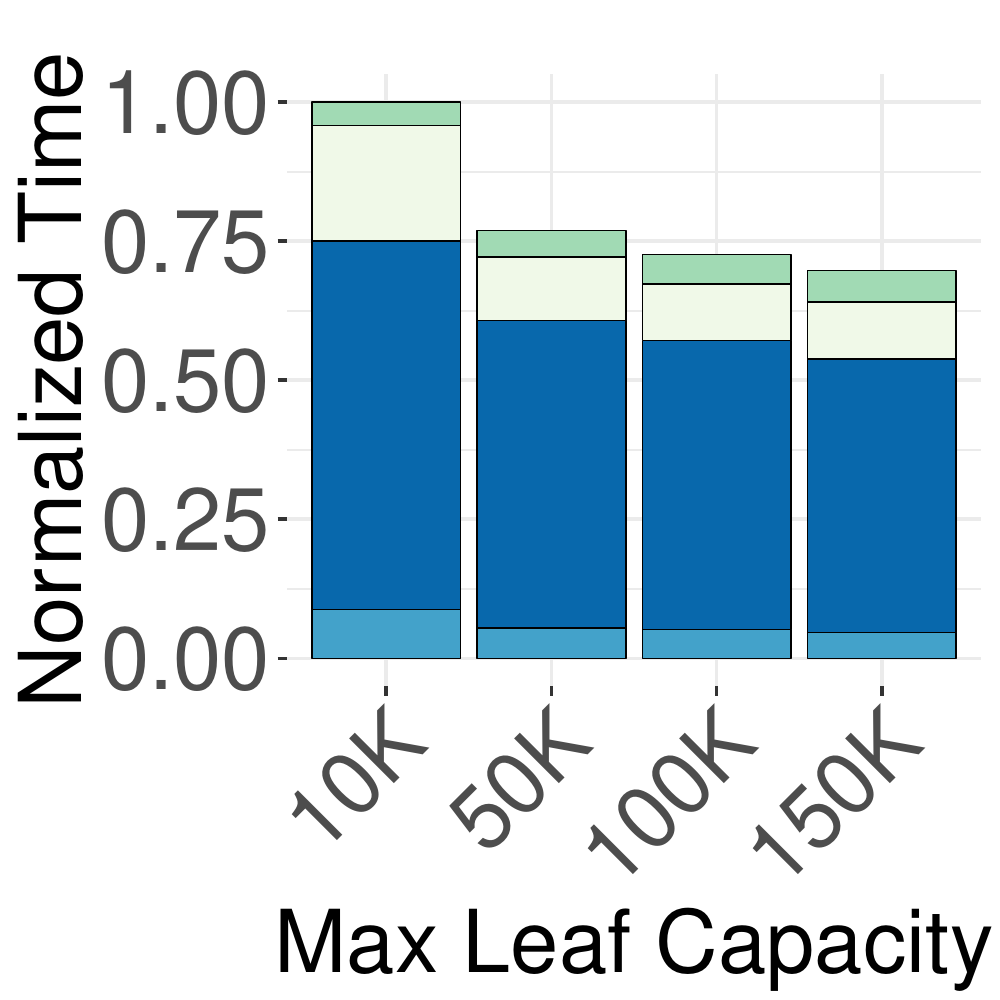}
	\caption{DSTree\\
	Dataset = 100GB}	
	\label{fig:exact:leafsize:time:idxproc:dstree}
\end{subfigure}
\begin{subfigure}{0.16\textwidth}
	\centering
	\includegraphics[width=\textwidth] {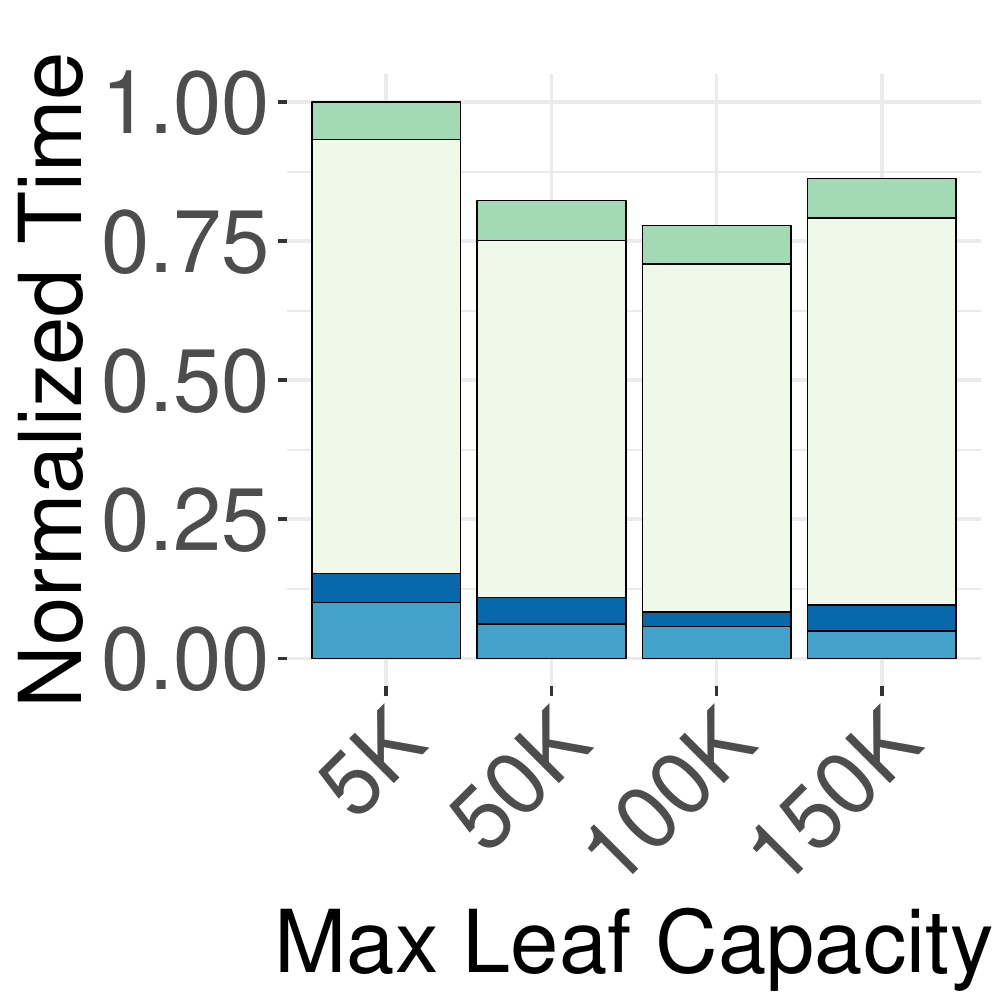}
	\caption{iSAX2+\\
	Dataset = 100GB}	
	\label{fig:exact:leafsize:time:idxproc:iSAX2+}
\end{subfigure}
\begin{subfigure}{0.16\textwidth}
	\centering
	\includegraphics[width=\textwidth] {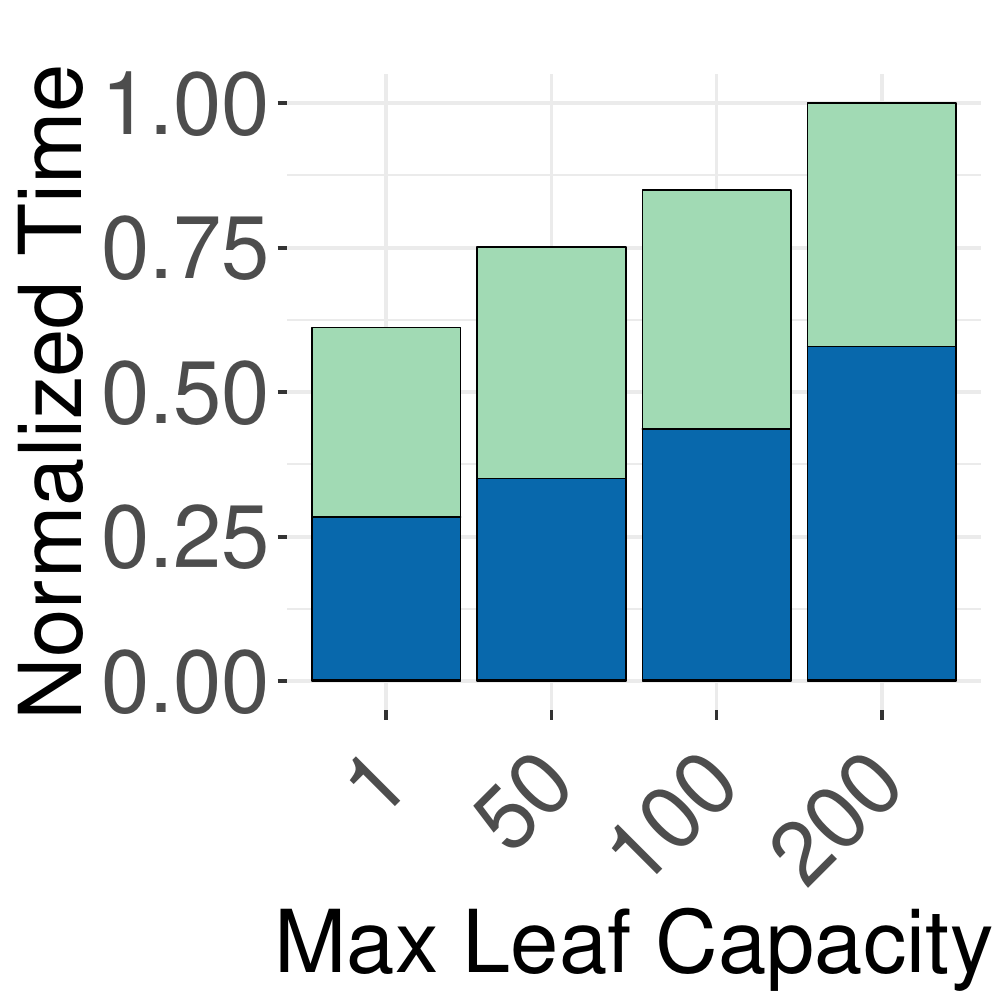}
	\caption{M-tree \\
	Dataset = 50GB}			
	\label{fig:exact:leafsize:time:idxproc:mtree}
\end{subfigure}
\begin{subfigure}{0.16\textwidth}
	\centering
	\includegraphics[width=\textwidth] {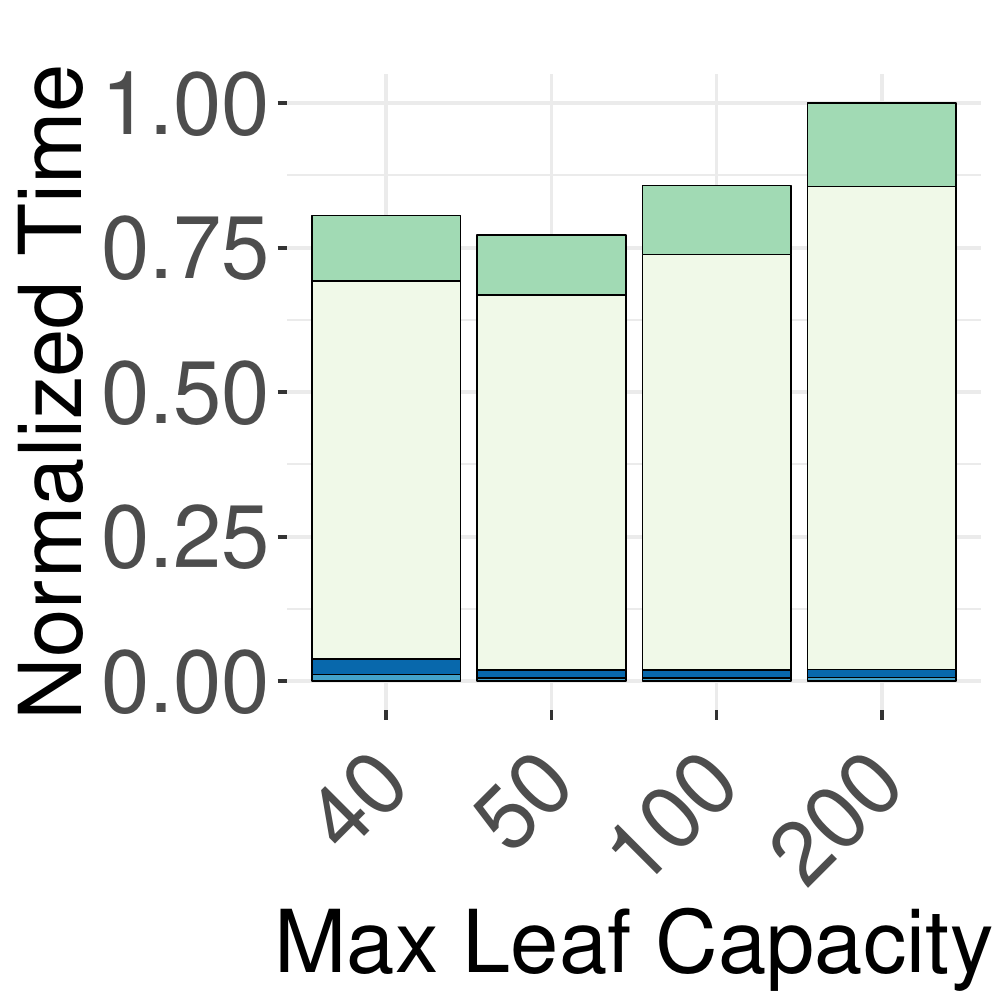}
	\caption{R*-tree\\
	Dataset = 50GB}			
	\label{fig:exact:leafsize:time:idxproc:rstree}
\end{subfigure}
\begin{subfigure}{0.16\textwidth}
	\centering
	\includegraphics[width=\textwidth] {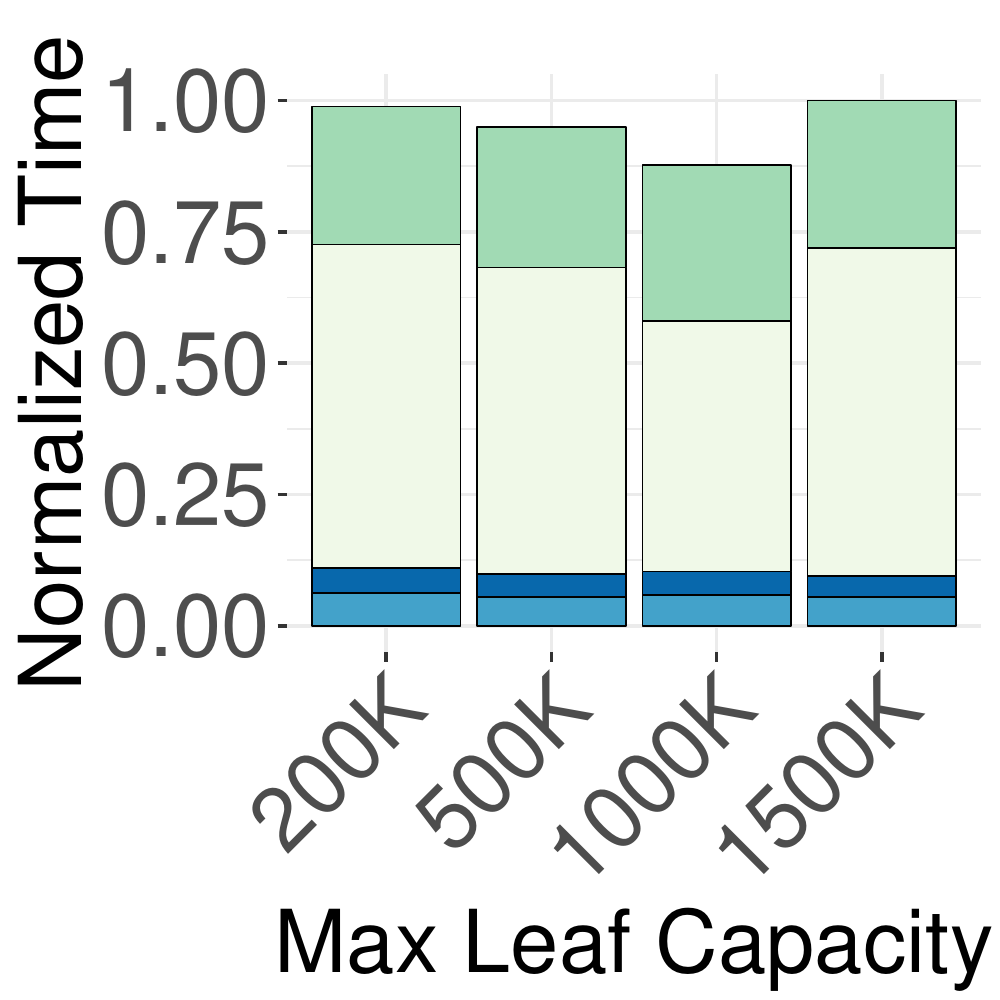}
	\caption{SFA trie \\
		Dataset = 100GB}			
	\label{fig:exact:leafsize:time:idxproc:sfa}
\end{subfigure}

\caption{Leaf size parametrization
}
\label{fig:exact:leafsize:time:idxproc}
\end{figure*}
\noindent{\textbf{Procedure.}}
Unless otherwise stated, experiments refer to answering 100 exact queries.
Experiments with query workloads of 10,000 queries report extrapolated values.
The extrapolation consists of discarding the best and worst five queries (of the original 100) in terms of total execution time, and multiplying the average of the 90 remaining queries by 10,000.  
Experiments involving an indexing method include a first step of building the index (or re-organizing the data as in the case of Stepwise).
Caches are fully cleared before each experiment.
During each experiment, the caches are warm, i.e., not cleared between indexing/preprocessing and query answering, nor after each query.

\ifJournal
{\color{blue}{\bf \\
 SCOPE: These items are only relevant for k-NN and range queries \\
	}
	{\it
	vary the radius \\
	vary k in kNN \\
}}
\fi

\subsection{Results}
\label{subsec:results}


\subsubsection{Parametrization}
\label{ssec:parametrization}

We start our experimentation by fine tuning each method.
Methods that do not support parameters are ran with their default values.
The methods that support parameters are ADS+, DSTree, iSAX2+, M-tree, R*-tree and SFA trie.
We use a synthetic dataset of 100GB with data series of length 256.
The only exceptions are M-tree and R*-tree, where we parametrize using 50GB, since experiments with 100GB, or above, take more than 24 hours to complete.

The most critical parameter for these methods is the leaf threshold, i.e., the maximum number of data series that an index leaf can hold.
We thus vary the leaf size and study the tradeoffs of index construction and query answering for each method.
Figure~\ref{fig:exact:leafsize:time:idxproc} reports indexing and querying execution times for each method, normalized by the largest total cost.
The ratio is broken down into CPU and I/O times.
Figure~\ref{fig:exact:leafsize:time:idxproc:ADS+} shows that the performance of ADS+ is the same across leaf sizes.
The leaf size affects indexing time, but not query answering.
This is not visible in the figure, because index construction time is minimal compared to query answering time.  
This behavior is expected, since ADS+ is an adaptive index,
which during querying splits the nodes until a minimal leaf size is reached.
For M-tree, larger leaves cause both indexing and querying times to deteriorate.
For all other methods, increasing the leaf size improves indexing time (because trees are smaller) and querying time (because several series are read together), but once the leaf size goes beyond the optimal leaf size, querying slows down (because some series are unnecessarily read and processed).
For DSTree,
the experiments execution logs indicate that querying is faster with the 100K leaf size.
The optimal leaf size for iSAX2+ is also 100K, for SFA is 1M, and for M-tree and R*-tree are 1 and 50, respectively.

SFA takes two other parameters: the alphabet size and the binning method. We ran experiments with both equi-depth and equi-width binning, and alphabet sizes from 8 (default value), to 256 (default alphabet size of iSAX2+ and ADS+).
Alphabet size 8 and equi-depth binning provided the best performance and were thus used for subsequent experiments.


Some of the evaluated methods also use internal buffers to manage raw data that do not fit in memory during index building and query processing.
We ran experiments varying these buffer sizes from 5GB to 60GB.
The maximum was set to 60GB (recall that total RAM was 75GB).
All methods benefit from a larger buffer size except ADS+.
This is because a smaller buffer size allows the OS to use extra memory for file caching during query processing, since ADS+ accesses the raw data file directly.

\subsubsection{\textbf{Evaluation of Individual Methods}}
\label{ssec:evaluation}
We now evaluate the indexing and search efficiency of the methods
by varying the dataset size. 
We used two datasets of size 25GB and 50GB that fit in memory and two datasets of size 100GB and 250GB that do not fit in memory (total RAM was 75GB), with the $Synth$-$Rand$ query workload.

\begin{figure*}[tb]
	\captionsetup{justification=centering}
	\captionsetup[subfigure]{justification=centering}
	\begin{subfigure}{0.16\textwidth}
		\centering
		\includegraphics[width=\textwidth]{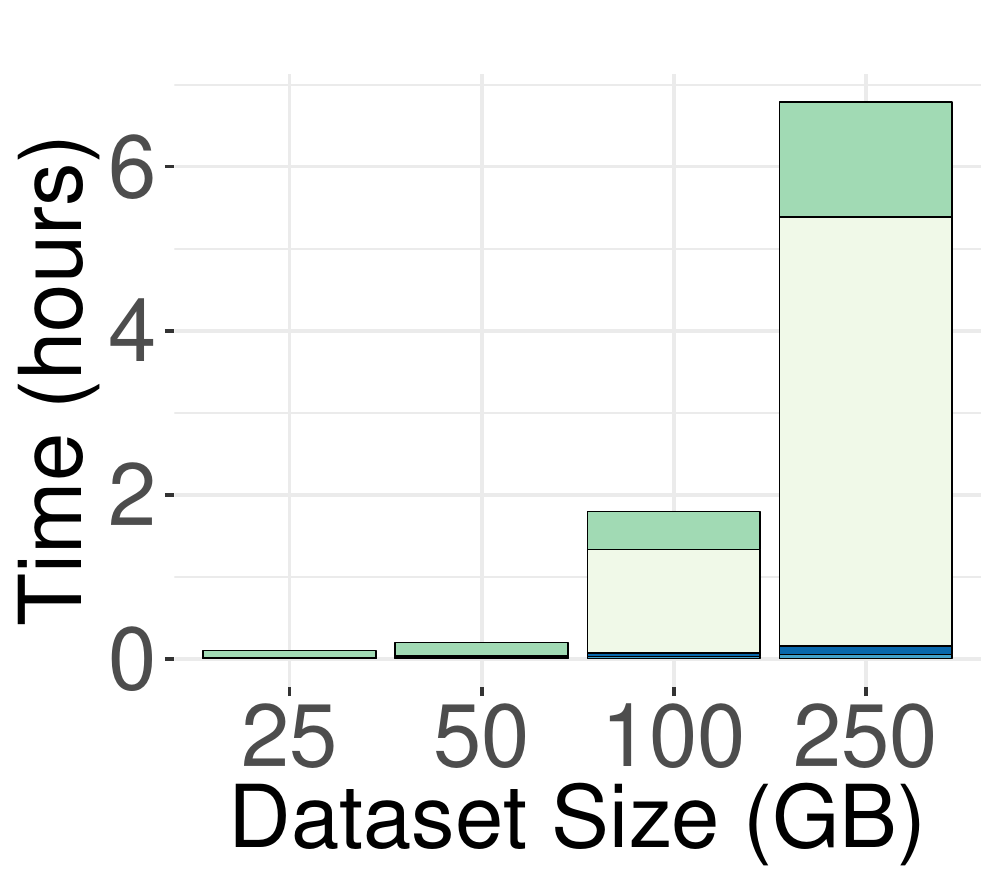}
		\caption{ADS+}
		\label{fig:exact:datasize:time:idxproc:cache:ADS+:1}
	\end{subfigure}
	\begin{subfigure}{0.16\textwidth}
		\centering
		\includegraphics[width=\textwidth]{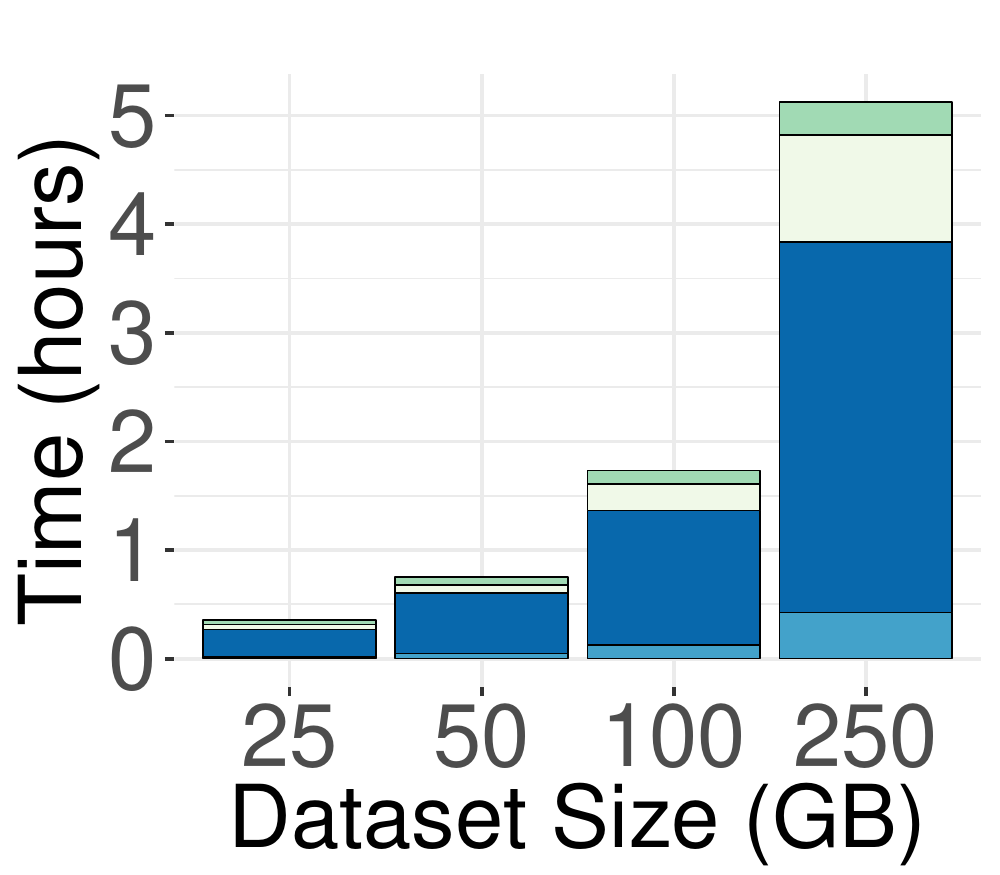}
		\caption{DSTree}
		\label{fig:exact:datasize:time:idxproc:cache:dstree}
	\end{subfigure}
	\begin{subfigure}{0.16\textwidth}
		\centering
		\includegraphics[width=\textwidth]{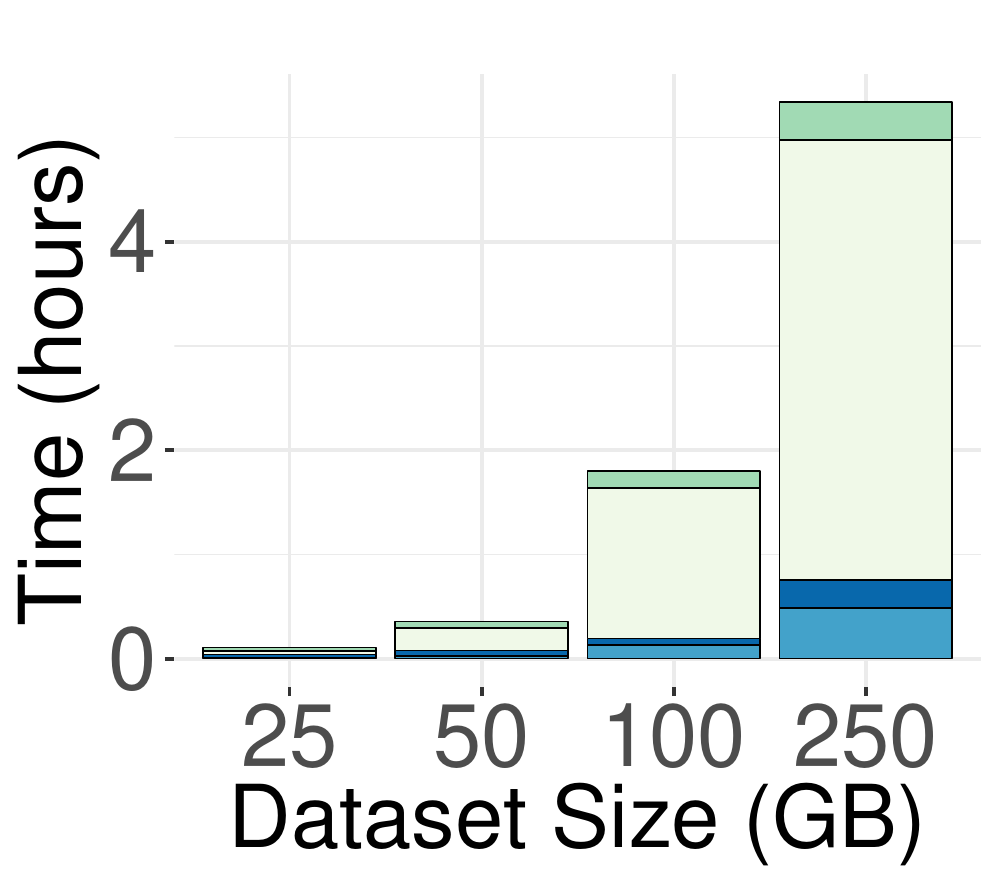}
		\caption{iSAX2+}
		\label{fig:exact:datasize:time:idxproc:cache:iSAX2+}
	\end{subfigure}
	\begin{subfigure}{0.16\textwidth}
		\centering
		\includegraphics[width=\textwidth]{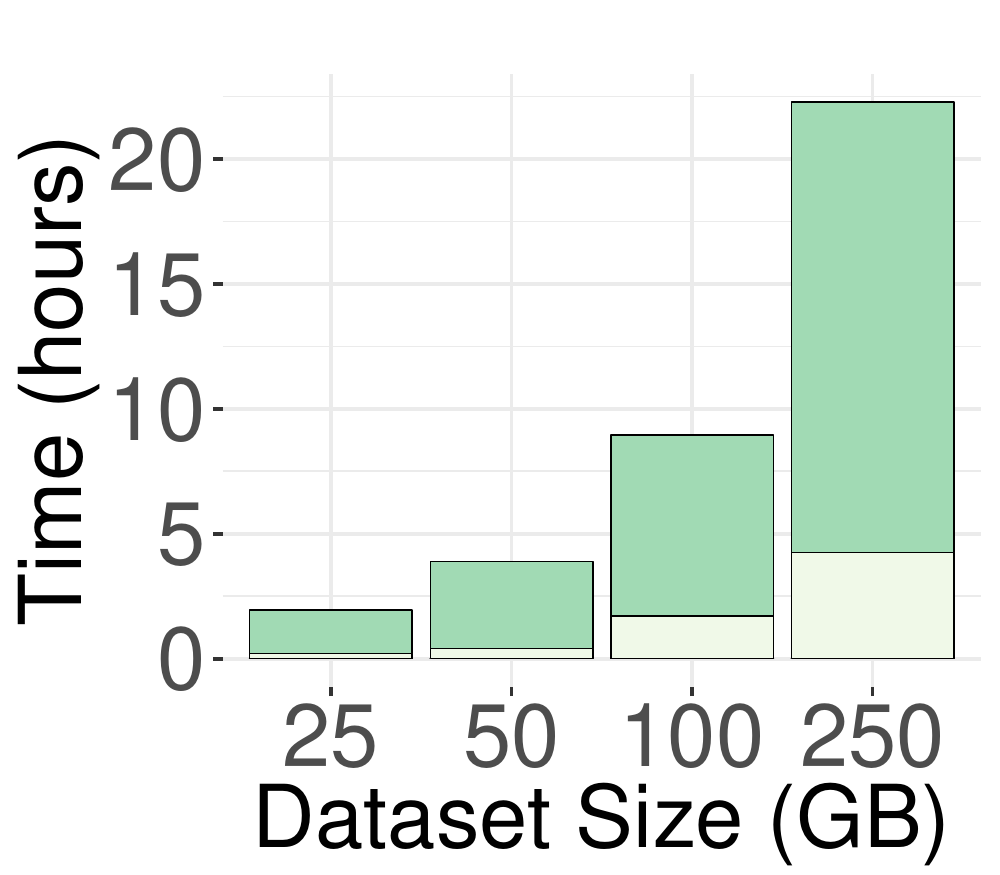}
		\caption{MASS}
		\label{fig:exact:datasize:time:idxproc:cache:mass}
	\end{subfigure}
	\begin{subfigure}{0.16\textwidth}
		\centering
		\includegraphics[width=\textwidth]{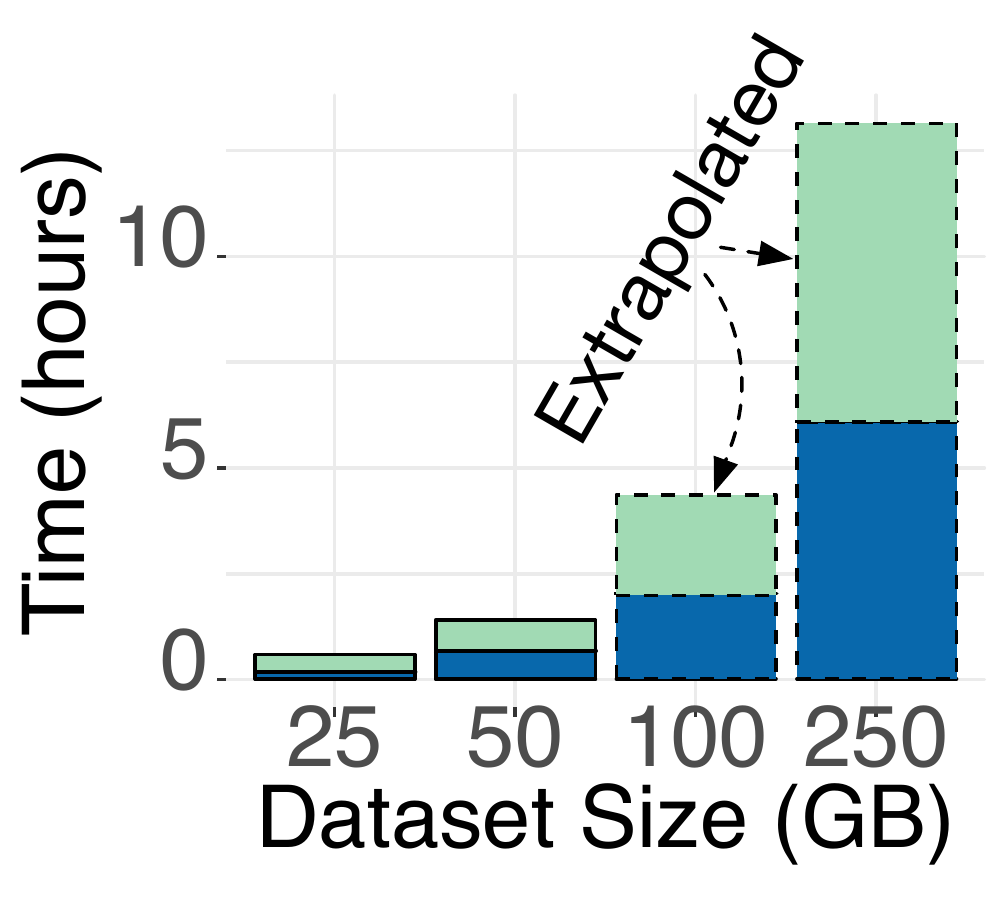}
		\caption{M-tree}
		\label{fig:exact:datasize:time:idxproc:cache:mtree}
	\end{subfigure}
	\begin{subfigure}{0.16\textwidth}
		\centering
		\includegraphics[width=\textwidth]{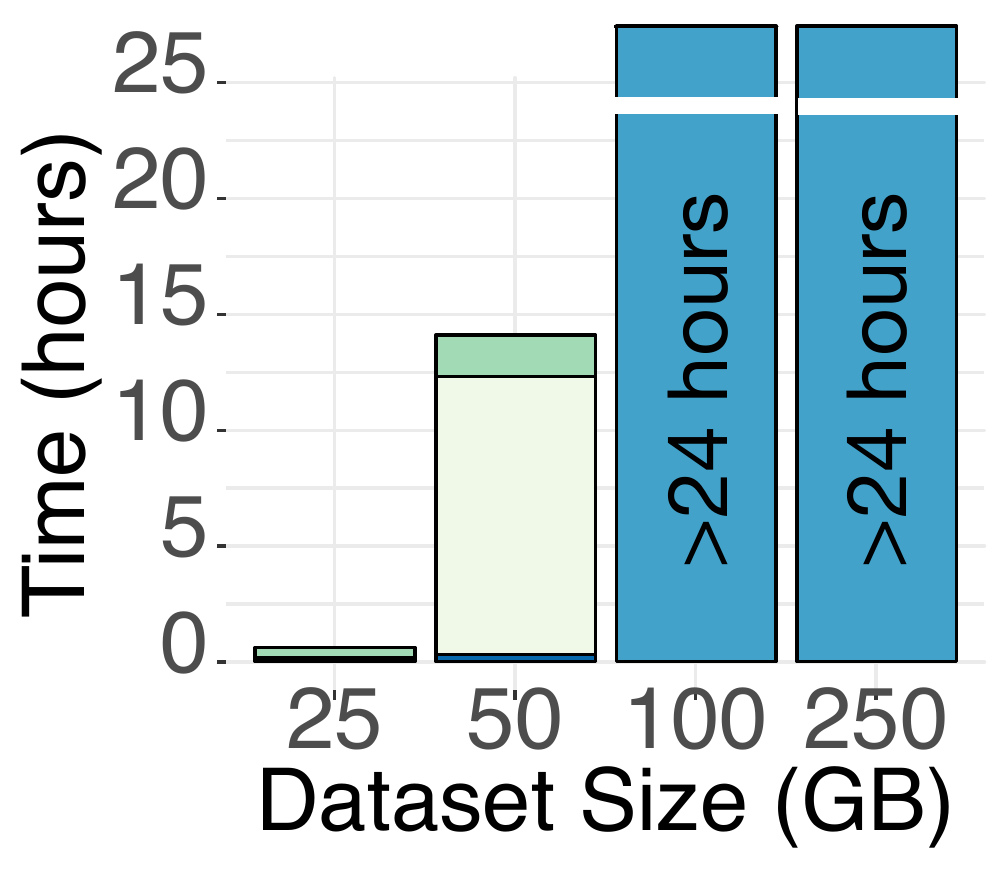}
		\caption{R*-tree}
		\label{fig:exact:datasize:time:idxproc:cache:rstree}
	\end{subfigure}
	\hspace*{0.2cm}
	\begin{subfigure}{0.17\textwidth}
		\centering
		\includegraphics[width=\textwidth]{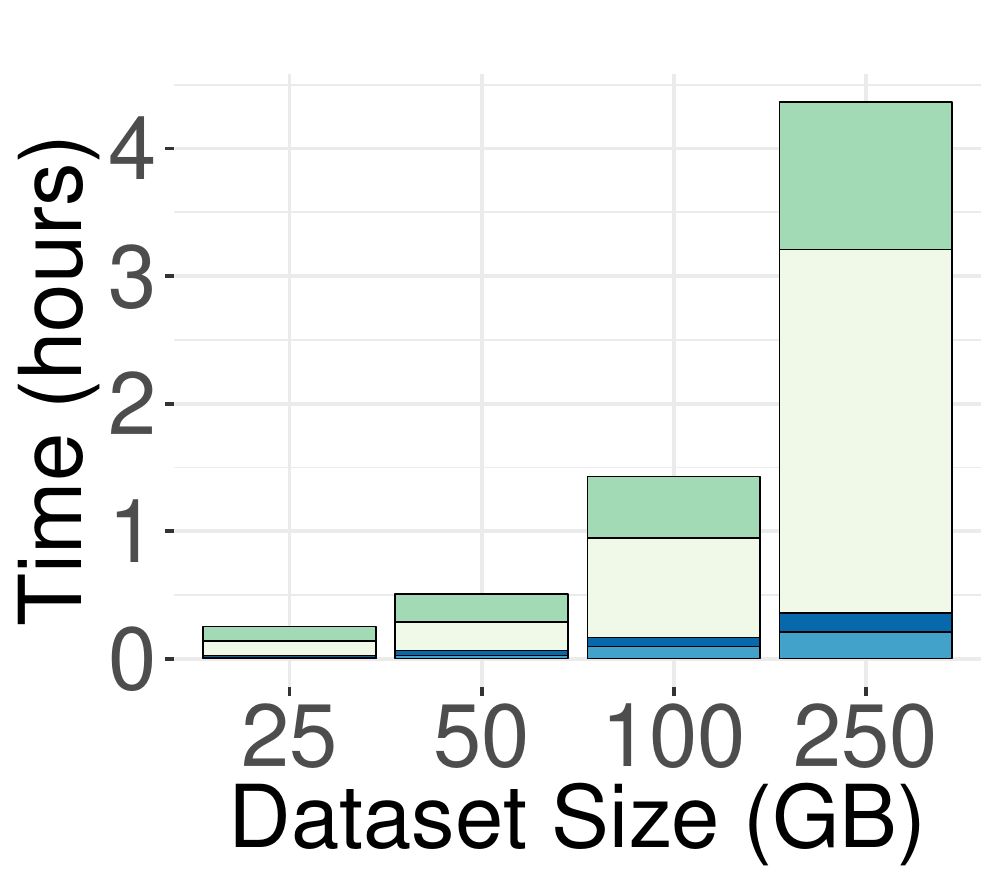}
		\caption{SFA trie}
		\label{fig:exact:datasize:time:idxproc:cache:sfa}
	\end{subfigure}
	\hspace*{0.2cm}
	\begin{subfigure}{0.17\textwidth}
		\centering
		\includegraphics[width=\textwidth]{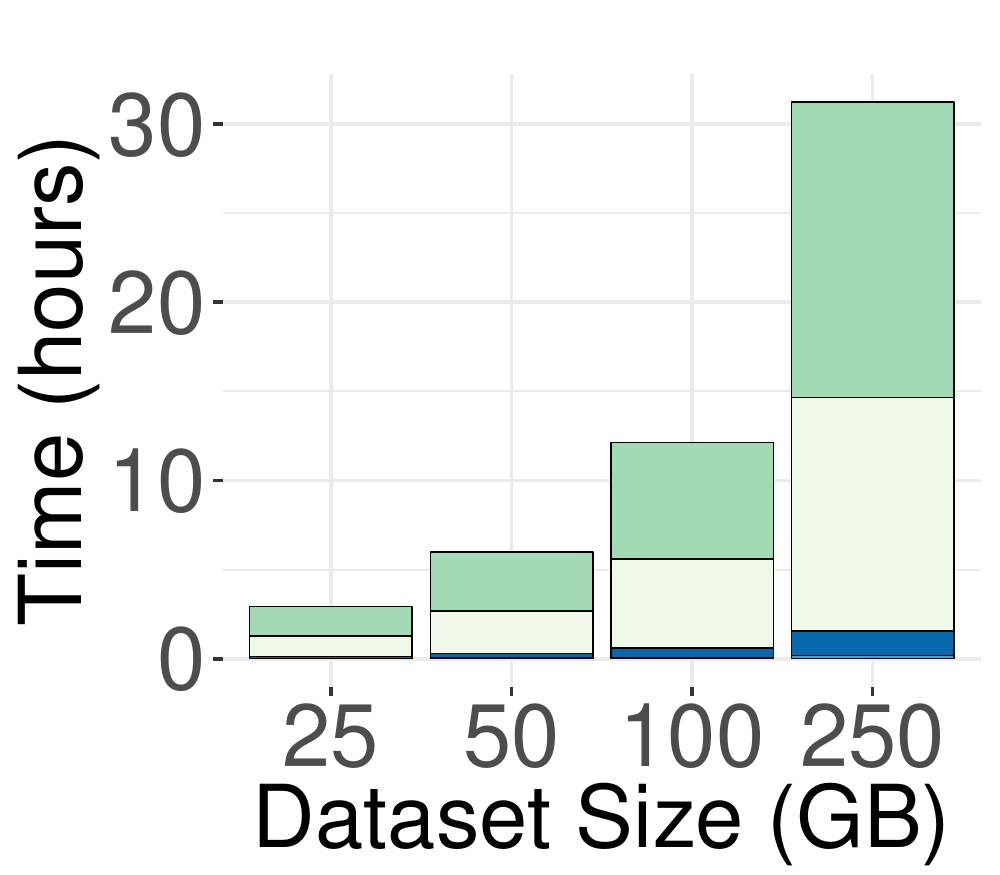}
		\caption{Stepwise}
		\label{fig:exact:datasize:time:idxproc:cache:Stepwise}
	\end{subfigure}
	\hspace*{0.2cm}
	\begin{subfigure}{0.17\textwidth}
		\centering
		\includegraphics[width=\textwidth]{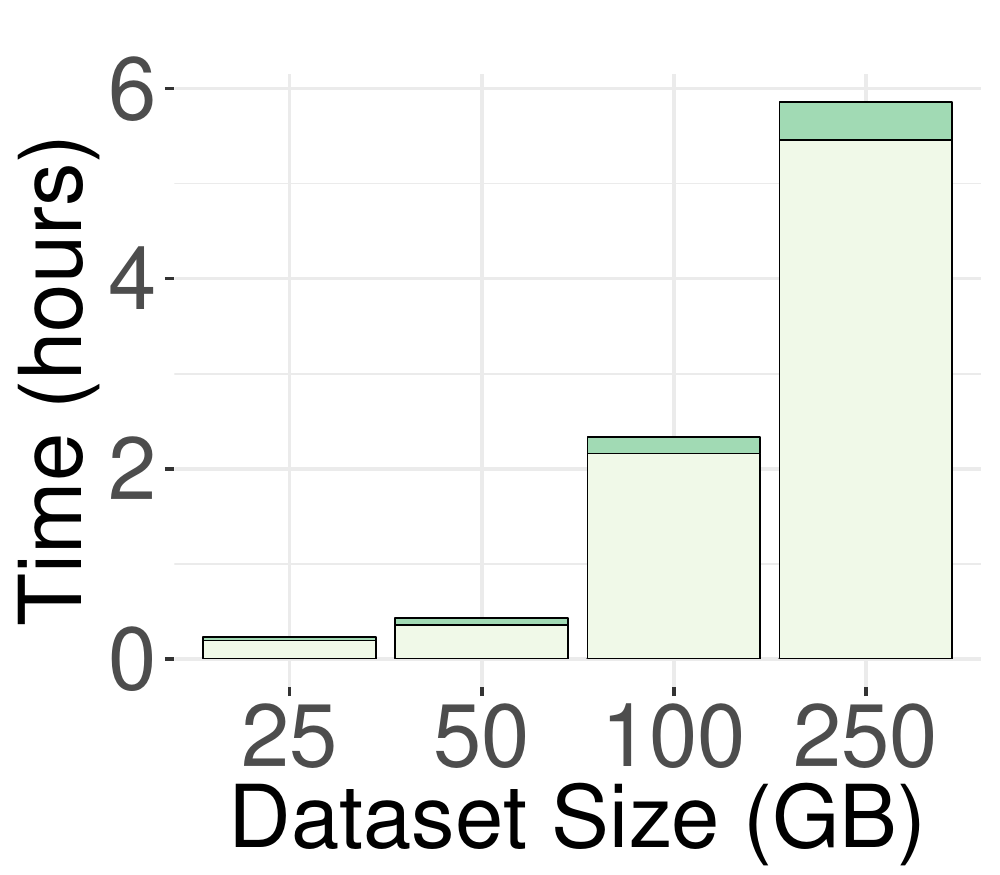}
		\caption{UCR Suite}
		\label{fig:exact:datasize:time:idxproc:cache:ucr}
	\end{subfigure}
	\hspace*{0.2cm}
	\begin{subfigure}{0.17\textwidth}
		\centering
		\includegraphics[width=\textwidth]{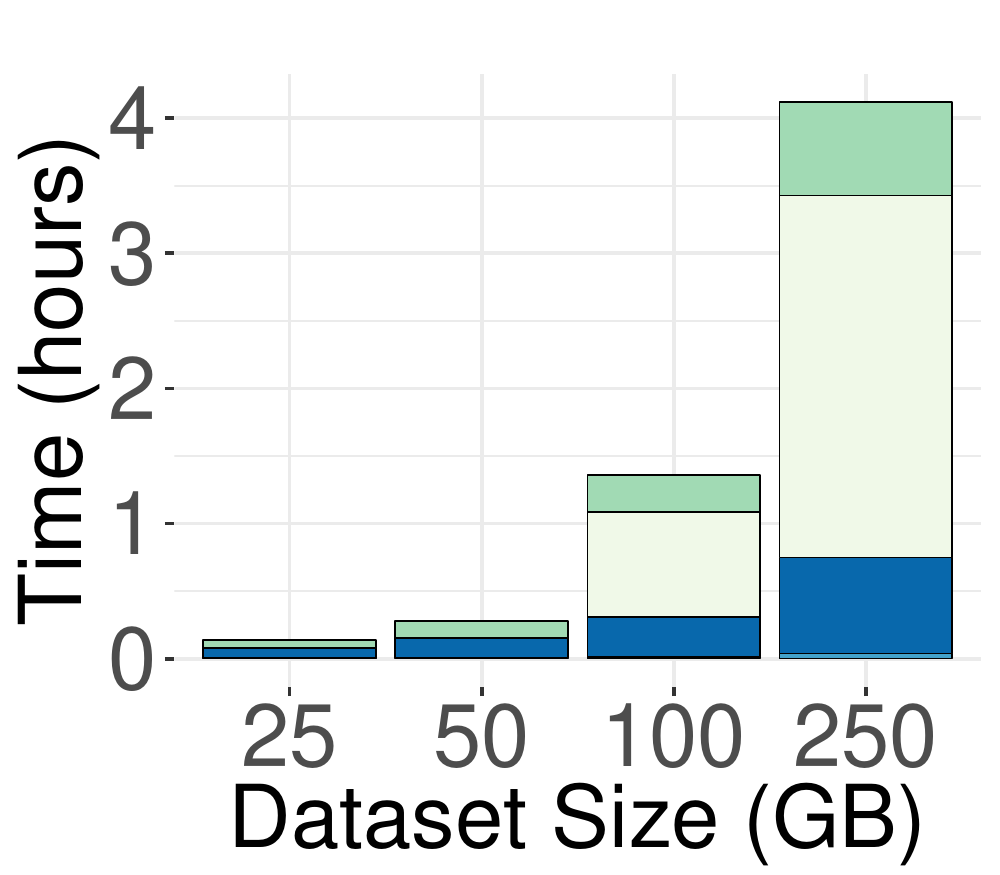}
		\caption{VA+file}
		\label{fig:exact:datasize:time:idxproc:cache:va+file}
	\end{subfigure}
	\begin{subfigure}{0.17\textwidth}
		\centering
		\hspace*{1cm}
		\includegraphics[scale=0.3]{{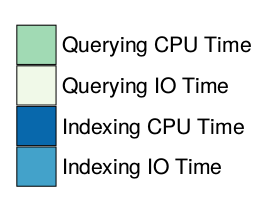}}
	\end{subfigure}			
	\caption{ Scalability with increasing dataset sizes 
	}
	\label{fig:exact:datasize:time:idxproc:cache}
\end{figure*}

\begin{figure}[t]
	\captionsetup{justification=centering}
	\captionsetup[subfigure]{justification=centering}
	\begin{subfigure}{0.5\textwidth}
		\centering
		\includegraphics[width=0.7\textwidth]{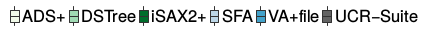}
	\end{subfigure}
	\begin{subfigure}{0.2365\textwidth}
		\centering
		\includegraphics[width=\textwidth]{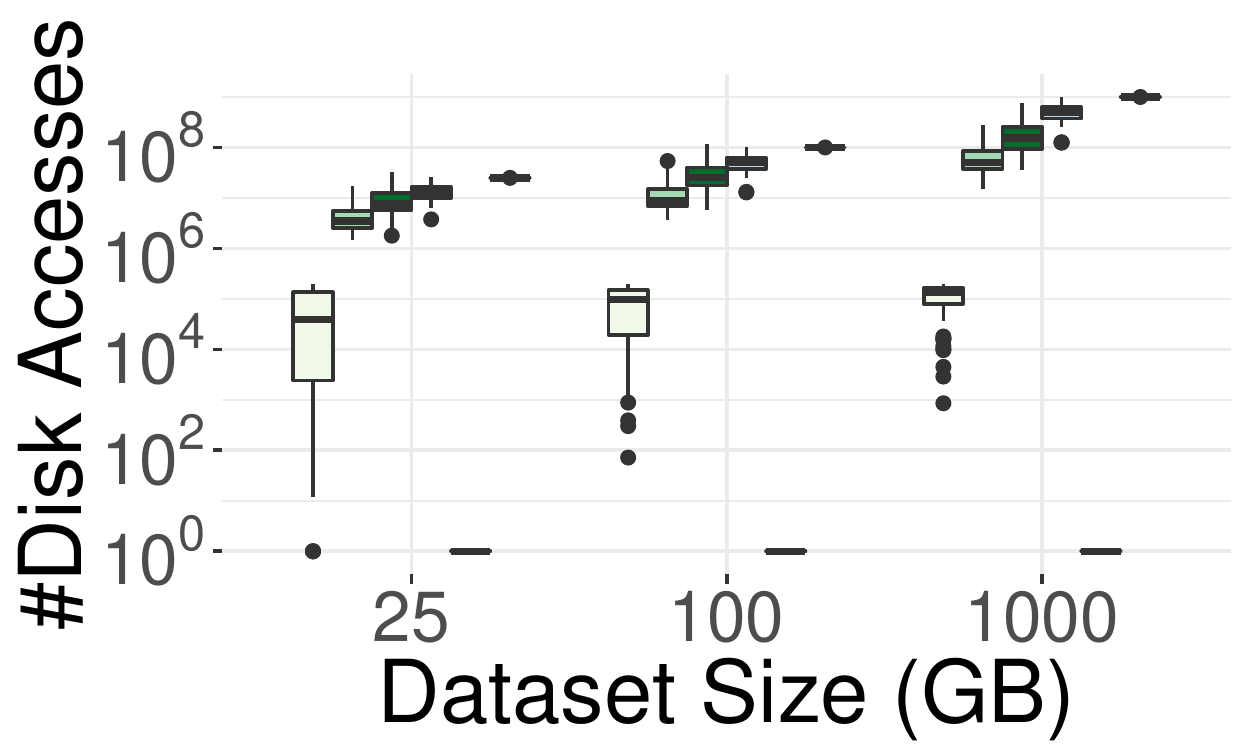}
		\caption{Sequential accesses \\
			Varying dataset sizes 
		}
		\label{fig:exact:synthetic:datasize:disk:sequential:cache:combined}
	\end{subfigure}
	\begin{subfigure}{0.2365\textwidth}
		\centering
		\includegraphics[width=\textwidth]{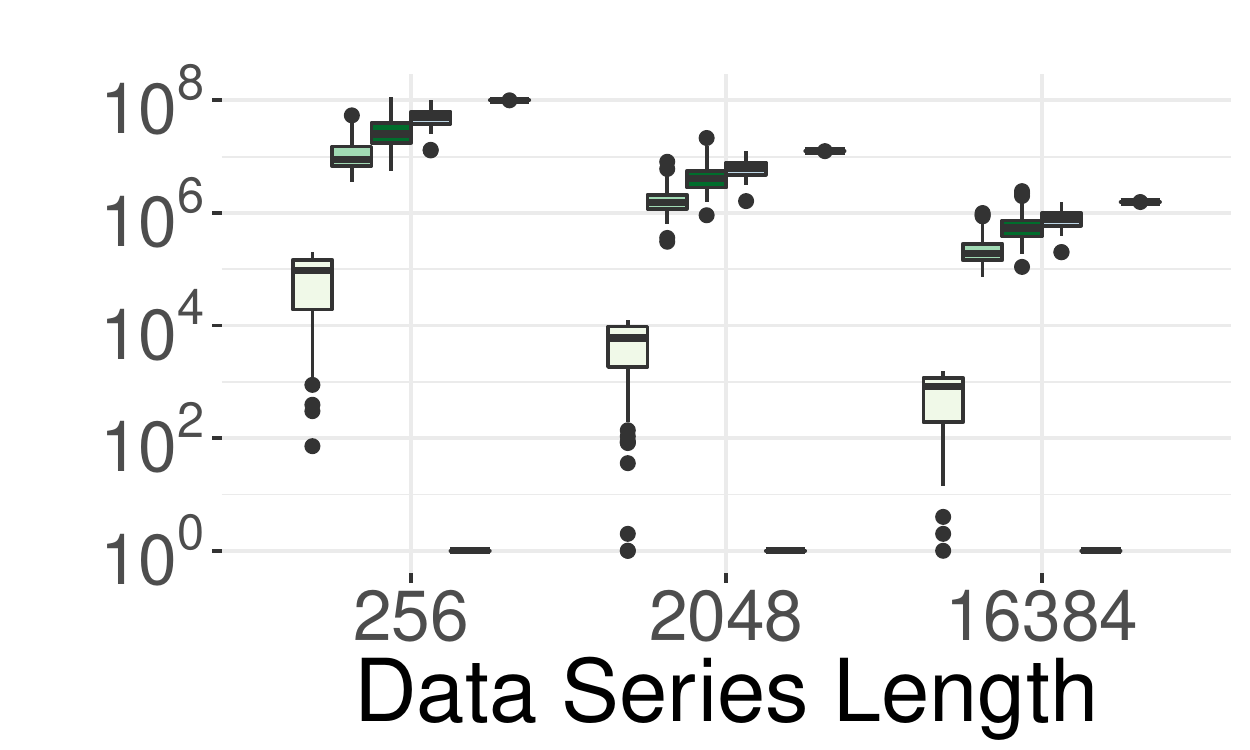}
	\caption{Sequential accesses \\
		Varying series lengths 
	}
	\label{fig:exact:synthetic:length:disk:sequential:cache:combined}
\end{subfigure}	
\begin{subfigure}{0.2365\textwidth}
	\centering
	\includegraphics[width=\textwidth]{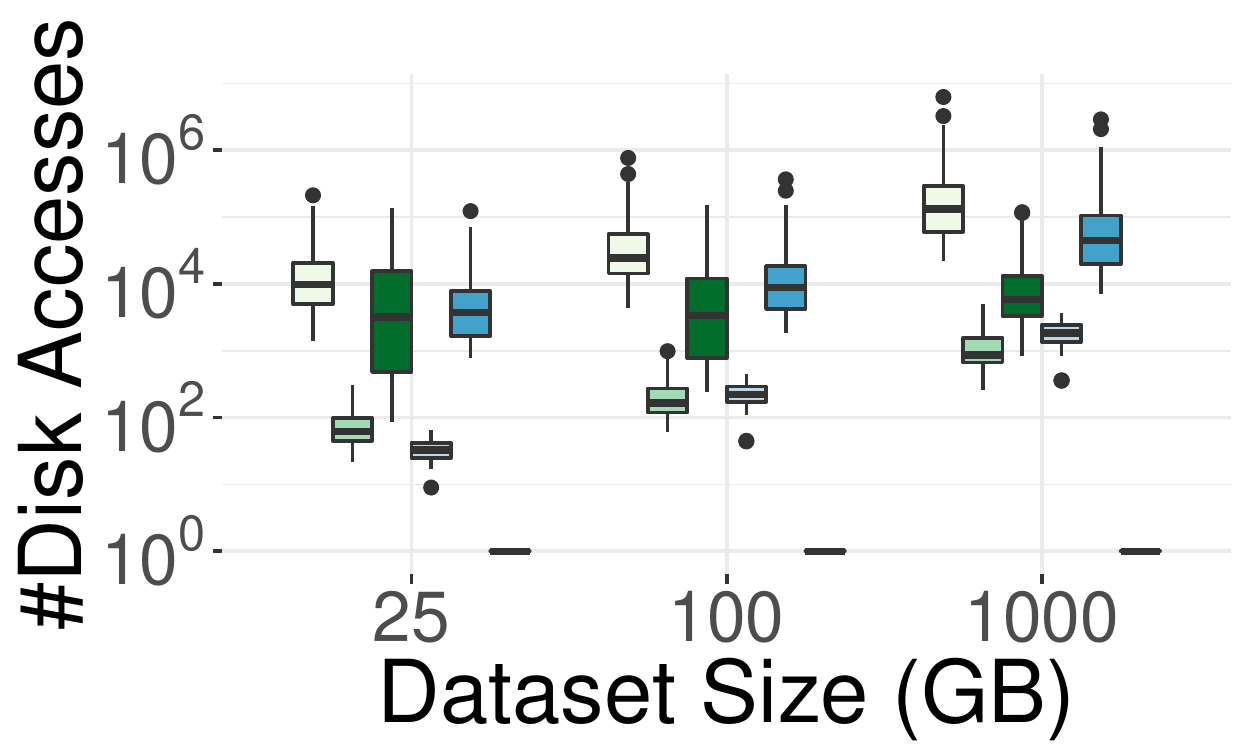}
	\caption{Random accesses \\
		Varying dataset sizes 
	}
	\label{fig:exact:synthetic:datasize:disk:random:cache:combined}
\end{subfigure}
\begin{subfigure}{0.2365\textwidth}
	\centering
	\includegraphics[width=\textwidth]{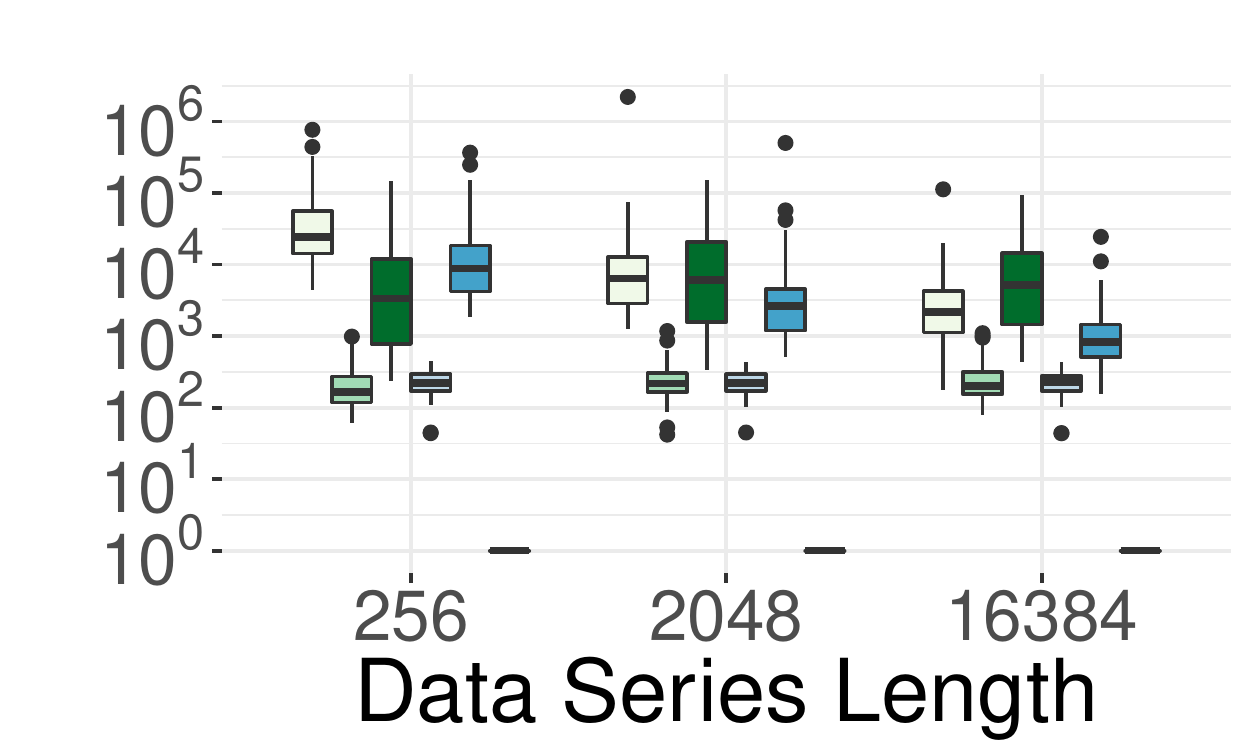}
\caption{Random accesses \\
	Varying series lengths \\
}
\label{fig:exact:synthetic:length:disk:random:cache:combined}
\end{subfigure}		

\caption{Number of disk accesses}
\label{fig:exact:synthetic:disk:cache:combined}
\end{figure}

\noindent\textbf{ADS+.} Figure~\ref{fig:exact:datasize:time:idxproc:cache:ADS+:1} shows that ADS+ is very efficient at index building, spending most of the cost for query answering, itself dominated by the input time. The reason is that ADS+ performs skip sequential accesses on the raw data file,
performing a skip almost every time a data series is pruned.

\noindent\textbf{DSTree.} In contrast, DSTree answers queries very fast whereas index building is costly (Figure~\ref{fig:exact:datasize:time:idxproc:cache:dstree}).
DSTree's cost for index building is mostly CPU, thus, offering great opportunities for parallelization.

\noindent\textbf{iSAX2+.} Figure~\ref{fig:exact:datasize:time:idxproc:cache:iSAX2+} summarizes the results for iSAX2+, which is slower to build the index compared to ADS+, but faster compared to DSTree.
Query answering is faster than ADS+ and slower than the DSTree.

\noindent\textbf{MASS.} Figure~\ref{fig:exact:datasize:time:idxproc:cache:mass} reports the results for MASS, which has been designed mainly for subsequence matching queries, but we adapted it for whole matching.
The very high CPU cost is due to the large number of operations involved in calculating Fourier transforms and the dot product cost.

\noindent\textbf{M-tree.} For the M-tree, we were only able to run experiments with in-memory datasets, because the only implementation we could use is a main memory index. The disk-aware implementations did not scale beyond 1GB. 
Figure~\ref{fig:exact:datasize:time:idxproc:cache:mtree} shows the M-tree experimental results for the 25GB and 50GB datasets, and the (optimistic) extrapolated results for the 100GB and 250GB datasets.
Note that going from 25GB to 50GB, the M-tree performance deteriorates by a factor of 3, even though both datasets fit in memory. 
(The M-tree experiments for the 100GB and 250GB datasets were not able to terminate, so we report extrapolated values in the graph, by multiplying the 50GB numbers by 3 and 9, respectively, which is an optimistic estimation.)
These results indicate that M-tree cannot scale to large dataset sizes.

\noindent\textbf{R*-tree.} Figure~\ref{fig:exact:datasize:time:idxproc:cache:rstree} shows the results for the R*-tree. Its performance deteriorates rapidly as dataset sizes increase.
Even using the best implementation among the ones we tried, when the dataset reaches half the available memory, swapping causes performance to degrade. Experiments on the 100GB and 250GB datasets were stopped after 24 hours. 

\noindent\textbf{SFA Trie.} Figure~\ref{fig:exact:datasize:time:idxproc:cache:sfa} reports the cost of index building and query processing for SFA.
We observe that query processing dominates the total cost and that query cost is mostly I/O, due to the optimal leaf size being rather large.

\noindent\textbf{Stepwise.} Figure~\ref{fig:exact:datasize:time:idxproc:cache:Stepwise} indicates the time it takes for Stepwise to build the DWHT tree and execute the workload.
The total cost is high and is dominated by query answering.
This is because answering one query entails filtering the data level by level and requires locating the remaining candidate data corresponding to higher resolutions through random I/O.

\noindent\textbf{UCR Suite.} Figure~\ref{fig:exact:datasize:time:idxproc:cache:ucr} shows the time it takes for the UCR-Suite to execute the workload.
Its cost is naturally dominated by input time, being a sequential scan algorithm.

\noindent\textbf{VA+file.} We observe in Figure~\ref{fig:exact:datasize:time:idxproc:cache:va+file} that VA+file is efficient at index building, spending most of the cost for query answering.
The indexing and querying costs are dominated by CPU and input time, respectively. The CPU cost is due to the time spent for determining the bit allocation and decision intervals for each dimension; the input time is incurred when accessing the non-pruned raw data series. 

\noindent\textbf{Summary.}
Overall, Figure~\ref{fig:exact:datasize:time:idxproc:cache} shows that it takes Stepwise, MASS, the M-tree and the R*-tree over 12 hours to complete the workload for the 250GB dataset, whereas the other methods need less than 7 hours.
Therefore, in the subsequent experiments, we will only include ADS+, DSTree, iSAX2+, SFA, the UCR suite and the VA+file. 

\subsubsection{\textbf{Comparison of the Best Methods}}
\label{ssec:comparison}

In the following experiments, we use the best methods as identified above, and compare them in more detail.

\noindent\textbf{Disk Accesses vs Dataset Size/Sequence Length.}
Figure~\ref{fig:exact:synthetic:disk:cache:combined}
shows the number of sequential and random disk accesses incurred by the 100 exact queries of the $Synth$-$Rand$ workload for increasing dataset sizes and increasing lengths. When the dataset size is varied, the length of the data series is kept constant at 256, whereas the dataset size is kept at 100GB when the length is varied. We can observe that the VA+file and ADS+ perform the smallest number of sequential disk accesses across dataset sizes and data series lengths, with the VA+ performing virtually none. As expected, the UCR-Suite performs the largest number of sequential accesses regardless of the length of the series, or the size of the dataset. This number is also steady across queries, thus its boxplot is represented by a flat line. There is not a significant difference between the number of sequential operations needed by the DSTree, SFA or iSAX2+ (DSTree does the least, and SFA the most). SFA requires more sequential accesses, because its optimal leaf size is 1M, as opposed to 100K for DSTree and iSAX2+.

As far as random I/O for different dataset sizes is concerned, ADS+ performs the largest number of random accesses, followed by the VA+file. The DSTree and SFA incur almost the same number of operations. However, the DSTree has a good balance between the number of random and sequential I/O operations. It is interesting to point out that as the dataset size increases, the number of random operations for iSAX2+ becomes less skewed across queries. This is because of the fixed split-point nature of iSAX2+ that causes it to better distribute data among leaves when the dataset is large: in small dataset sizes, many leaves can contain very few series. 
 
When the dataset size is set to 100GB and the data series length is increased, we can observe a dramatic decrease of the number of random operations incurred by ADS+ and VA+file. The reason is that both methods use a skip-sequential algorithm, so even if the pruning ratio stays the same, when the data series is long, the algorithm skips larger blocks of data, thus the number of skips decreases. The random I/Os across lengths for the other methods remain quite steady, with SFA and DSTree performing the least.

\ifJournal

\begin{figure}[t]
	\captionsetup{justification=centering}
\captionsetup{justification=centering}
\captionsetup[subfigure]{justification=centering}
	
	\begin{subfigure}{\columnwidth}
		\centering
		\hspace{4.5cm}
		\includegraphics[width =\columnwidth]{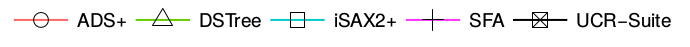}
	\end{subfigure}	
	\hspace*{\fill} 
	\hspace*{\fill} 
	
	\begin{subfigure}{0.48\columnwidth}
		\centering
		\includegraphics[width=\columnwidth]{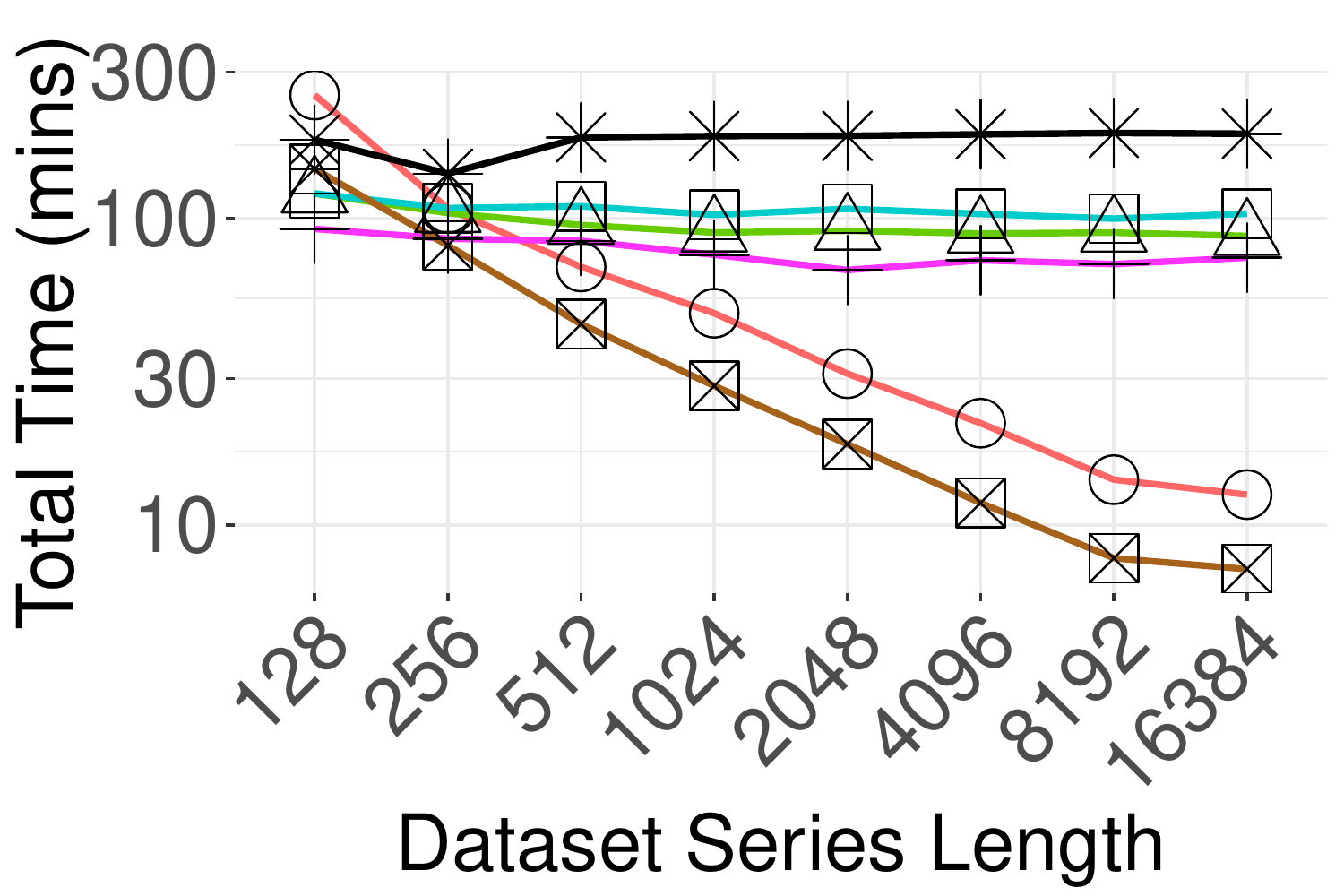}
		\caption{Idx+Exact100}
		\label{fig:exact:length:time:idxproc:cache:combined:100:exact}
	\end{subfigure}
	\begin{subfigure}{0.48\columnwidth}
		\centering
		\includegraphics[width=\columnwidth]{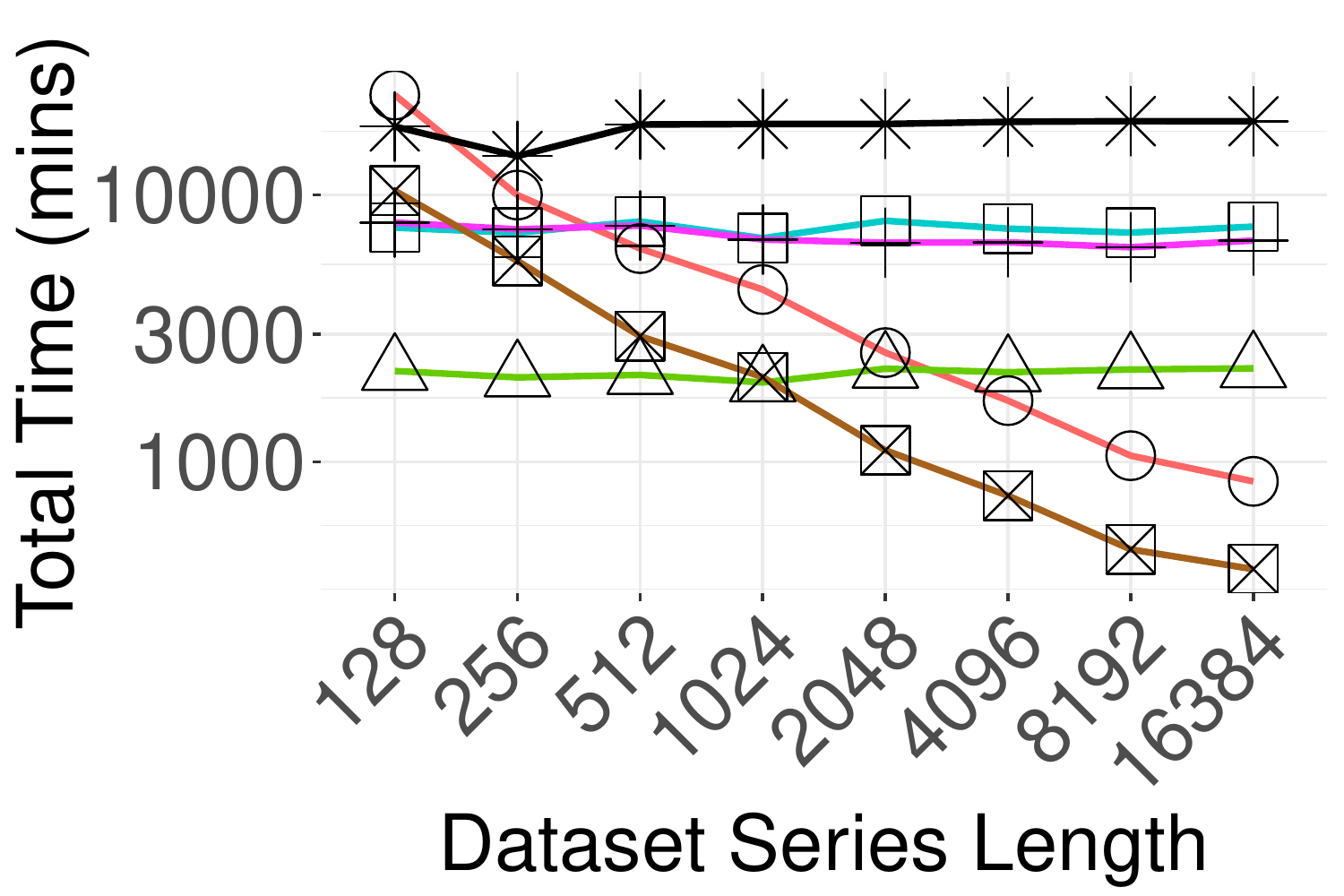}
		\caption{Idx+Exact10K}
		\label{fig:exact:length:time:idxproc:cache:combined:1000:exact}
	\end{subfigure}
	\caption{Scalability with increasing lengths }
	\label{fig:exact:length:time:idxproc:combined}
\end{figure}

\begin{figure*}[tb]
	\begin{minipage}{\textwidth}
		\captionsetup{justification=centering}
		\begin{subfigure}{\textwidth}
			\centering
			\includegraphics[width=0.6\textwidth]{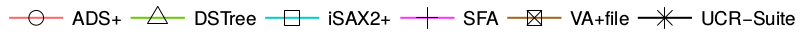}
		\end{subfigure}	\captionsetup[subfigure]{justification=centering}
		\hspace*{\fill} 
		\hspace*{\fill} 
		
		\captionsetup{justification=centering}
		\begin{subfigure}{0.16\textwidth}
			\centering
			\includegraphics[width=\textwidth] {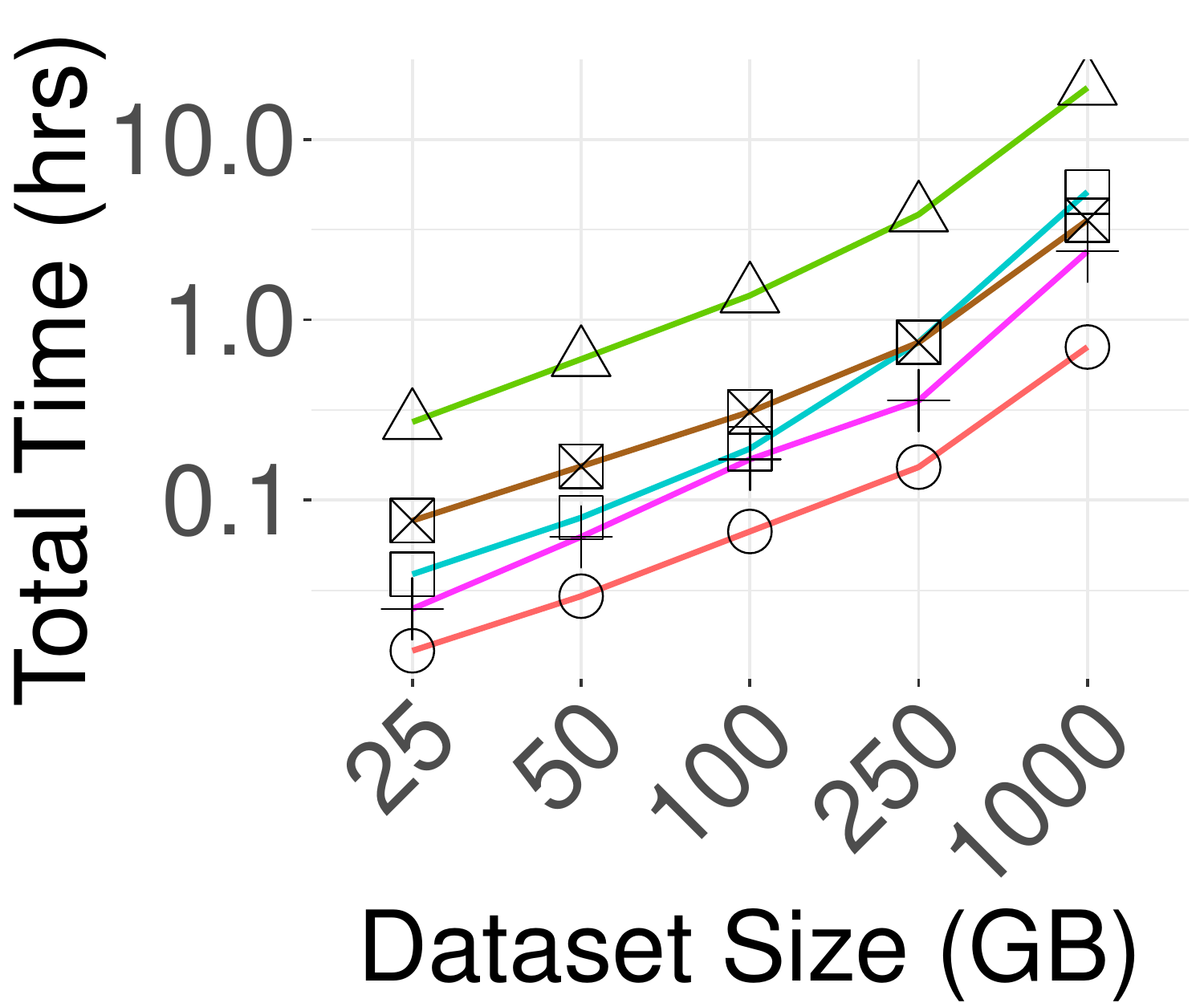}
			\caption{Idx}
			\label{fig:exact:datasize:time:idxproc:cache:combined:indexing}
		\end{subfigure}
		\begin{subfigure}{0.16\textwidth}
			\centering
			\includegraphics[width=\textwidth]{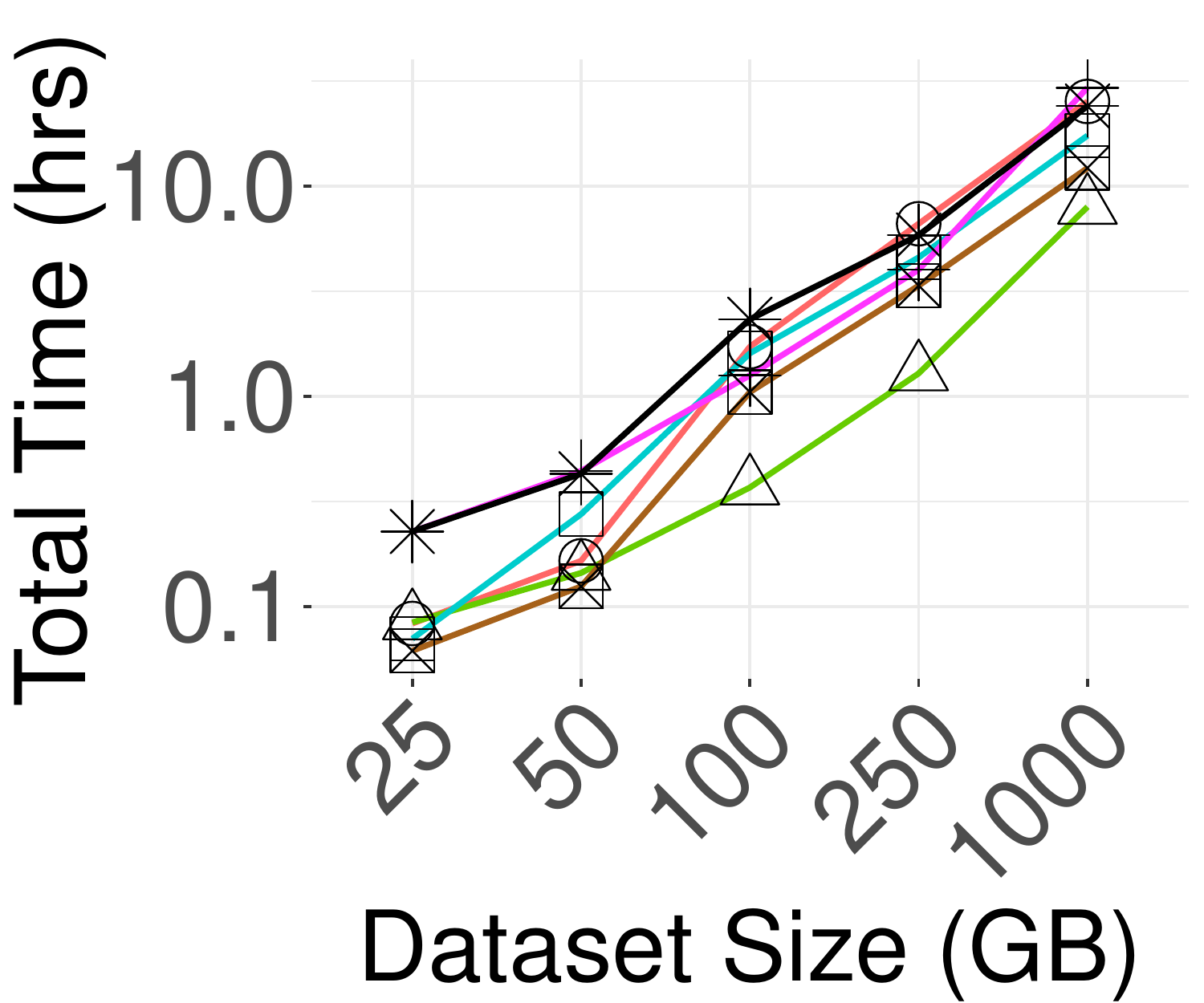}
			\caption{Exact100}	
			\label{fig:exact:datasize:time:idxproc:cache:combined:100exact}
		\end{subfigure}
		\begin{subfigure}{0.16\textwidth}
			\centering
			\includegraphics[width=\textwidth] {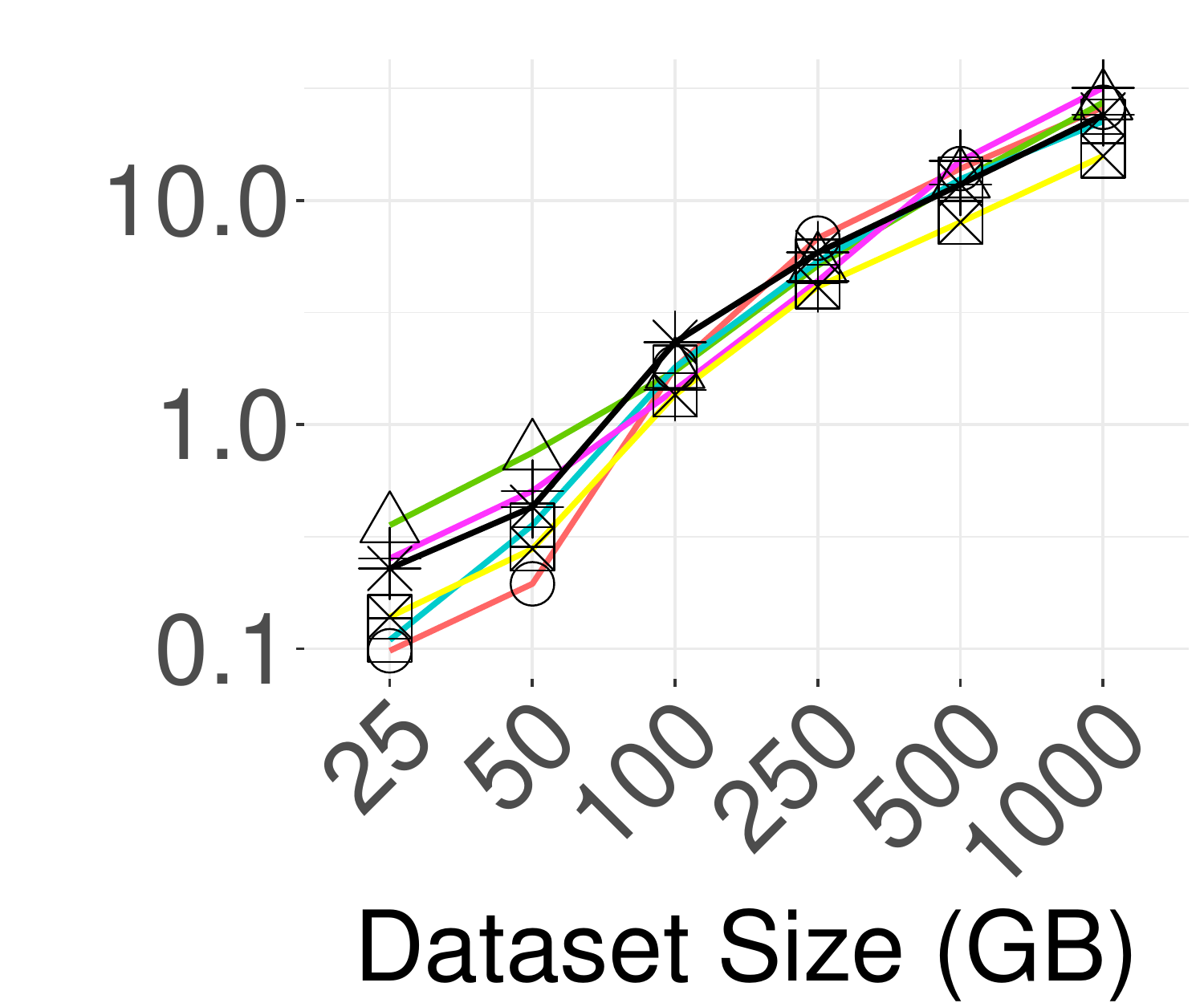}
			\caption{Idx+Exact100}			
			\label{fig:exact:datasize:time:idxproc:cache:combined:idx100exact}
		\end{subfigure}
		\begin{subfigure}{0.16\textwidth}
			\centering
			\includegraphics[width=\textwidth] {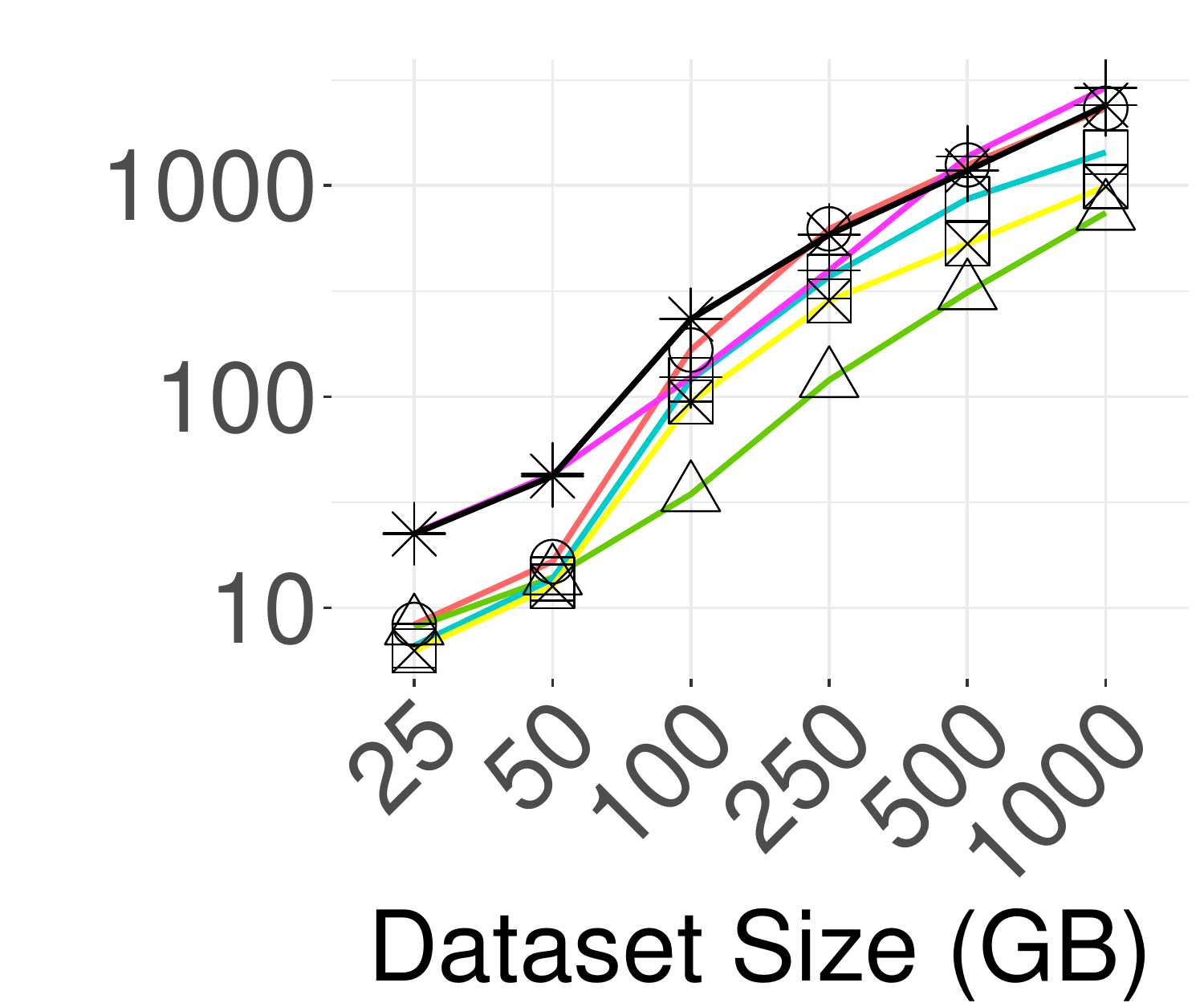}
			\caption{Idx+Exact10K}			
			\label{fig:exact:datasize:time:idxproc:cache:combined:idx10Kexact}
		\end{subfigure}
		\begin{subfigure}{0.16\textwidth}
			\centering
			\includegraphics[width=\textwidth]{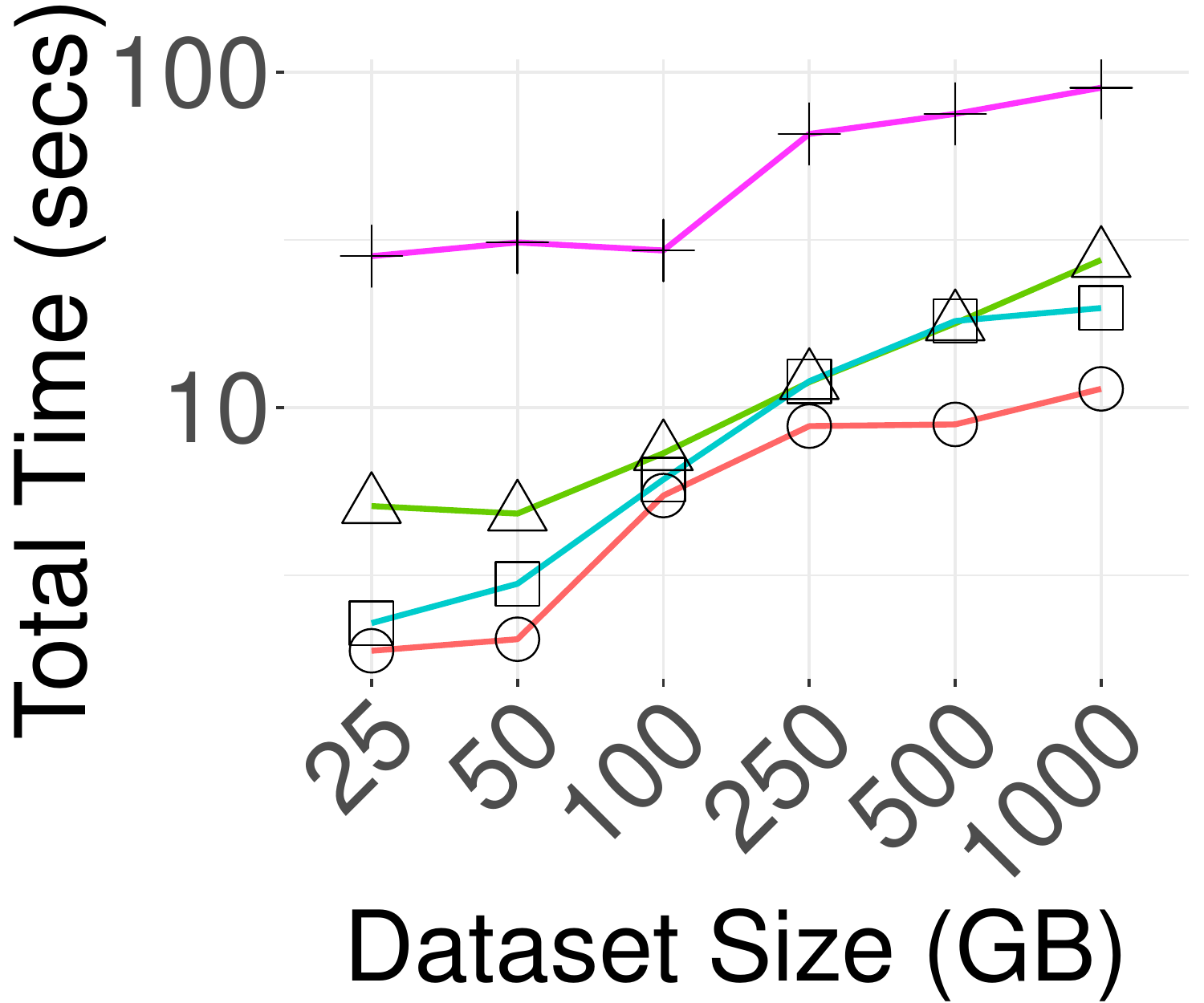}
			\caption{App100}	
			\label{fig:exact:datasize:time:idxproc:cache:combined:100approx}
		\end{subfigure}
		\begin{subfigure}{0.16\textwidth}
			\centering
			\includegraphics[width=\textwidth] {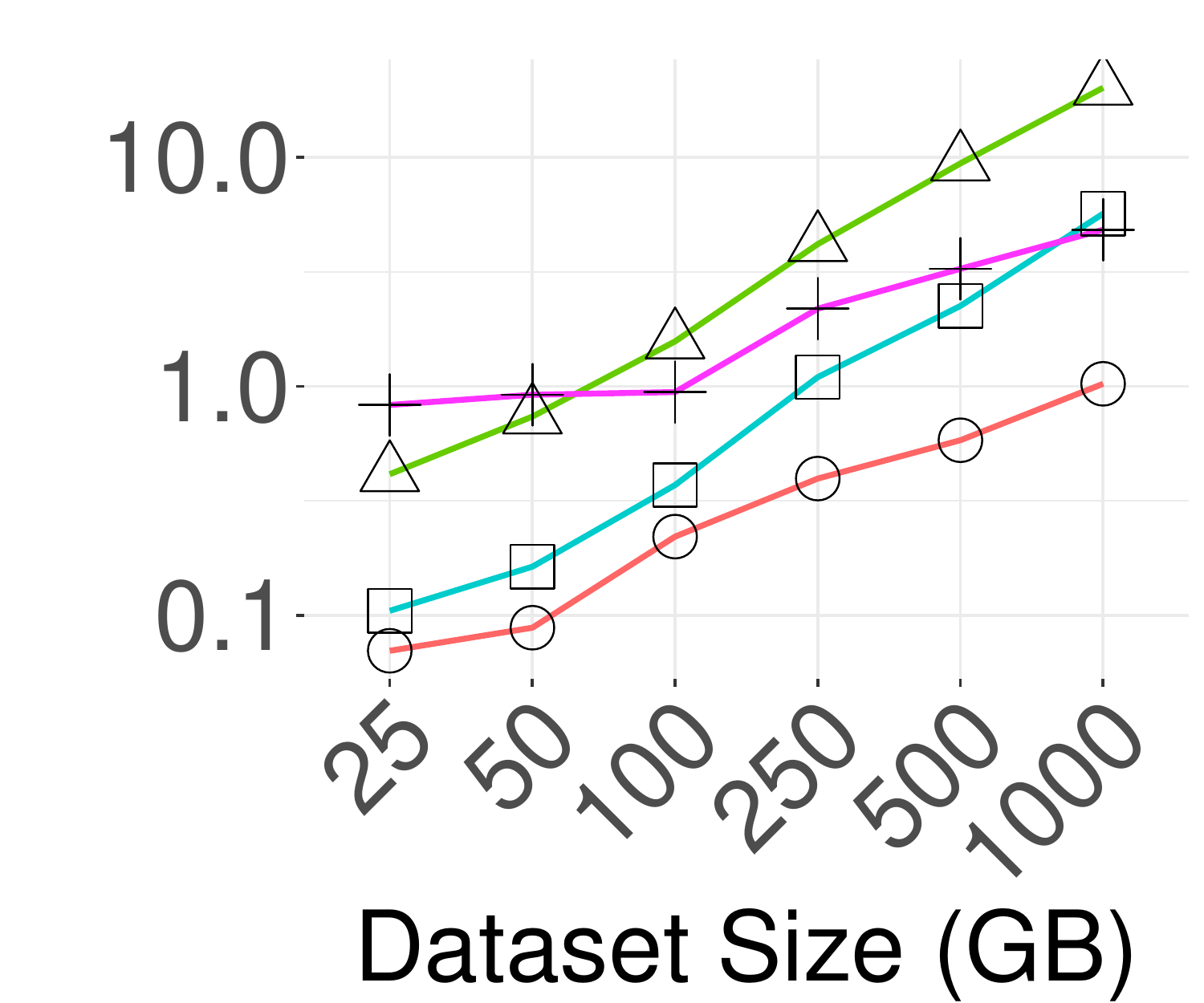}
			\caption{Idx+App10K}			
			\label{fig:exact:datasize:time:idxproc:cache:combined:idx10Kapprox}
		\end{subfigure}

		\caption{Scalability comparison (HDD)
	}
		\label{fig:exact:datasize:time:idxproc:cache:combined}
	\end{minipage}
	\begin{minipage}{\textwidth}
				\captionsetup{justification=centering}
				
		\begin{subfigure}{0.16\textwidth}
			\centering
			\includegraphics[width=\textwidth] {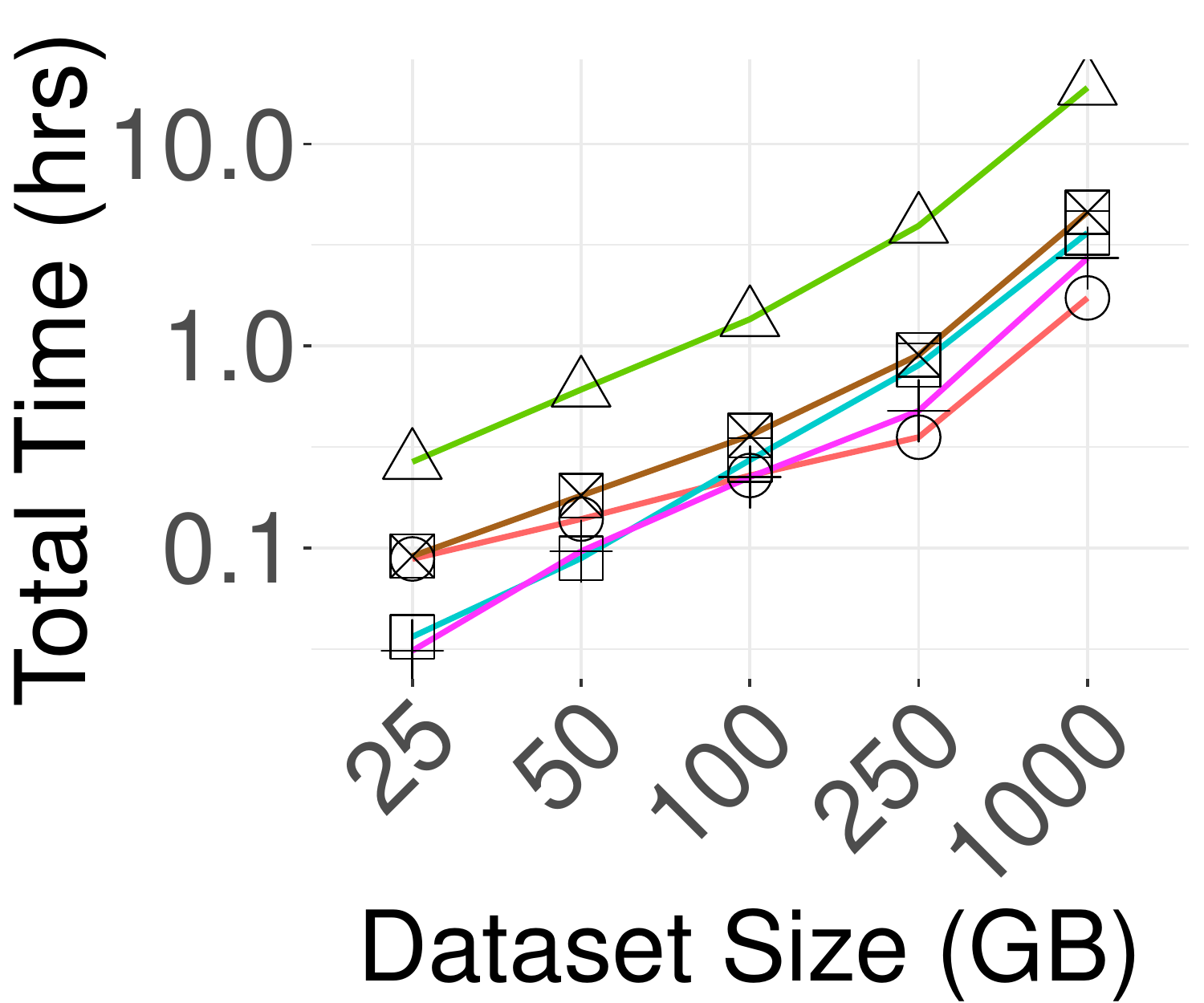}
			\caption{Idx}
			\label{fig:exact:datasize:time:idxproc		:cache:combined:indexing:nefeli}
		\end{subfigure}
		\begin{subfigure}{0.16\textwidth}
			\centering
			\includegraphics[width=\textwidth]{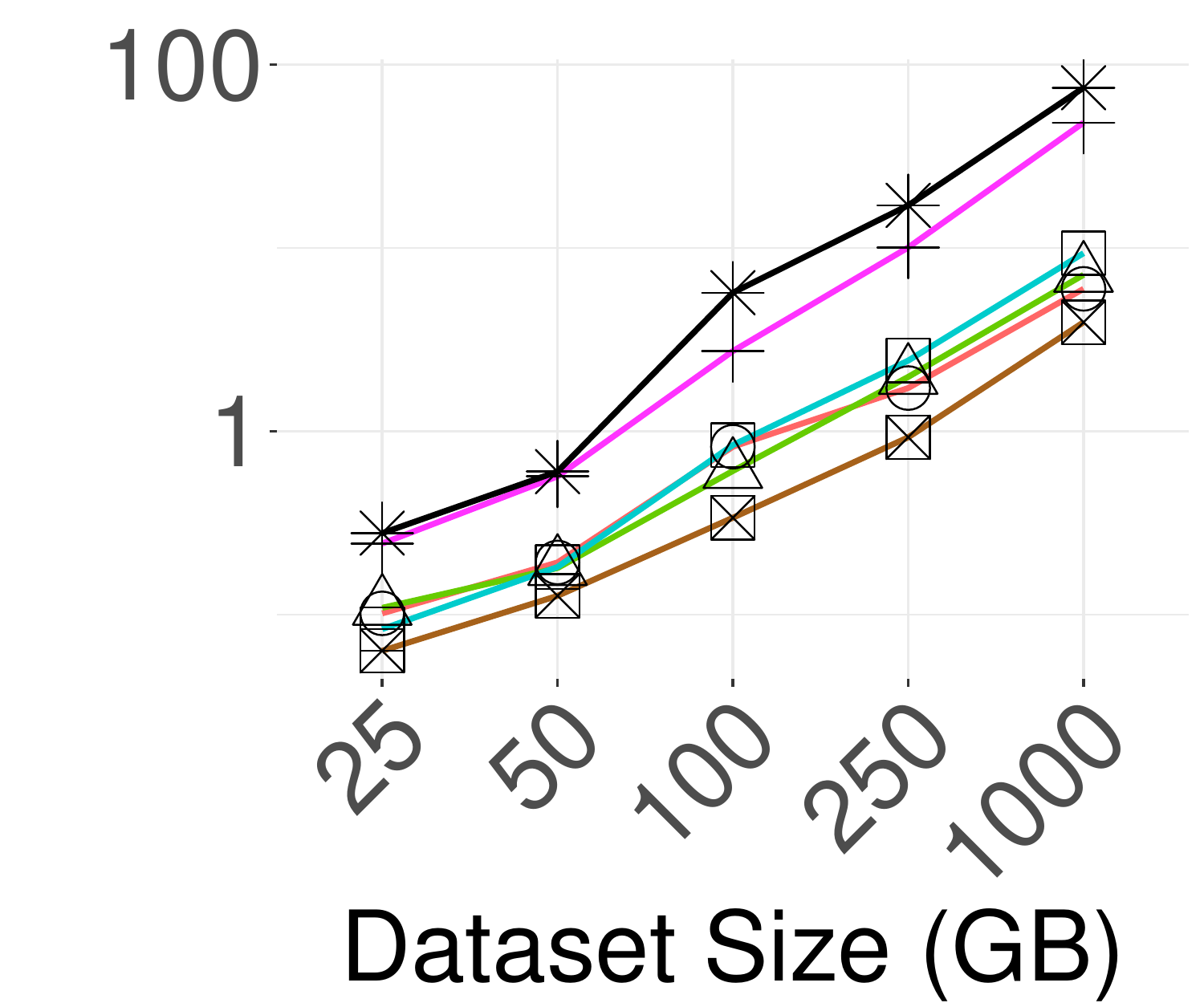}
			\caption{Exact100}	
			\label{fig:exact:datasize:time:idxproc		:cache:combined:100exact:nefeli}
		\end{subfigure}
		\begin{subfigure}{0.16\textwidth}
			\centering
			\includegraphics[width=\textwidth] {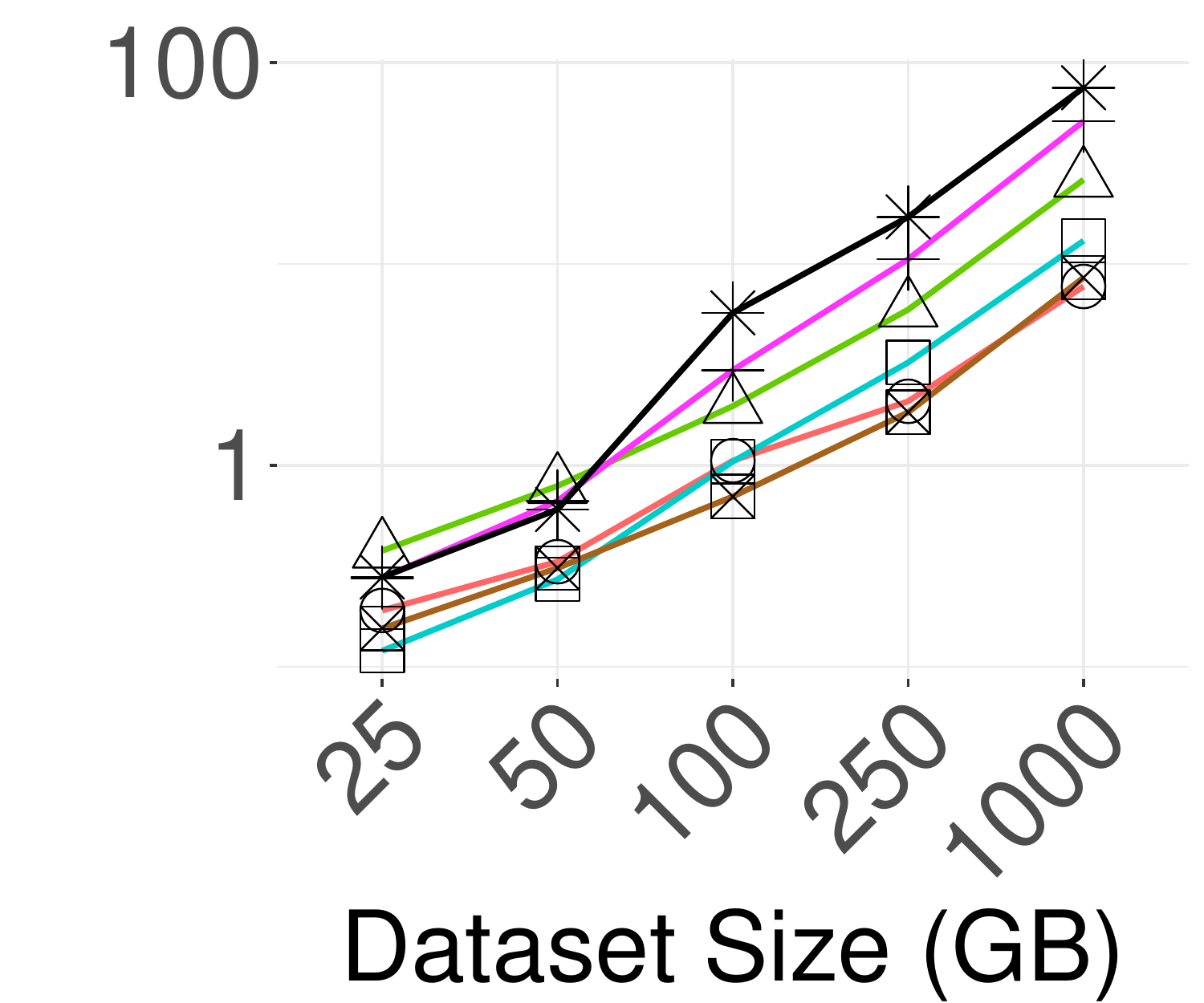}
			\caption{Idx+Exact100}			
			\label{fig:exact:datasize:time:idxproc		k:cache:combined:idx100exact:nefeli}
		\end{subfigure}
		\begin{subfigure}{0.16\textwidth}
			\centering
			\includegraphics[width=\textwidth] {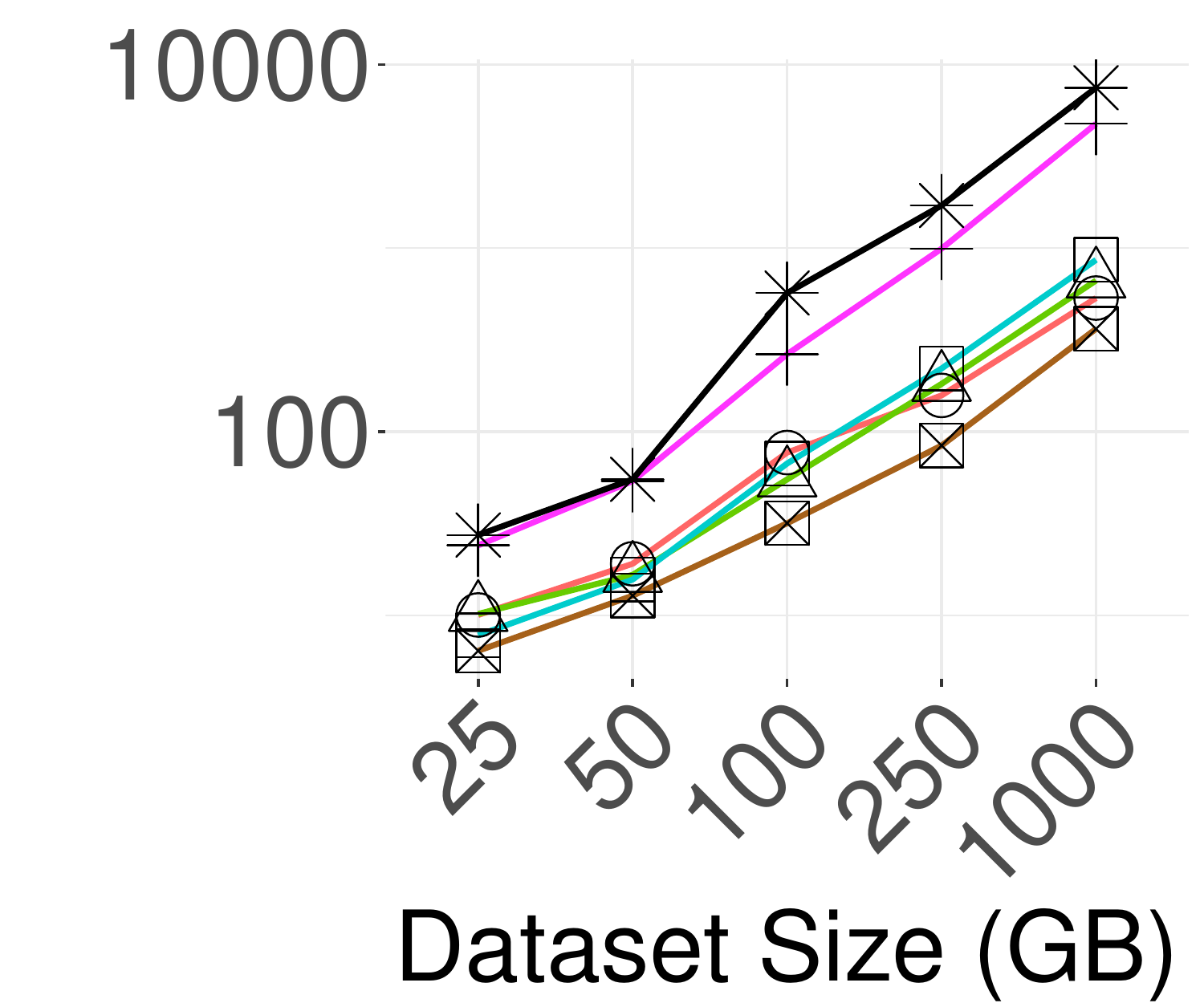}
			\caption{Idx+Exact10K}			
			\label{fig:exact:datasize:time:idxproc		:cache:combined:idx10Kexact:nefeli}
		\end{subfigure}		
		\begin{subfigure}{0.16\textwidth}
			\centering
			\includegraphics[width=\textwidth]{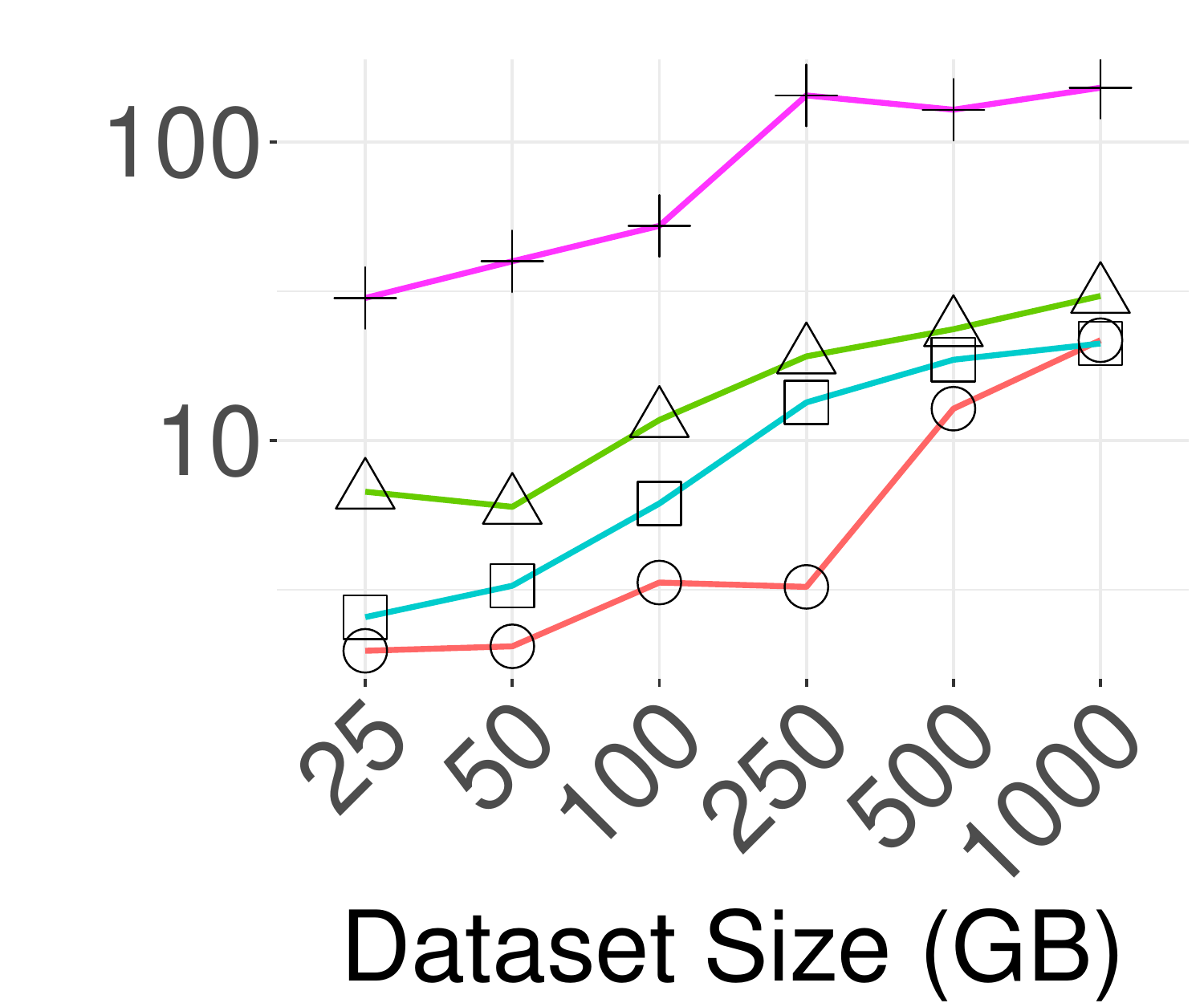}
			\caption{App100}	
			\label{fig:exact:datasize:time:idxproc		:cache:combined:100approx:nefeli}
		\end{subfigure}	
		\begin{subfigure}{0.16\textwidth}
			\centering
			\includegraphics[width=\textwidth]
			{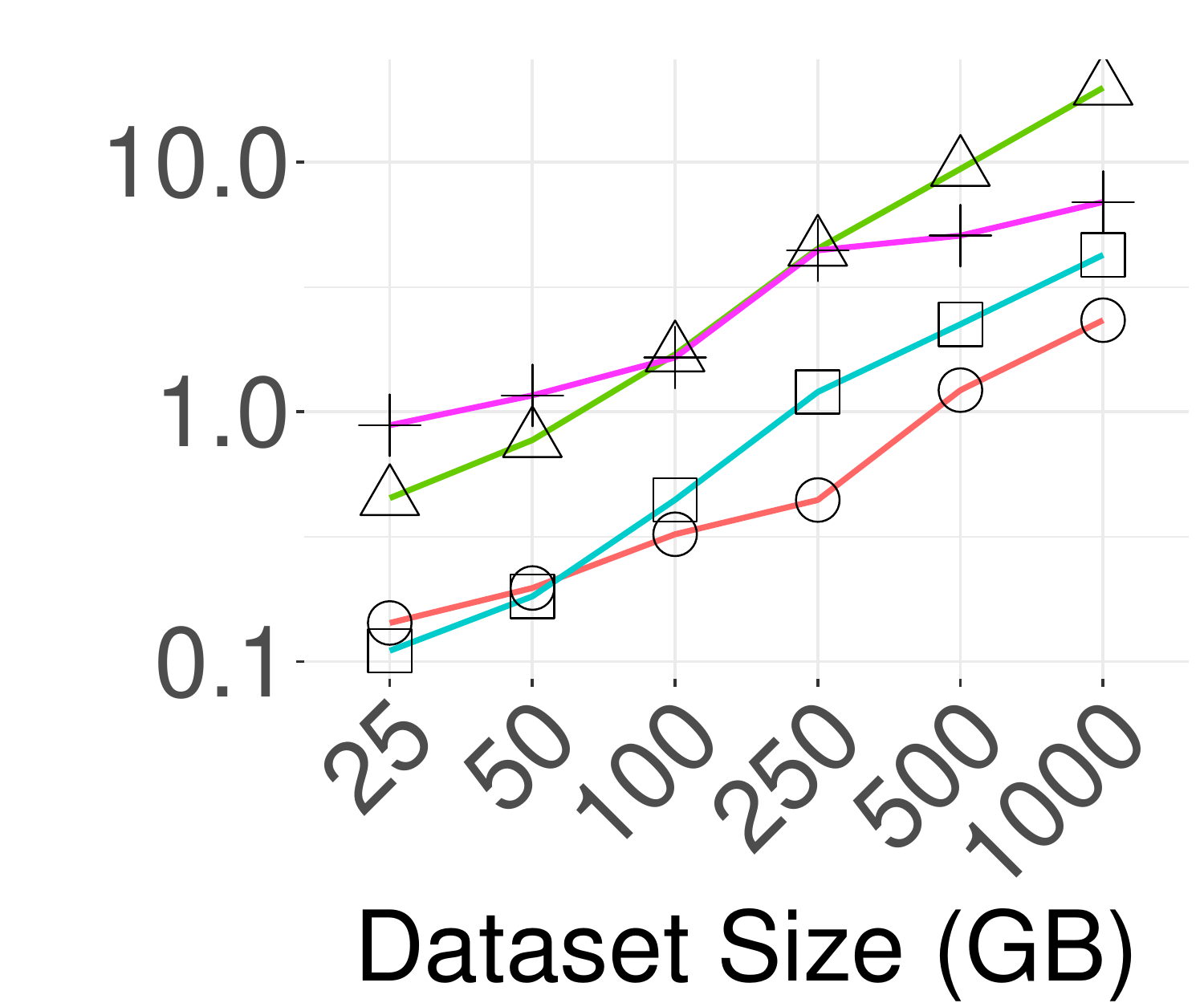}
			\caption{Idx+App10K}			
			\label{fig:exact:datasize:time:idxproc		:cache:combined:idx100approx:nefeli}
		\end{subfigure}
		
		\caption{Scalability comparison (SSD)
		}
		\label{fig:exact:datasize:time:idxproc:cache:combined:nefeli}
	\end{minipage}
\end{figure*}
\fi

\begin{figure*}[tb]
	\captionsetup{justification=centering}
	\centering
	\begin{subfigure}{0.5\textwidth}
		\centering
		\includegraphics[width=\textwidth]{{line_plot_new_legend}}
	\end{subfigure}	\captionsetup[subfigure]{justification=centering}

	\begin{minipage}{0.23\textwidth}
		\captionsetup{justification=centering}
		\begin{subfigure}{\textwidth}
			\centering
			\includegraphics[width=\textwidth]{exact_length_time_idxproc_combined_100_exact}
			\caption{Idx+Exact100}
			\label{fig:exact:length:time:idxproc:cache:combined:100:exact}
		\end{subfigure}
		\begin{subfigure}{\textwidth}
			\centering
			\includegraphics[width=\textwidth]{exact_length_time_idxproc_combined_10000_exact}
			\caption{Idx+Exact10K}
			\label{fig:exact:length:time:idxproc:cache:combined:1000:exact}
		\end{subfigure}
		\caption{Scalability with increasing lengths}
		\label{fig:exact:length:time:idxproc:combined}
	\end{minipage}
	\begin{minipage}{0.7\textwidth}

		\captionsetup{justification=centering}
		\begin{subfigure}{0.24\textwidth}
			\centering
			\includegraphics[width=\textwidth] {{exact_datasize_time_indexing_cache_combined_100_exact}}
			\caption{Idx}
			\label{fig:exact:datasize:time:idxproc:cache:combined:indexing}
		\end{subfigure}
		\begin{subfigure}{0.24\textwidth}
			\centering
			\includegraphics[width=\textwidth]{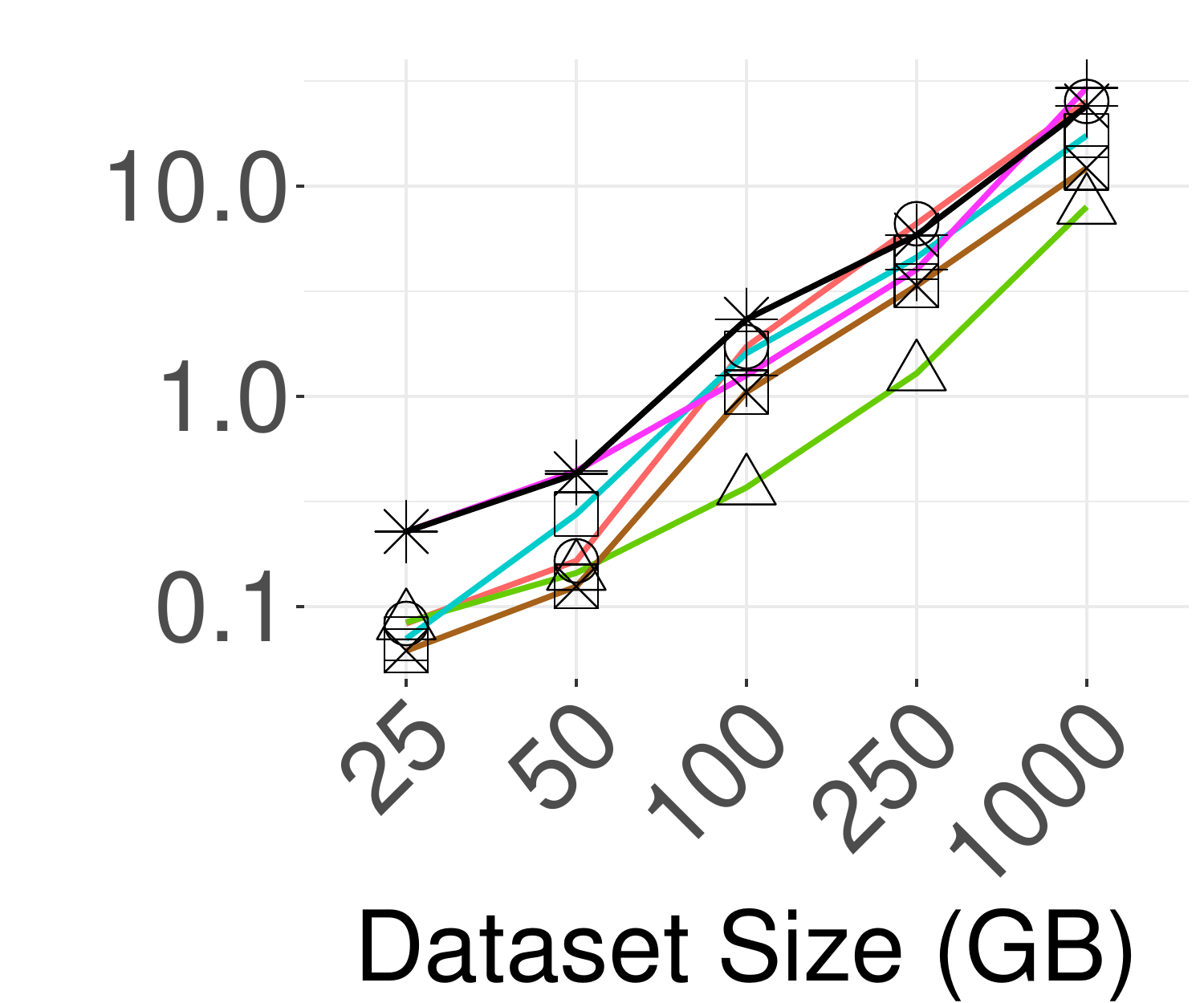}
			\caption{Exact100}	
			\label{fig:exact:datasize:time:idxproc:cache:combined:100exact}
		\end{subfigure}
		\begin{subfigure}{0.24\textwidth}
			\centering
			\includegraphics[width=\textwidth] {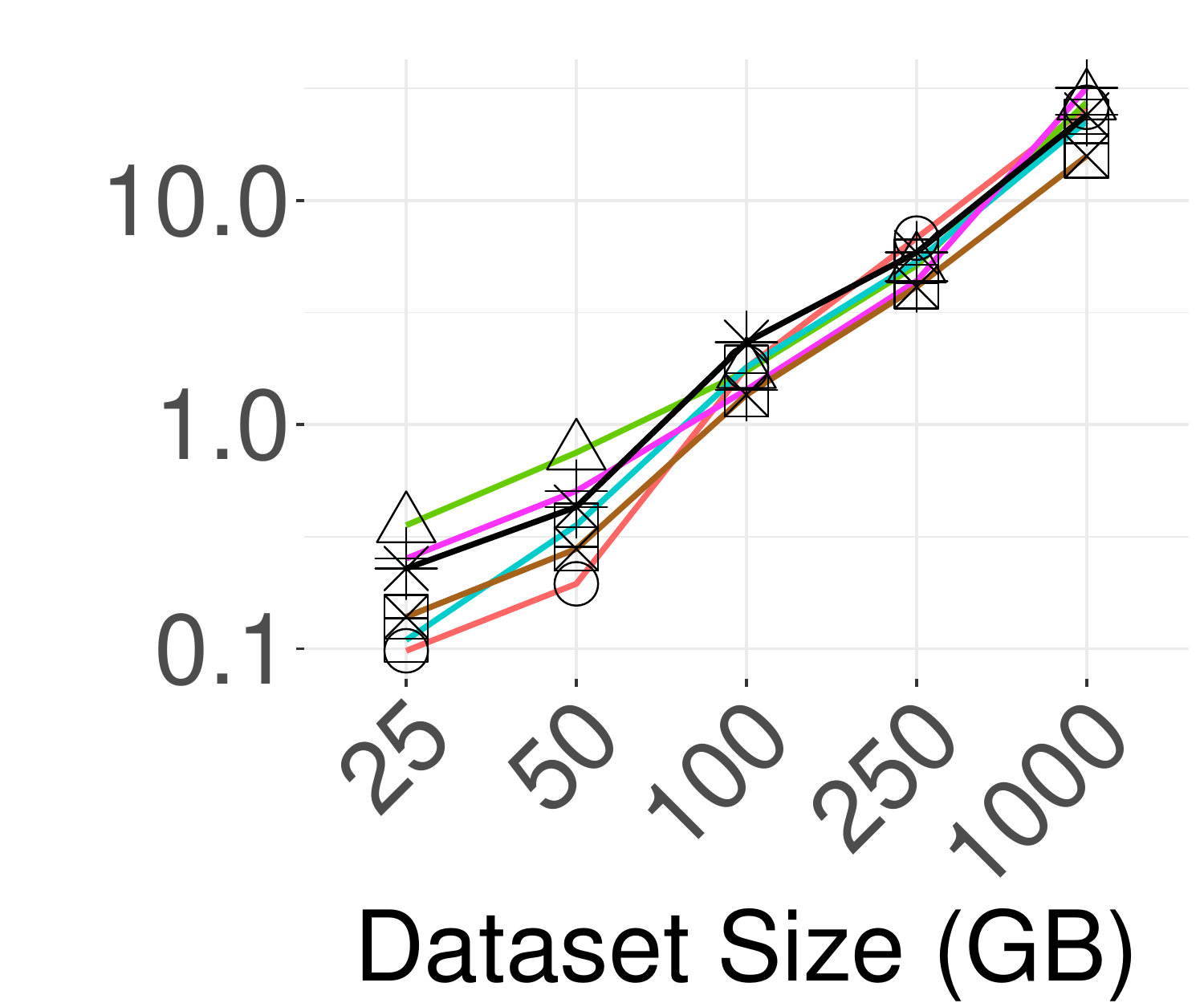}
			\caption{Idx+Exact100}			
			\label{fig:exact:datasize:time:idxproc:cache:combined:idx100exact}
		\end{subfigure}
		\begin{subfigure}{0.24\textwidth}
			\centering
			\includegraphics[width=\textwidth] {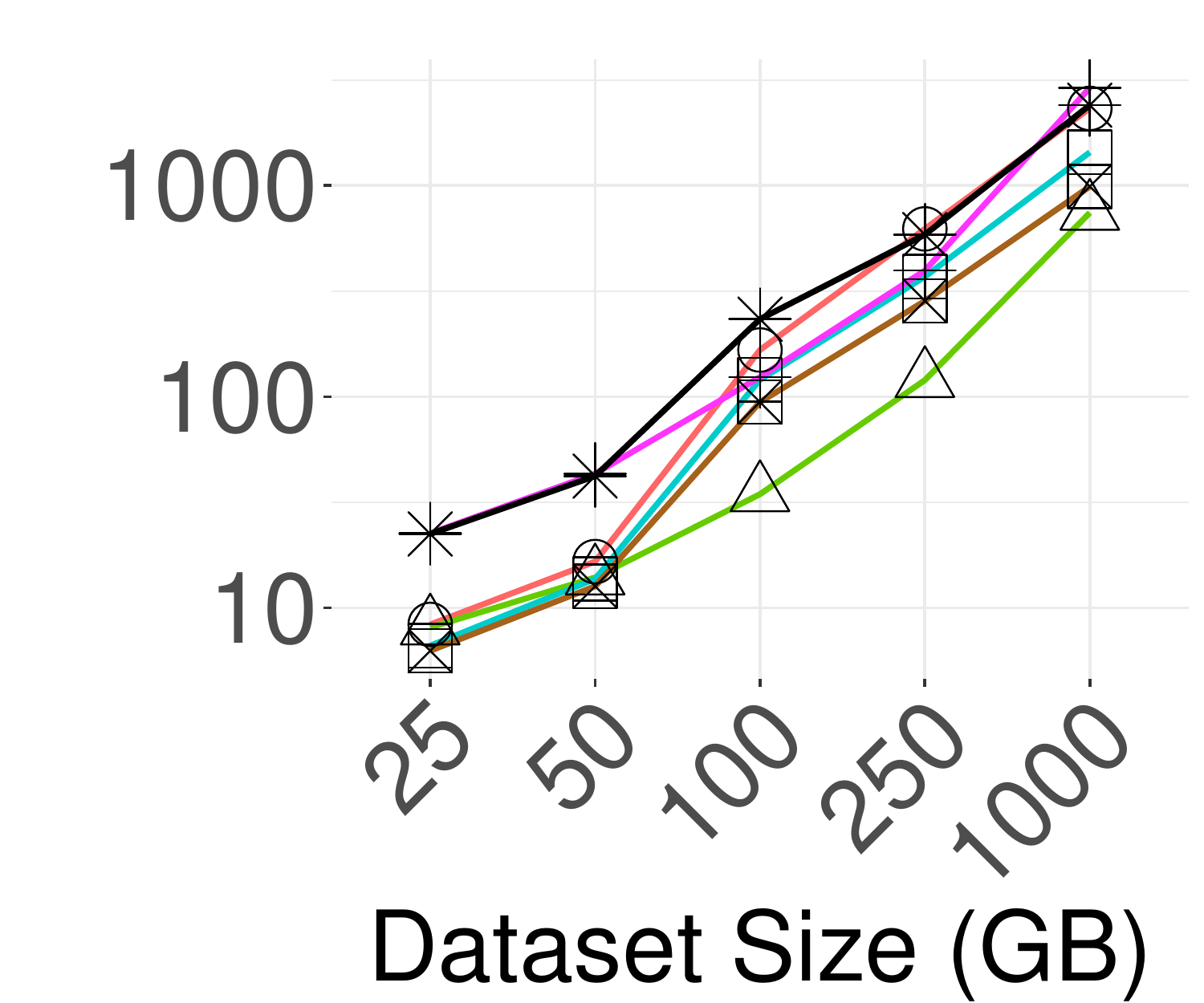}
			\caption{Idx+Exact10K}			
			\label{fig:exact:datasize:time:idxproc:cache:combined:idx10Kexact}
		\end{subfigure}		
		\caption{Scalability comparison (HDD) 
		}
		\label{fig:exact:datasize:time:idxproc:cache:combined}

\captionsetup[subfigure]{justification=centering}

\captionsetup{justification=centering}
		\centering
		\begin{subfigure}{0.24\textwidth}
			\centering
			\includegraphics[width=\textwidth] {{exact_datasize_time_indexing_cache_combined_100_exact_nefeli}}
			\caption{Idx}
			\label{fig:exact:datasize:time:idxproc		:cache:combined:indexing:nefeli}
		\end{subfigure}
		\begin{subfigure}{0.24\textwidth}
			\centering
			\includegraphics[width=\textwidth]{exact_datasize_time_exactproc_cache_combined_100_exact_nefeli_noylabel}
			\caption{Exact100}	
			\label{fig:exact:datasize:time:idxproc		:cache:combined:100exact:nefeli}
		\end{subfigure}
		\begin{subfigure}{0.24\textwidth}
			\centering
			\includegraphics[width=\textwidth] {{exact_datasize_time_idxproc_cache_combined_100_exact_nefeli_noylabel}}
			\caption{Idx+Exact100}			
			\label{fig:exact:datasize:time:idxproc		:cache:combined:idx100exact:nefeli}
		\end{subfigure}
		\begin{subfigure}{0.24\textwidth}
			\centering
			\includegraphics[width=\textwidth] {{exact_datasize_time_idxproc_cache_combined_10000_exact_nefeli_noylabel}}
			\caption{Idx+Exact10K}			
			\label{fig:exact:datasize:time:idxproc		:cache:combined:idx10Kexact:nefeli}
		\end{subfigure}		
		
		\caption{Scalability comparison (SSD) 
		}
		\label{fig:exact:datasize:time:idxproc:cache:combined:nefeli}
	\end{minipage}
\end{figure*}

\noindent{\textbf{Scalability/Search Efficiency vs Sequence Length.}}
Figure~\ref{fig:exact:length:time:idxproc:combined} depicts the performance of the different methods with increasing data series lengths. In order to factor out other parameters, we fix the dataset size to 100GB, and the dimensionality of the methods that use summarizations to 16, for all data series lengths. We observe that the indexing and querying costs for ADS+ and VA+file plummet as the data series length increases, whereas the cost of the other methods remains relatively steady across all lengths. This is because with increasing lengths, both algorithms perform larger sequential reads on the raw data file and fewer, contiguous skips. VA+file performs better than ADS+ since it incurs less random and almost no sequential I/Os (Figure~\ref {fig:exact:synthetic:disk:cache:combined}).

\noindent\textbf{Scalability/Search Efficiency vs Dataset Size - HDD.}
Figure~\ref{fig:exact:datasize:time:idxproc:cache:combined} compares the scalability and search efficiency of the best methods on the HDD platform for the $Synth$-$Rand$ workload on synthetic datasets ranging from 25GB to 1TB. There are 4 scenarios: indexing (Idx), answering 100 exact queries (Exact100), indexing and answering 100 exact queries (Idx+Exact100), and indexing and answering 10,000 queries (Idx+Exact10K). 
Times are shown in log scale to reveal the performance on smaller datasets.

Figure~\ref{fig:exact:datasize:time:idxproc:cache:combined:indexing} indicates only the indexing times. ADS+  outperforms all other methods and is an order of magnitude faster than the slowest, DSTree. 
Figure~\ref{fig:exact:datasize:time:idxproc:cache:combined:100exact} shows the times for running 100 exact queries. We observe two trends in this plot. For in-memory datasets, VA+file surpasses the other methods. For the larger datasets, the DSTree is a clear winner, followed by VA+file, while the performance of the other methods converge to that of sequential scan.
Figure~\ref{fig:exact:datasize:time:idxproc:cache:combined:idx100exact} refers to indexing and answering the 100 exact queries. For in-memory datasets, ADS+ shows the best performance, with iSAX2+ performing equally well on the 25GB dataset. However, for larger datasets, VA+file outperforms all other methods.

Figure~\ref{fig:exact:datasize:time:idxproc:cache:combined:idx10Kexact} shows the time for indexing and answering 10K exact queries. The trends now change. For in-memory datasets, iSAX2+ and VA+file outperform all other methods, in particular ADS+. Both iSAX2+ and VA+file are slower than ADS+ in index building, but this high initial cost is amortized over the large query workload. 

The DSTree is the best contender for large data sets that do not fit in memory, followed by VA+file and iSAX2+. 
The other methods perform similar to a sequential scan. The DSTree has the highest indexing cost among these methods, but once the index is built, query answering is very fast, thus being amortized for large query workloads. The strength of the DSTree is based on its sophisticated splitting policy, the upper/lower bounds used in query answering, and its parameter-free summarization algorithm.

\ifJournal
Recall that exact search using ADS+, DSTree, iSAX2+ and SFA always involves an initial approximate query answering step, which provides a first estimate of the answer. Figure~\ref{fig:exact:datasize:time:idxproc:cache:combined:100approx} reports the approximate answering times for the 100 queries in the workload, while Figure~\ref{fig:exact:datasize:time:idxproc:cache:combined:idx10Kapprox} reports the combined indexing time and 
approximate answering times for 10K queries. In both scenarios, we can observe that ADS+ is the fastest across all datasets, beating the rest of the methods by up to an order of magnitude. In particular, for the 1TB dataset, it takes ADS+ less than 1 hour to build the index, and only 10 seconds to answer 100 approximate queries. 

\fi

Our results for in-memory datasets 
corroborate earlier studies~\cite{journal/vldb/Zoumpatianos2016} (i.e., ADS+ outperforms alternative methods), yet, we additionally bring in the picture VA+file, which is very competitive and had not been considered in earlier works.
Moreover, for 
out-of-memory data, our results show that ADS+ is not faster than sequential scan, as was previously reported.
The reason for this discrepancy in results lies with the different hardware characteristics, which can significantly affect the performance of different algorithms, both in relative, as well as in absolute terms.
More specifically, the disks used in~\cite{journal/vldb/Zoumpatianos2016} had 60\% of the sequential throughput of the disks used in this paper.
As a result, ADS+ can be outperformed by a sequential scan of the data when the disk throughput is high and the length of the sequences is small enough, where ADS+ is forced to perform multiple disk seeks. Figures~\ref{fig:exact:synthetic:datasize:disk:sequential:cache:combined} and~\ref{fig:exact:synthetic:datasize:disk:random:cache:combined} clearly show that ADS+ performs the smallest number of sequential disk operations and the largest number of random disk operations across all datasets.
In main-memory, SSDs, and with batched I/Os, ADS+ is expected to perform significantly better.
\ifJournal
\begin{figure}[!htb]
	\captionsetup{justification=centering}
	\captionsetup[subfigure]{justification=centering}
  \begin{subfigure}{\columnwidth}
  	\hspace{0.3cm}
  	\includegraphics[width=0.9\columnwidth]{{exact_datasize_time_idxproc_cache_legend}}
  \end{subfigure}
	
	\begin{subfigure}{0.48\columnwidth}
		\centering
		\includegraphics[width=\columnwidth]{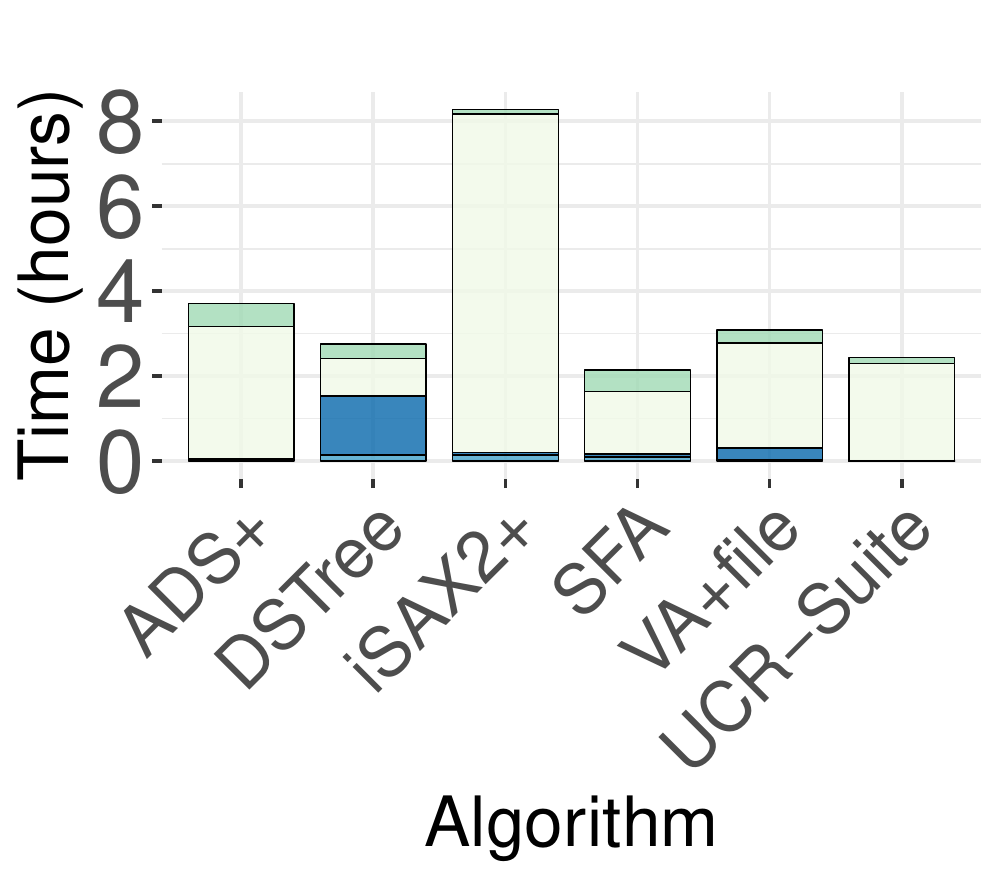}
		\caption{100 Exact Queries}
		\label{fig:exact:real:time:idxproc:cache:seismic:100}
	\end{subfigure}
	\hspace*{\fill} 
	\begin{subfigure}{0.48\columnwidth}
		\centering
		\includegraphics[width=\columnwidth]{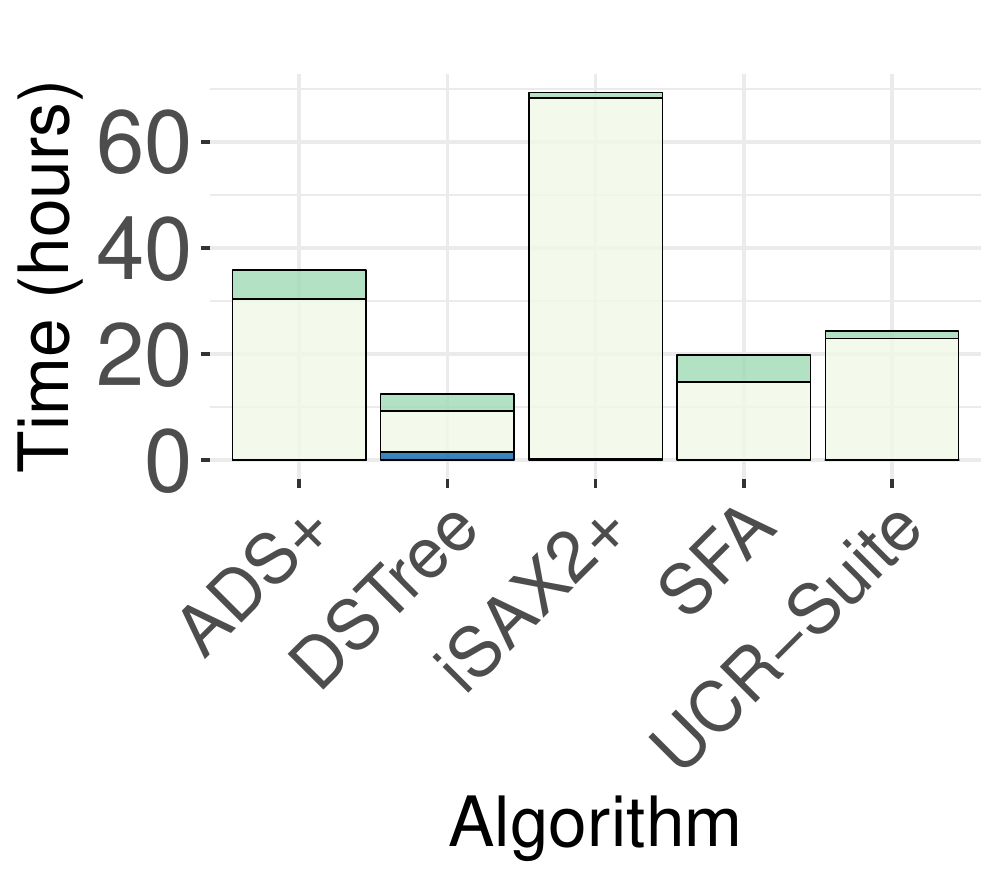}
		\caption{1K Exact Queries}
		\label{fig:exact:real:time:idxproc:cache:seismic:100}	
	\end{subfigure}
	\hspace*{\fill} 
	\caption{Combined Indexing and Exact Querying for the Seismic Dataset}
	\label{fig:exact:real:time:idxproc:cache:seismic}
}
\end{figure}

\begin{figure}[!htb]
	\captionsetup{justification=centering}
	\captionsetup[subfigure]{justification=centering}
  \begin{subfigure}{\columnwidth}
  	\hspace{0.3cm}
  	\includegraphics[width=0.9\columnwidth]{{exact_datasize_time_idxproc_cache_legend}}
  \end{subfigure}
  
	\begin{subfigure}{0.48\columnwidth}
		\centering
		\includegraphics[width=\columnwidth]{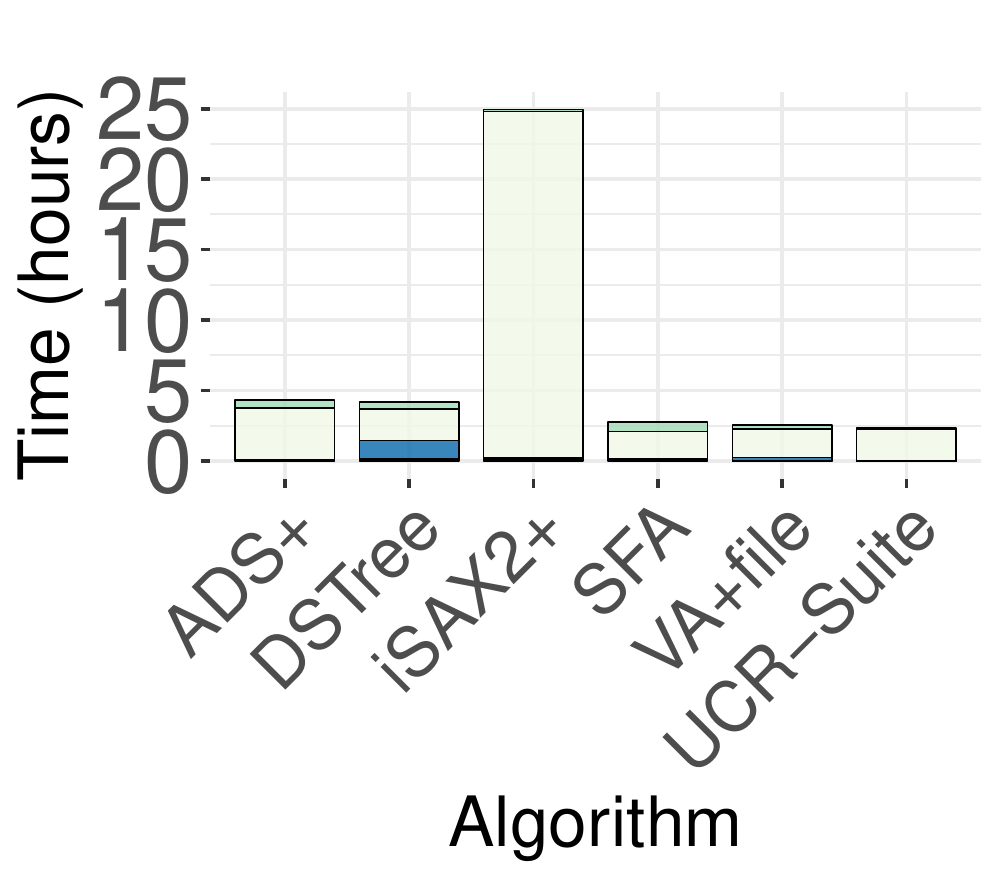}
		\caption{100 Exact Queries}
		\label{fig:exact:real:time:idxproc:cache:astronomy:100}
	\end{subfigure}
	\hspace*{\fill} 
	\begin{subfigure}{0.48\columnwidth}
		\centering
		\includegraphics[width=\columnwidth]{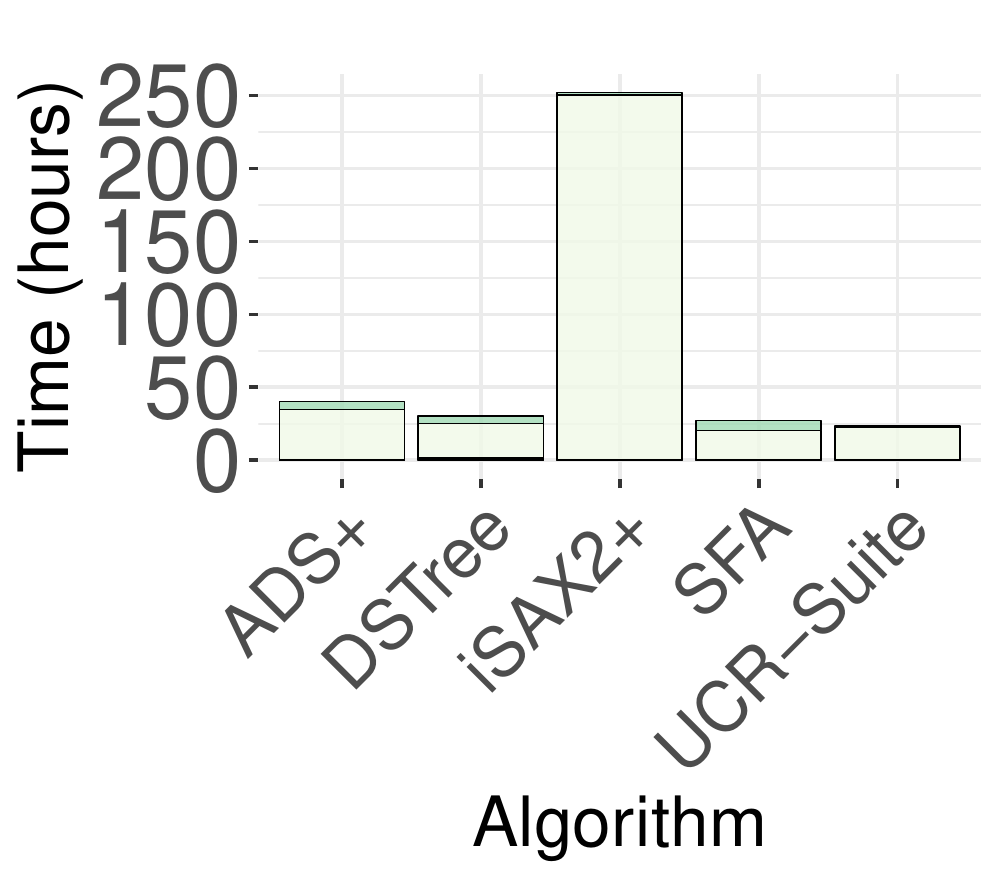}
		\caption{1K Exact Queries}
		\label{fig:exact:real:time:idxproc:cache:astronomy:100}
	\end{subfigure}
	\hspace*{\fill} 
	\caption{Combined Indexing and Exact Querying for the Astronomy Dataset}
	\label{fig:exact:real:time:idxproc:cache:astronomy}
}
\end{figure}

\begin{figure}[!htb]
	\captionsetup{justification=centering}
	\captionsetup[subfigure]{justification=centering}
  \begin{subfigure}{\columnwidth}
  	\hspace{0.3cm}
  	\includegraphics[width=0.9\columnwidth]{{exact_datasize_time_idxproc_cache_legend}}
  \end{subfigure}
  
	\begin{subfigure}{0.48\columnwidth}
		\centering
		\includegraphics[width=\columnwidth]{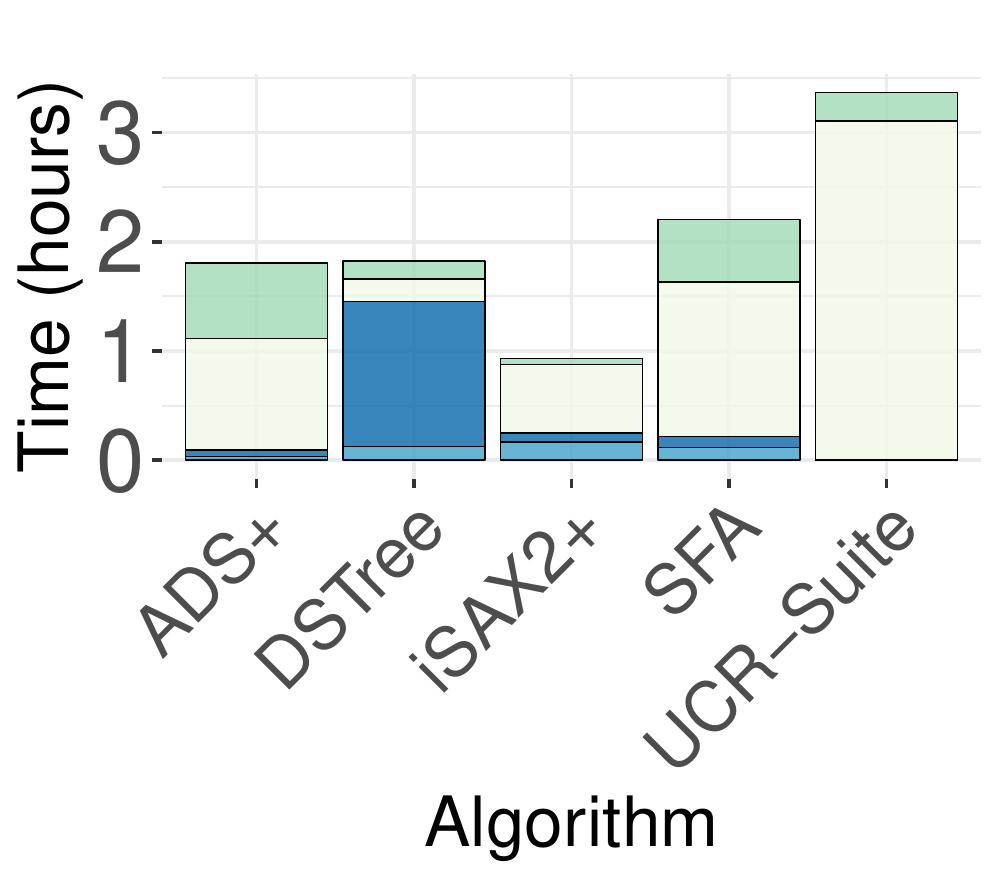}
		\caption{100 Exact Queries}
		\label{fig:exact:real:time:idxproc:cache:eeg:100}
	\end{subfigure}
	\hspace*{\fill} 
	\begin{subfigure}{0.48\columnwidth}
		\centering
		\includegraphics[width=\columnwidth]{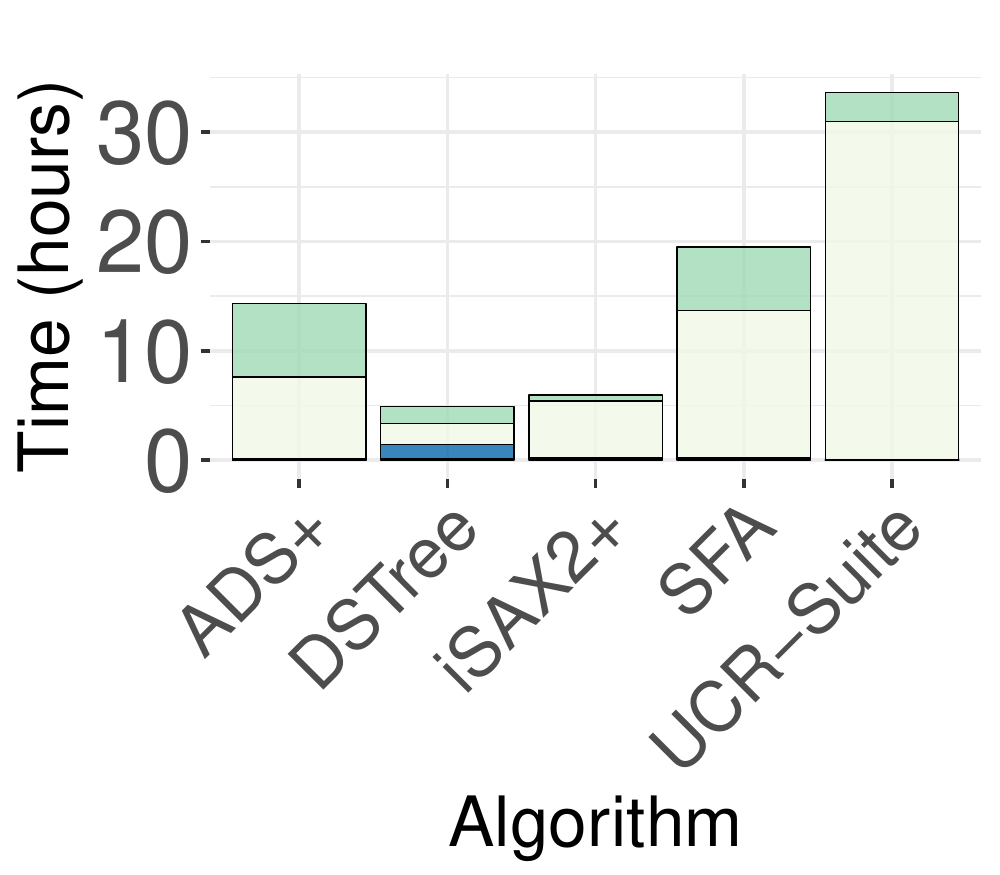}
		\caption{1K Exact Queries}
		\label{fig:exact:real:time:idxproc:cache:eeg:100}
	\end{subfigure}
	\hspace*{\fill} 
	\caption{Combined Indexing and Exact Querying for the EEG Dataset}
	\label{fig:exact:real:time:idxproc:cache:eeg}
}
\end{figure}

\begin{figure}[!htb]
	\captionsetup{justification=centering}
	\captionsetup[subfigure]{justification=centering}
  \begin{subfigure}{\columnwidth}
  	\hspace{0.3cm}
  	\includegraphics[width=0.9\columnwidth]{{exact_datasize_time_idxproc_cache_legend}}
  \end{subfigure}
  
	\begin{subfigure}{0.48\columnwidth}
		\centering
		\includegraphics[width=\columnwidth]{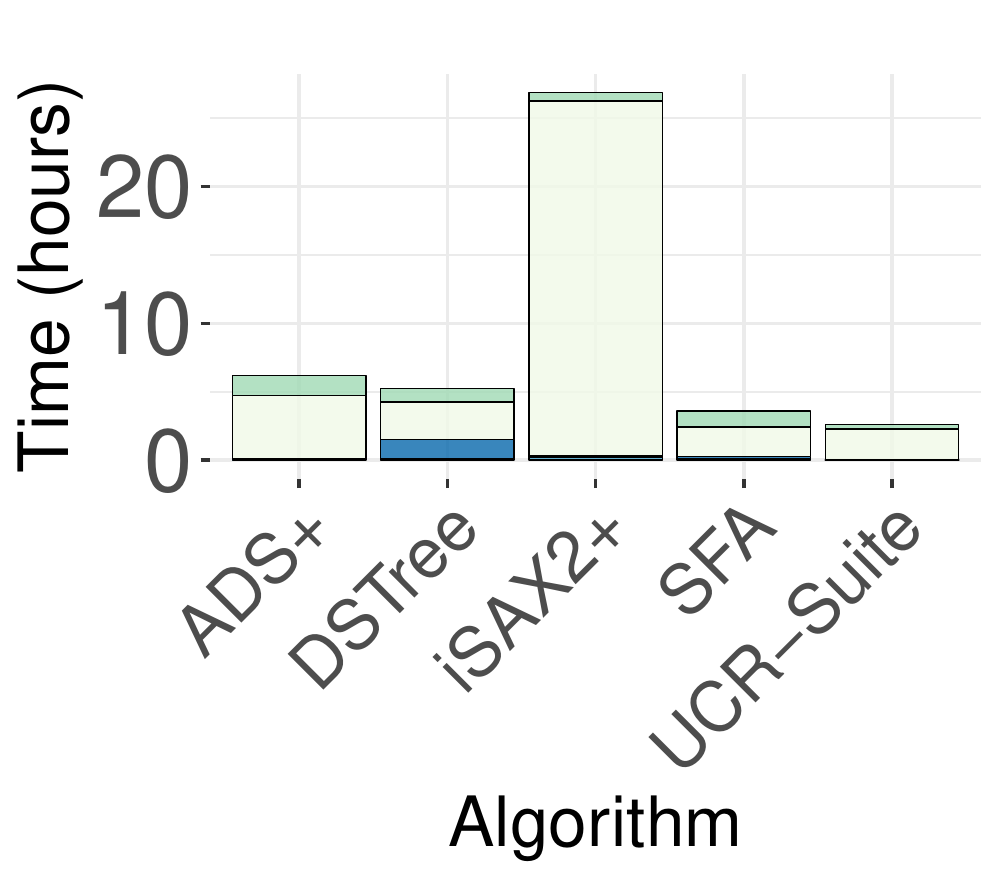}
		\caption{100 Exact Queries}
		\label{fig:exact:real:time:idxproc:cache:deep1b:100:orig}
	\end{subfigure}
	\hspace*{\fill} 
	\begin{subfigure}{0.48\columnwidth}
		\centering
		\includegraphics[width=\columnwidth]{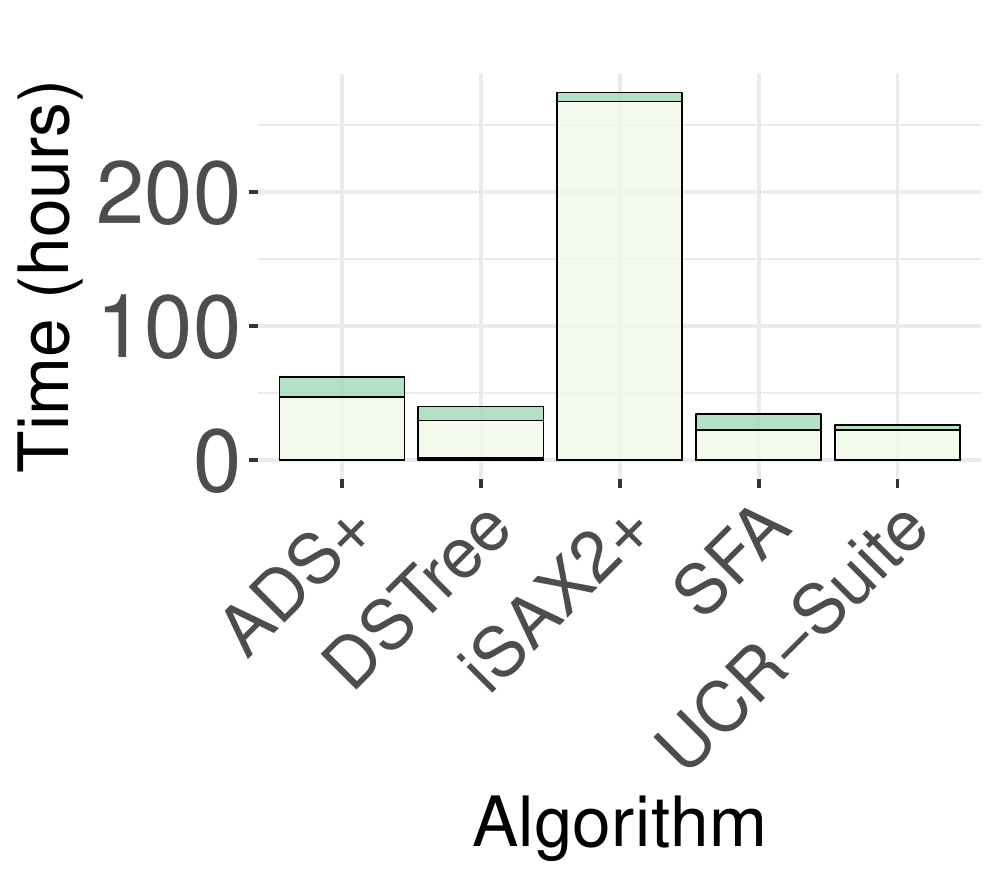}
		\caption{1K Exact Queries}
		\label{fig:exact:real:time:idxproc:cache:deep1b:1000:orig}
	\end{subfigure}
	\hspace*{\fill} 
	\caption{Combined Indexing and Exact Querying for the Deep1B Dataset with Original Queries}
	\label{fig:exact:real:time:idxproc:cache:deep1b:orig}
}
\end{figure}

\begin{figure}[!htb]
	\captionsetup{justification=centering}
	\captionsetup[subfigure]{justification=centering}
	\begin{subfigure}{\columnwidth}
		\hspace{0.3cm}
		\includegraphics[width=0.9\columnwidth]{{exact_datasize_time_idxproc_cache_legend}}
	\end{subfigure}
	
	\begin{subfigure}{0.48\columnwidth}
		\centering
		\includegraphics[width=\columnwidth]{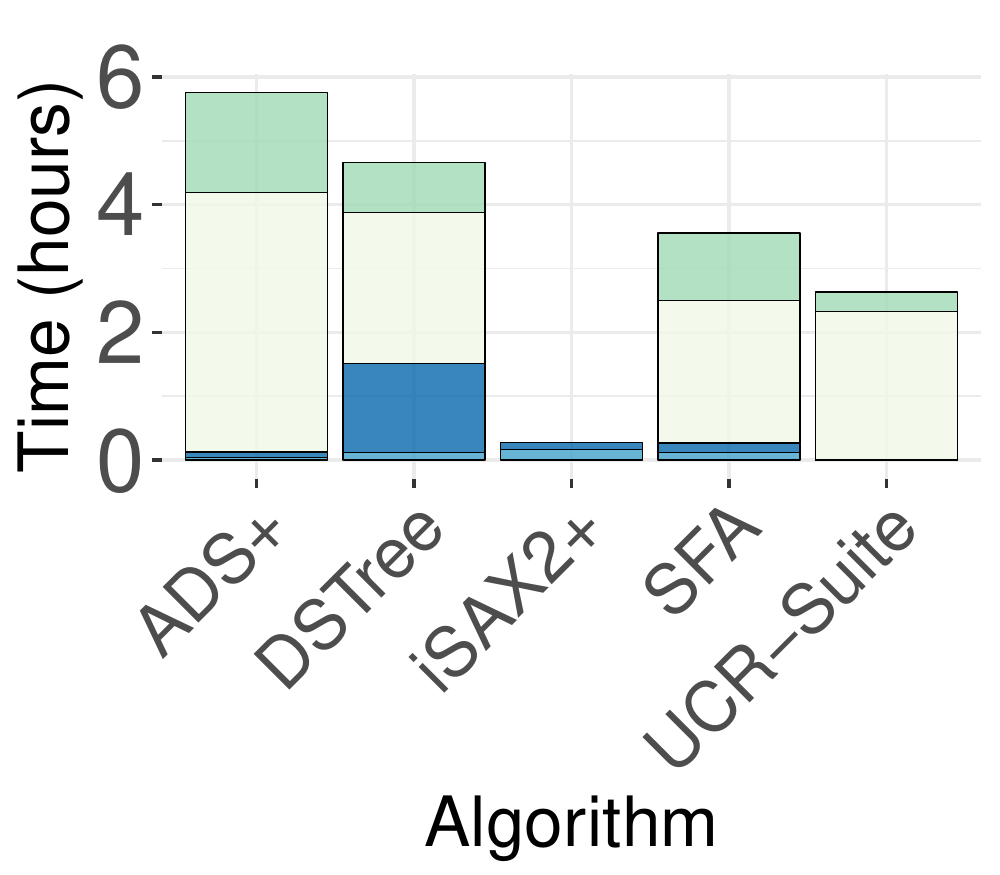}
		\caption{100 Exact Queries}
		\label{fig:exact:real:time:idxproc:cache:deep1b:100:our}
	\end{subfigure}
	\hspace*{\fill} 
	\begin{subfigure}{0.48\columnwidth}
		\centering
		\includegraphics[width=\columnwidth]{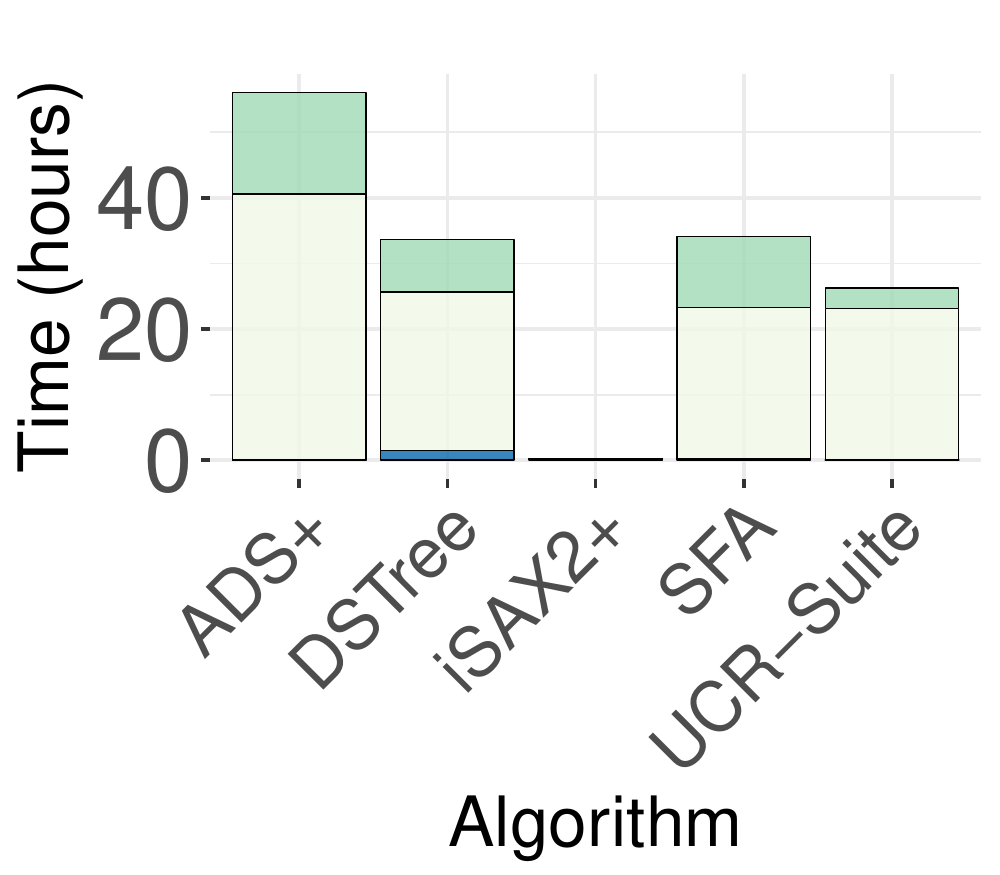}
		\caption{1K Exact Queries}
		\label{fig:exact:real:time:idxproc:cache:deep1b:1000:our}
	\end{subfigure}
	\hspace*{\fill} 
	\caption{Combined Indexing and Exact Querying for the Deep1B Dataset with Our Queries}
	\label{fig:exact:real:time:idxproc:cache:deep1b:our}
}
\end{figure}
\fi

\noindent{\textbf{Scalability/Search Efficiency vs Dataset Size - SSD.}}
In order to further study the effect of different hardware on the performance of similarity search methods, we repeated the experiments described in the last paragraph on the SSD machine. 
We once again tuned each index on the 100GB dataset to find the optimal leaf threshold, which this time was an order of magnitude smaller than the optimal leaf size for the HDD platform. However, we were not able to perform experiments with our larger datasets with these smaller leaf sizes, because the maximum number of possible split points was reached before indexing the entire dataset. Although small leaf sizes can improve performance on smaller datasets, they cannot be used in practice, since the index itself cannot be constructed.
Therefore, we iteratively increased the leaf sizes, and picked the ones that worked for all datasets in our experiments: these leaf sizes proved to be the same as the ones for the HDD platform. 
We note that the SFA trie was particularly sensitive to parametrization.

There are two main observations on these results (see Figure~\ref{fig:exact:datasize:time:idxproc:cache:combined:nefeli}). The first is that VA+file and ADS+ are now the best performers on most scenarios. The only exceptions are iSAX2+ surpassing ADS+ on the 25GB workload, and iSAX2+/SFA being faster in indexing the in-memory datasets. As discussed earlier, the bottleneck of ADS+ and VA+file is random I/O, so the fast performance of the SSD machine on random I/O explains why they both win over the other methods. ADS+ is faster than  VA+file at indexing, while the opposite is true for query answering. The indexing cost of VA+file is amortized in the 10K workload. The second observation is that UCR-Suite performs poorly, due to the low disk throughput of the SSD server.

\begin{table}[tb]
\caption{Controlled workloads experimental results summary (sequential scan algorithm is highlighted) } 
\scriptsize
\renewcommand{\arraystretch}{0.95}
{
\centering
\setlength{\tabcolsep}{4pt} 
\resizebox{\columnwidth}{!}{%
\begin{tabular*}{\columnwidth}{|*{9}{c|}} 		
	\cline{3-9}
\multicolumn{2}{c|}{} & \multicolumn{6}{c|}{Scenarios}  \\\hhline{---------} 	
&  Dataset & Idx &   & Idx+ & Idx+ &  &  \\
&    &  & Exact  & Exact & Exact  & Exact & Exact   \\
&    &  &   100 & 100 & 10K  & Easy-20 & Hard-20   
\\\hhline{---------}		 			 
\multicolumn{1}{|c|}{\multirow{6}{*}{\rotatebox[origin=c]{90}{ HDD }}} & Small &A&D&S&D&D&D\\ \hhline{~--------}		 			 		
&Large  & A & D  &S & D  & D & D \\\hhline{~--------}
&\multicolumn{1}{c|}{Astro} & A & \cellcolor{lightgray!40}U  & \cellcolor{lightgray!40}U &V&V  & \cellcolor{lightgray!40}U \\\hhline{~--------}		 			 
&\multicolumn{1}{c|}{Deep1B}  & A & \cellcolor{lightgray!40}U  & \cellcolor{lightgray!40}U &  \cellcolor{lightgray!40}U   & D & \cellcolor{lightgray!40}U \\\hhline{~--------}			 			 
&\multicolumn{1}{c|}{SALD}  & A & D & I & D  & D & D \\\hhline{~--------}		 		 
&\multicolumn{1}{c|}{Seismic} & A & D   & S & D & D &  \cellcolor{lightgray!40}U\\\hhline{---------}			 				
\multicolumn{1}{|c|}{\multirow{6}{*}{\rotatebox[origin=c]{90}{ SSD }}} & Small  &S&D & I &D & I &D\\\hhline{~--------}			 			 		
&\multicolumn{1}{c|}{Large} &S&D&I&D& I & D \\\hhline{~--------}			 			 
&\multicolumn{1}{c|}{Astro}   &I&V&V &V&V&V \\\hhline{~--------}	 			 
&\multicolumn{1}{c|}{Deep1B}   &S  & I  & I & V& I   & \cellcolor{lightgray!40}U   \\\hhline{~--------}			 			 
&\multicolumn{1}{c|}{SALD} &S& I    & I & I & I &V\\\hhline{~--------}	 			 
&\multicolumn{1}{c|}{Seismic}  &A  &V   &V  &V& D &V\\\hhline{---------}
\multicolumn{8}{c}{\multirow{1}{*}{\textbf{A:} ADS, \textbf{D:} DSTree, \textbf{I:} iSAX2+}}\\
\multicolumn{8}{c}{\textbf{S:} SFA, \textbf{U:} UCR-Suite, \textbf{V:} VA+file}		 			 
\end{tabular*}
}
} 
\label{tab:summary}
\end{table}


\begin{figure*}[tb]
	\captionsetup{justification=centering}
	\captionsetup[subfigure]{justification=centering}
	\begin{subfigure}{0.5\textwidth}
		\centering
		\includegraphics[scale=0.35]{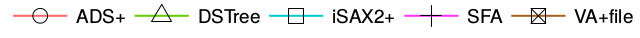}
	\end{subfigure}
	\begin{subfigure}{0.5\textwidth}
			\centering
			\includegraphics[scale=0.43]{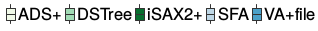}

	\end{subfigure}

	\begin{subfigure}{0.135\textwidth}
		\centering
		\includegraphics[width=\textwidth]{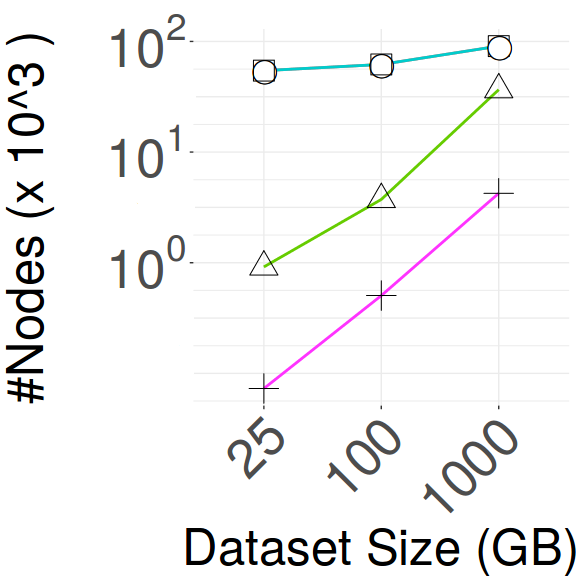}
		\caption{Nodes}
		\label{fig:exact:datasize:nodes:combined}
	\end{subfigure}
	\begin{subfigure}{0.135\textwidth}
		\centering
		\includegraphics[width=\textwidth]{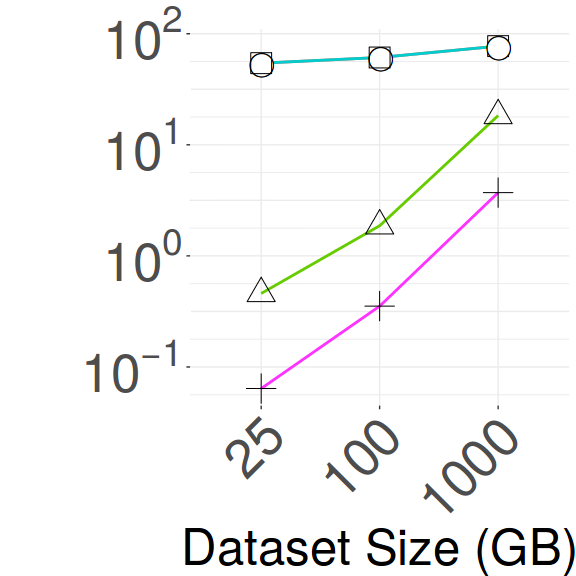}
		\caption{Leaf Nodes}
		\label{fig:exact:datasize:leaves:combined}
	\end{subfigure}
	\begin{subfigure}{0.135\textwidth}
		\centering
		\includegraphics[width=\textwidth]{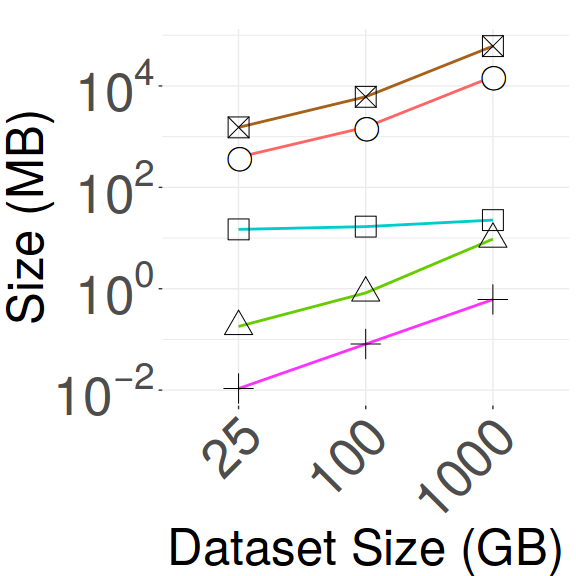}
		\caption{Mem. Size}
		\label{fig:exact:datasize:memory:combined}
	\end{subfigure}
	\begin{subfigure}{0.135\textwidth}
		\centering
		\includegraphics[width=\textwidth]{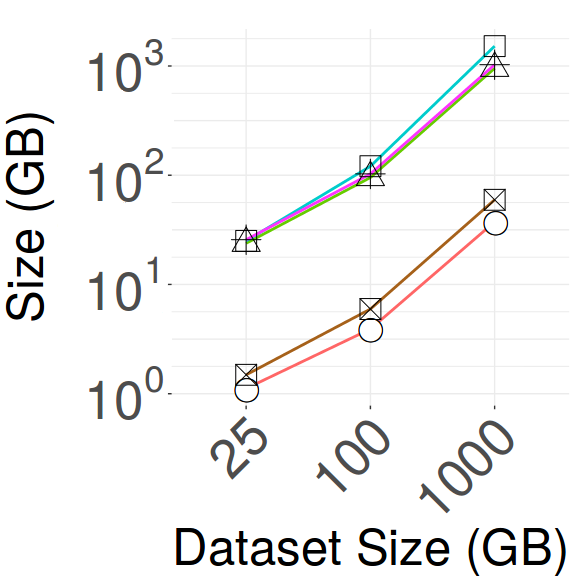}
		\caption{Disk Size}
		\label{fig:exact:datasize:disk:combined}
	\end{subfigure}
	\begin{subfigure}{0.21\textwidth}
		\centering
		\includegraphics[scale=0.235]{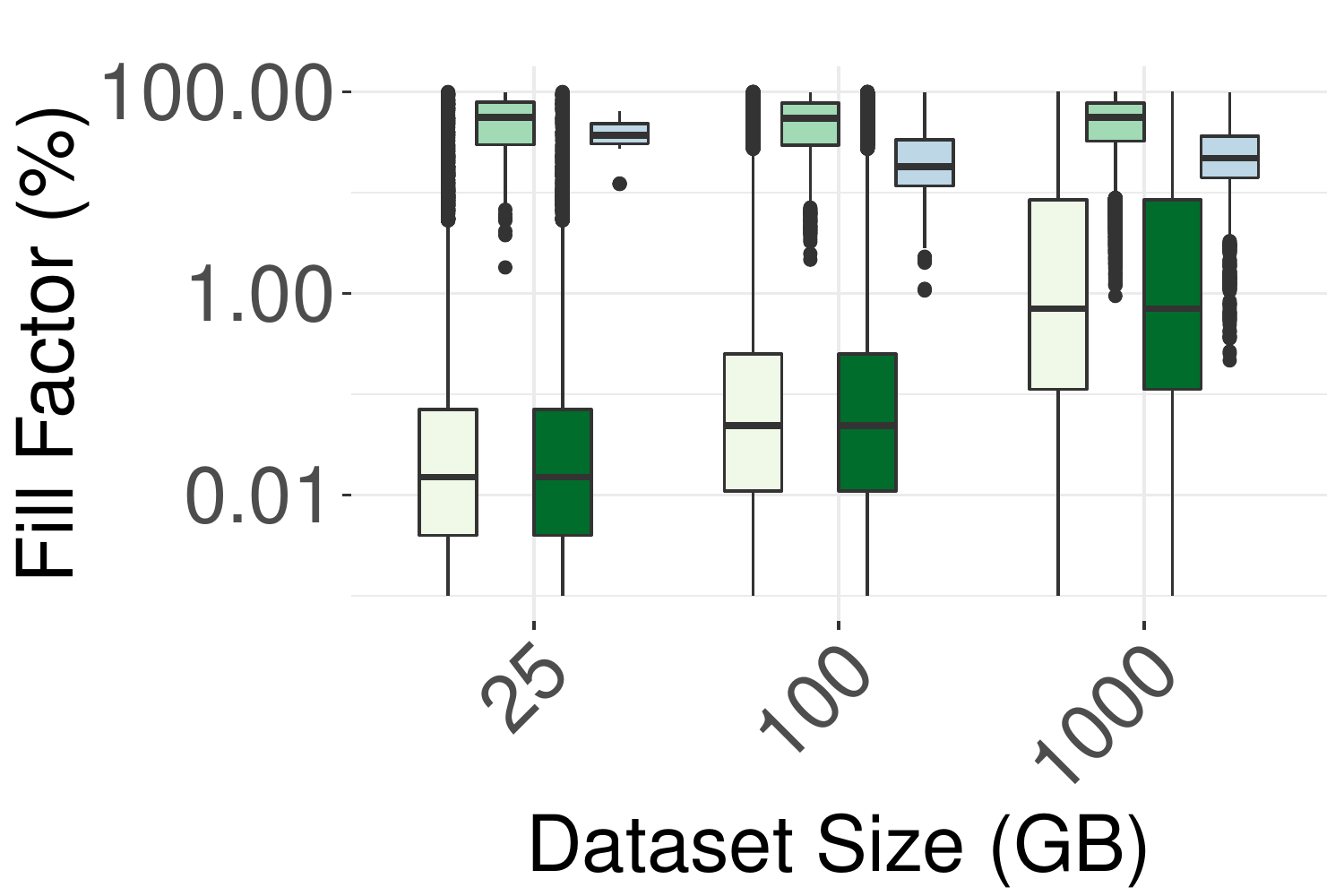}
		\caption{Leaf fill factor}
		\label{fig:exact:datasize:fill:combined}
	\end{subfigure}
	\begin{subfigure}{0.22\textwidth}
		\includegraphics[scale=0.235]{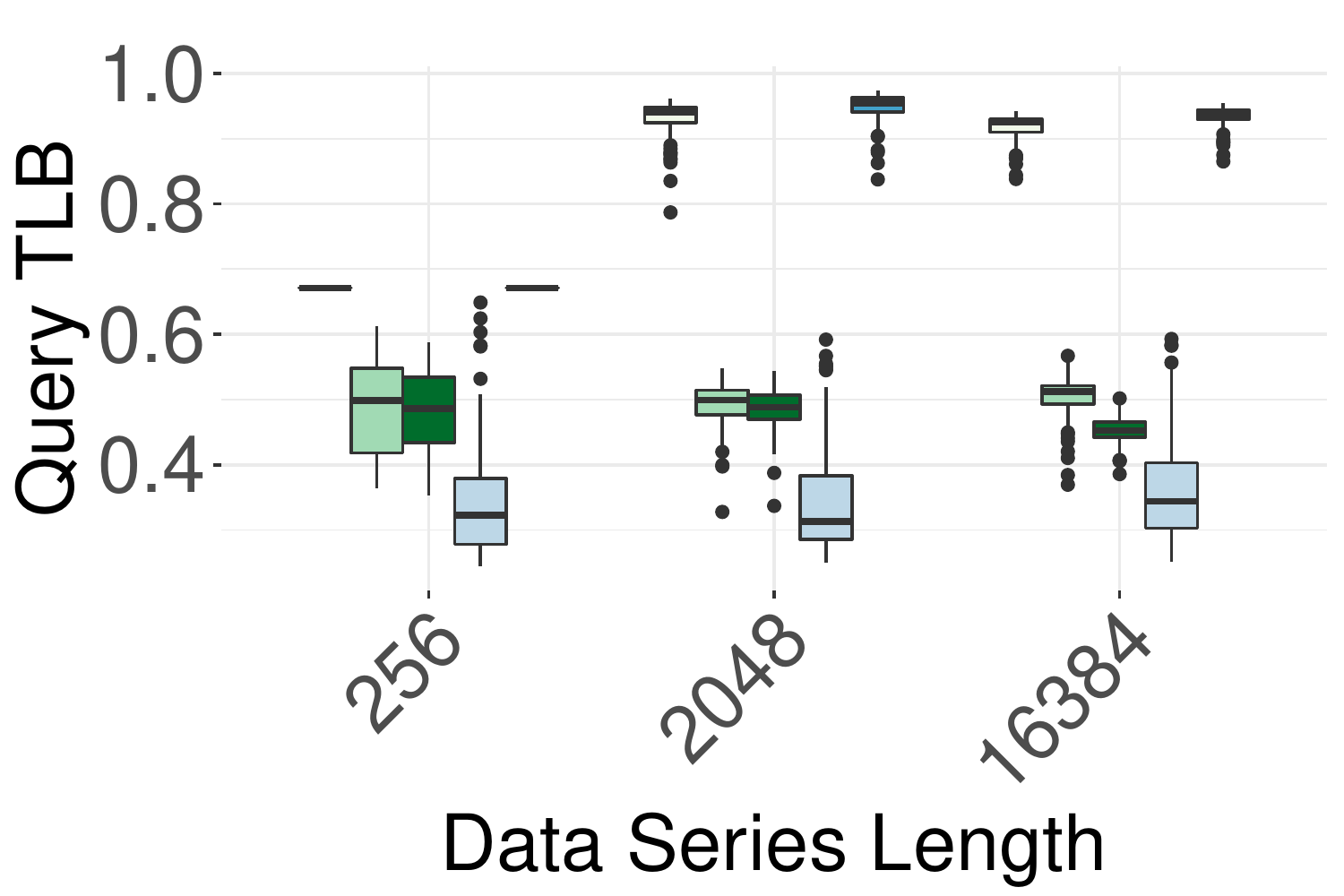}
		\caption{TLB}
		\label{fig:exact:tlb:combined}
	\end{subfigure}
\caption{Exact methods footprint and TLB for synthetic datasets}
\label{fig:exact:datasize:footprint:combined}
\end{figure*}

\ifJournal
{\color{blue}{\bf \\
		SCOPE: This is relevant for the journal version
	}
	{\it \\
    We should extend the vary\_length experiment to vary both length and num of segments: heat map
	}}
\fi

\ifFuture
{\color{blue}{\bf \\
		SCOPE: This is relevant for subsequence matching
	}
	{\it \\
		In this experiment, we compare the best indexes with the UCR-Suite scans subsequent matching. We fix the data set size. We decrease the number of sequences, and we increase the length of sequences. One extreme is good for indexes and one for UCR-Suite scans. \\

	}}
\fi

\noindent{\textbf{Memory/Disk Footprint vs Dataset Size.}}
In this set of experiments, we compare the disk and memory footprints of all methods.
Figure~\ref{fig:exact:datasize:nodes:combined} shows that SAX-based indexes have the largest number of nodes. SFA has a very low number of nodes, because the leaf size we use is 1,000,000 (refer to Figure~\ref{fig:exact:leafsize:time:idxproc}), whereas the leaf sizes for DSTree and iSAX2+ are both 100,000.
The ADS+ index is indifferent to leaf size so we set its initial value to 100,000.
For all methods, most nodes are leaves, as shown in Figure~\ref{fig:exact:datasize:leaves:combined}.
Note that ADS+ and iSAX2+ have the same tree structure with en equal number of nodes, since the leaf size is the same.

As shown in Figures~\ref{fig:exact:datasize:memory:combined} and~\ref{fig:exact:datasize:disk:combined}, the size of the indexes in memory and on disk follows the same trend as the number of nodes. Although ADS+ and iSAX2+ have the same tree shape, some of the data types and structures they use are not the same, thus the different sizes in memory. 
For the VA+file, we only report the size of the approximation file on disk, since it does not build an auxiliary tree structure.

We use two measures to compare the overall structure of the indexes. The first is the leaf nodes fill factor, which measures the percentage of the leaf that is full, and gives a good indication of whether the index distributes evenly the data among leaves.
The second measure is the depth of the leaves, which can help evaluate how balanced an index is.
While none of the best performing index trees is truly height-balanced, some are better balanced in practice than others.
Figure~\ref{fig:exact:datasize:fill:combined} shows the leaf nodes fill factor for different dataset sizes and methods.
(Note that VA+file is missing, since it has no tree; though, if we consider as leaves the pages, where it stores the data, then the fill factor of these pages is 100\%.)
We observe that SFA offers the least variability in the fill factor for the small datasets (as indicated by the size of the boxplot), but the median fill factor fluctuates as the data set changes. DSTree provides the highest median fill factor (as indicated by the line in the middle of the boxplot), which also remains steady with increasing data set sizes.
DSTree also displays the least skew and virtually no outliers, which means that this index effectively partitions the dataset and distributes the series across all its leaves. The SAX-based indexes have many outliers, with some leaves being full and others being empty. 
The graph showing the depth of the indexes can be found elsewhere~\cite{url/DSSeval}.

\ifJournal

\begin{figure*}[tb]
	\captionsetup{justification=centering}
	\includegraphics[height={\setHeight},width={\linewidth}]{exact_realsynth}
	\caption{
		{\color{red}{\bf DO THIS\\
				Choose which experiments to repeat with Real and Synthetic Non Normalized Data Sets.\\
				{\color{red} x-axis:} TBD\\
				{\color{red} y-axis:} TBD\\
				{\color{red} z-axis:} TBD\\
				{\color{red} curves:} TBD)
			}}
		}
	\label{fig:exact:realsynth}
\end{figure*}
\fi
\ifJournal
\begin{figure*}[tb]
	\captionsetup{justification=centering}

	\includegraphics[height={\setHeight},width={\linewidth}]{exact_datasize_tlb_buffersize}
	\caption{
		{\color{blue}{\bf This is relevant for the journal version\\
		Exact Methods Tightness of LowerBound for Increasing Data Series Length and Dimensionality\\
		{\color{red} x-axis:} Data Sets for each method on the same axis (similar to Figure 6 in \cite{conf/kdd/Shieh2008}) or represent as heat map\\
		{\color{red} y-axis:} TLB (0 to 1) \\
		{\color{red} z-axis:} Dimensionality Size in Bytes \\
		{\color{red} curves:} TLB value for each dataset and method (for 1NN only or also other queries?)
		}}
	}
	\label{fig:exact:datasize:tlb}
\end{figure*}
\fi

\ifJournal
\begin{figure*}[tb]
	\captionsetup{justification=centering}
	\includegraphics[height={\setHeight},width={\linewidth}]{exact_dtw}
\caption{
	{\color{blue}{\bf This is relevant for the journal version\\ Choose which experiments to repeat with DTW Whole Matching\\
			{\color{red} x-axis:} TBD\\
			{\color{red} y-axis:} TBD\\
			{\color{red} z-axis:} TBD\\
			{\color{red} curves:} TBD
		}}
	}
	\label{fig:exact:dtw}
\end{figure*}
\fi


\ifJournal
\begin{figure*}[tb]
	\captionsetup{justification=centering}
	\includegraphics[height={\setHeight},width={\linewidth}]{exact_range}
\caption{
	{\color{blue}{\bf This is relevant for the journal version\\
			Choose which experiments to repeat with Range Queries Whole Matching\\
			{\color{red} x-axis:} TBD\\
			{\color{red} y-axis:} TBD\\
			{\color{red} z-axis:} TBD\\
			{\color{red} curves:} TBD)
		}}
	}
	\label{fig:exact:range}
\end{figure*}
\fi

\ifFuture
\begin{figure*}[tb]
	\captionsetup{justification=centering}
	\includegraphics[height={\setHeight},width={\linewidth}]{exact_numseq_time_lenseq}
\caption{
	{\color{blue}{\bf This is relevant for subsequence matching\\
Choose the best indexes based on the previous figures and compare to UCR-Suite scans in the case of subsequence matching\\
{\color{red} x-axis:} Number of Sequences\\
{\color{red} y-axis:} Time in Hours\\
{\color{red} z-axis:} Length of Sequences\\
{\color{red} curves:} Detailed Times (Query and Pre-Query Times)
		}}
	}
	\label{fig:exact:numseq:time:lenseq}
\end{figure*}
\fi


\ifJournal
\begin{figure*}
	\begin{subfigure}{\textwidth}
		\centering
		\includegraphics[width=0.25\textwidth]{{boxplot_partial_legend_greyscale}}
	\end{subfigure}
	\begin{minipage}{0.5\textwidth}
		\captionsetup{justification=centering}
		\captionsetup[subfigure]{justification=centering}
		\begin{subfigure}{0.32\textwidth}
			\centering
			\includegraphics[width=\textwidth]{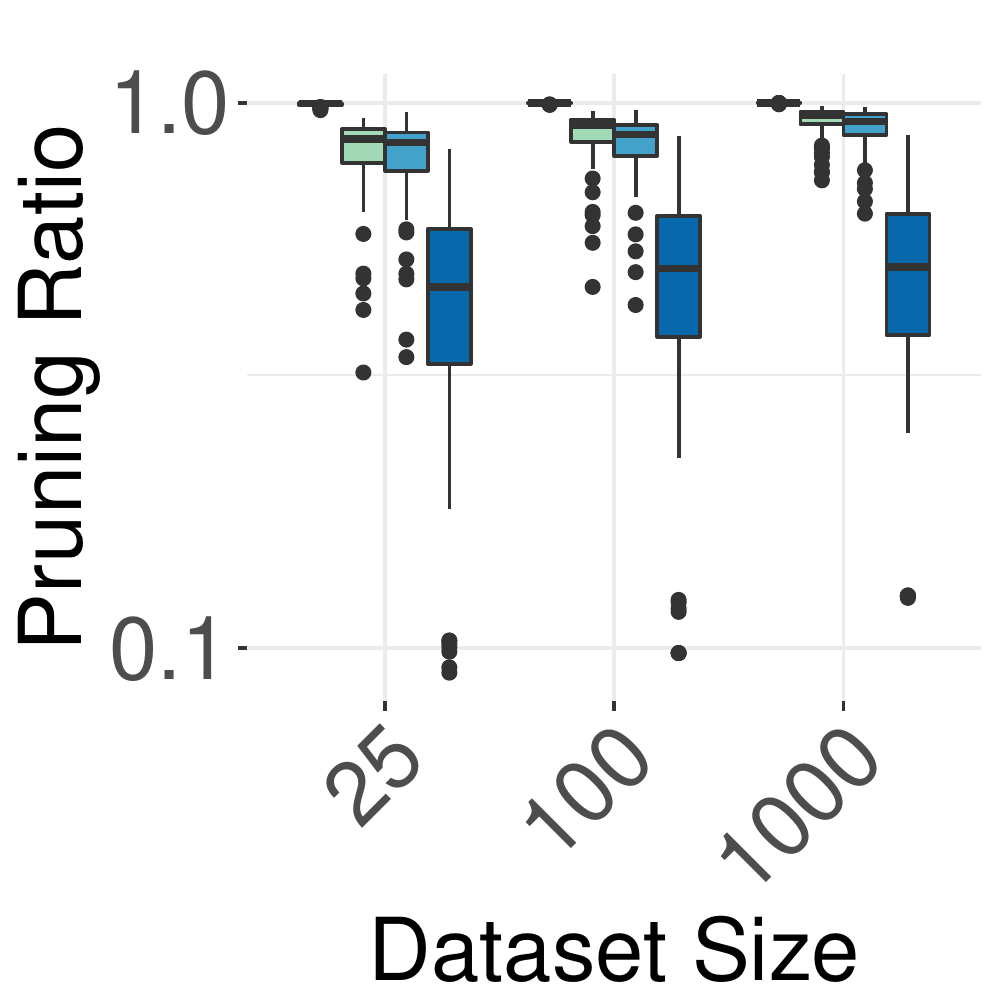}
			\caption{Synthetic}
			\label{fig:exact:synthetic:datasize:pruning:combined}
		\end{subfigure}
		\begin{subfigure}{0.32\textwidth}
			\centering
			\includegraphics[width=\textwidth]{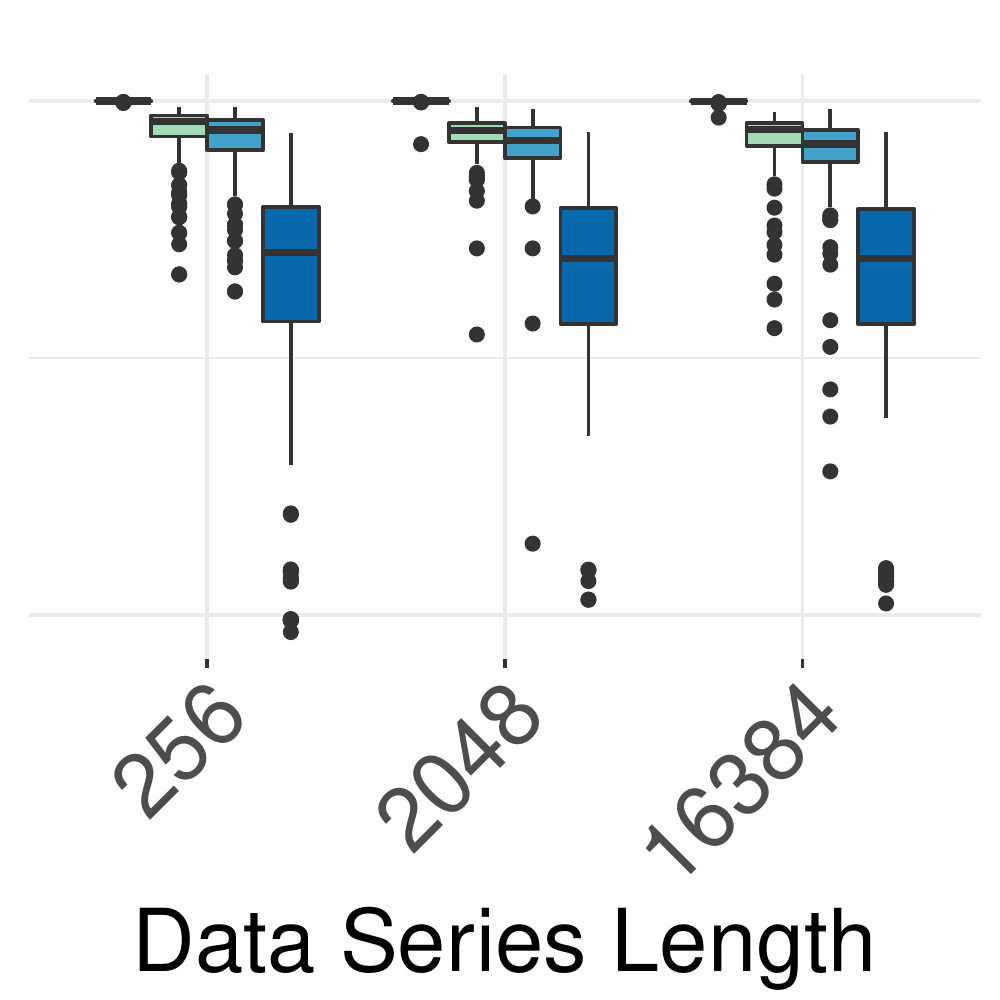}
			\caption{Synthetic}
			\label{fig:exact:synthetic:length:pruning:combined}
		\end{subfigure}	
		\begin{subfigure}{0.32\textwidth}
			\centering
			\includegraphics[width=\textwidth]{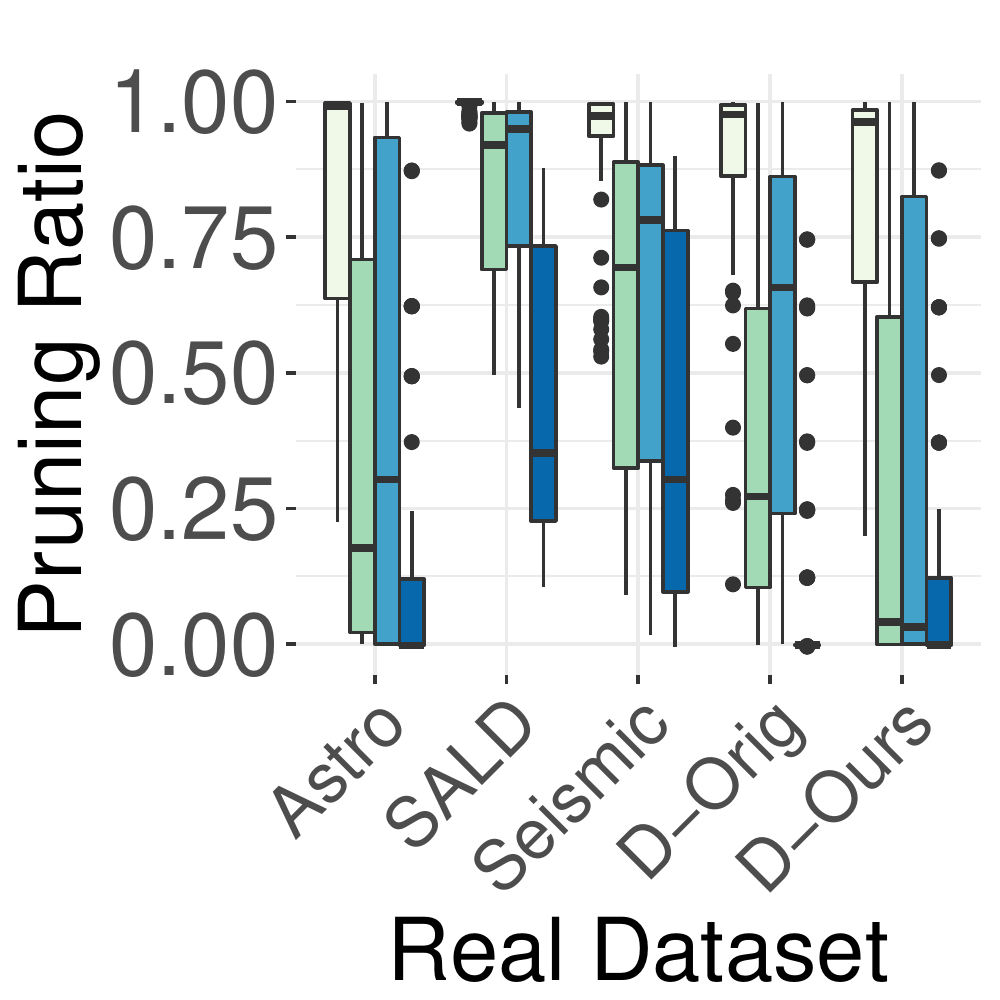}
			\caption{Real}
			\label{fig:exact:real:pruning:combined}
		\end{subfigure}
		\caption{Pruning ratio}
		\label{fig:exact:data::pruning}
	\end{minipage}
	\begin{minipage}{0.5\textwidth}
		\begin{subfigure}{0.32\textwidth}
			\centering
			\includegraphics[width=\textwidth]{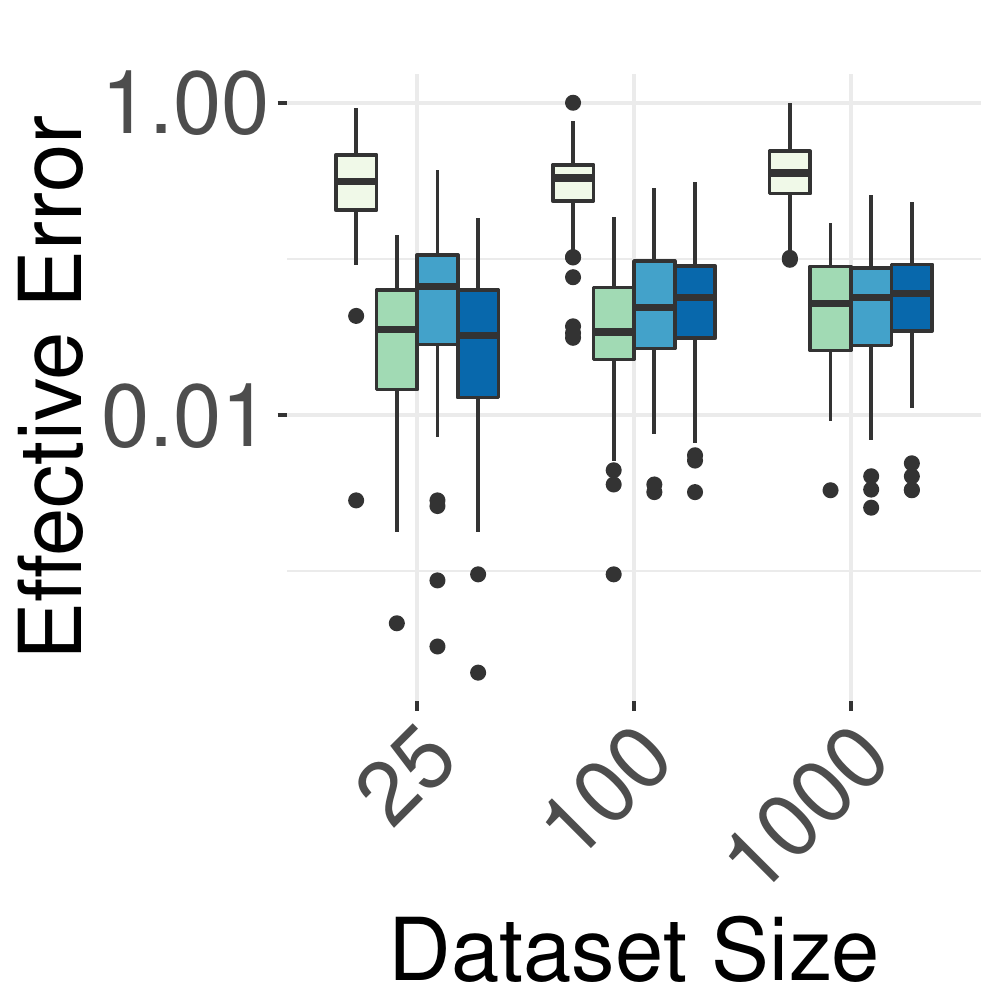}
			\caption{Synthetic}
			\label{fig:exact:synthetic:datasize:error:combined}
		\end{subfigure}
		\begin{subfigure}{0.32\textwidth}
			\centering
			\includegraphics[width=\textwidth]{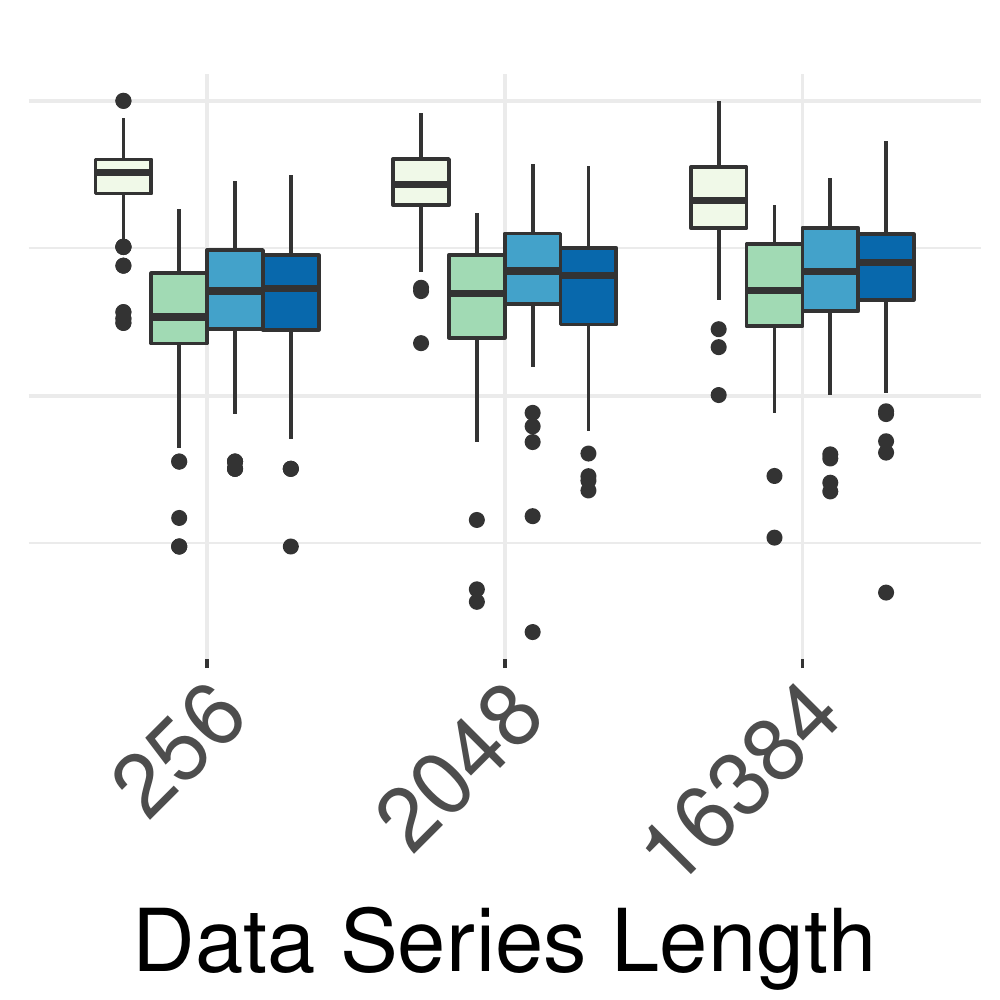}
			\caption{Synthetic}
			\label{fig:exact:synthetic:length:error:combined}
		\end{subfigure}
		\begin{subfigure}{0.32\textwidth}
			\centering
			\includegraphics[width=\textwidth]{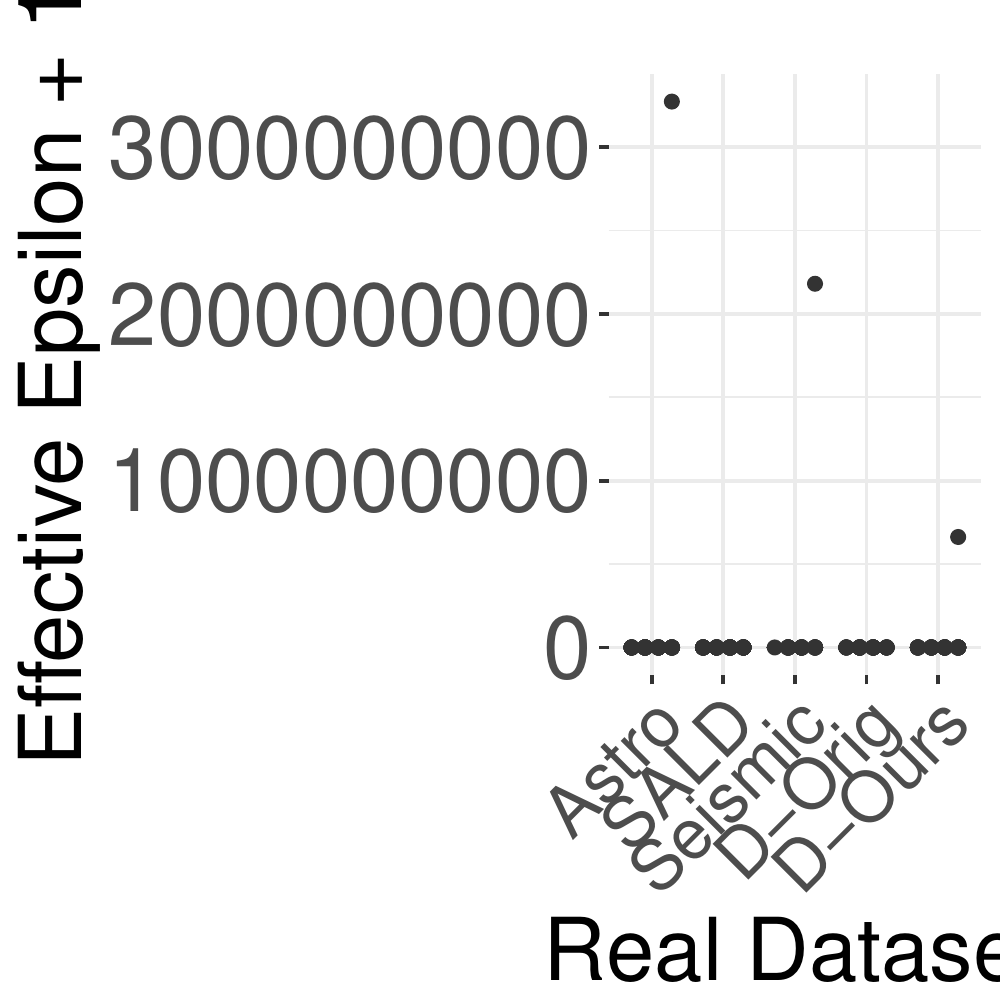}
			\caption{Real}
			\label{fig:exact:real:error:combined}
		\end{subfigure}
		\caption{Effective error} 
		\label{fig:exact:data::error}
	\end{minipage}
\end{figure*}

\fi

\begin{figure}[tb]
	\captionsetup{justification=centering}
	\centering
	\includegraphics[width =0.95\columnwidth]{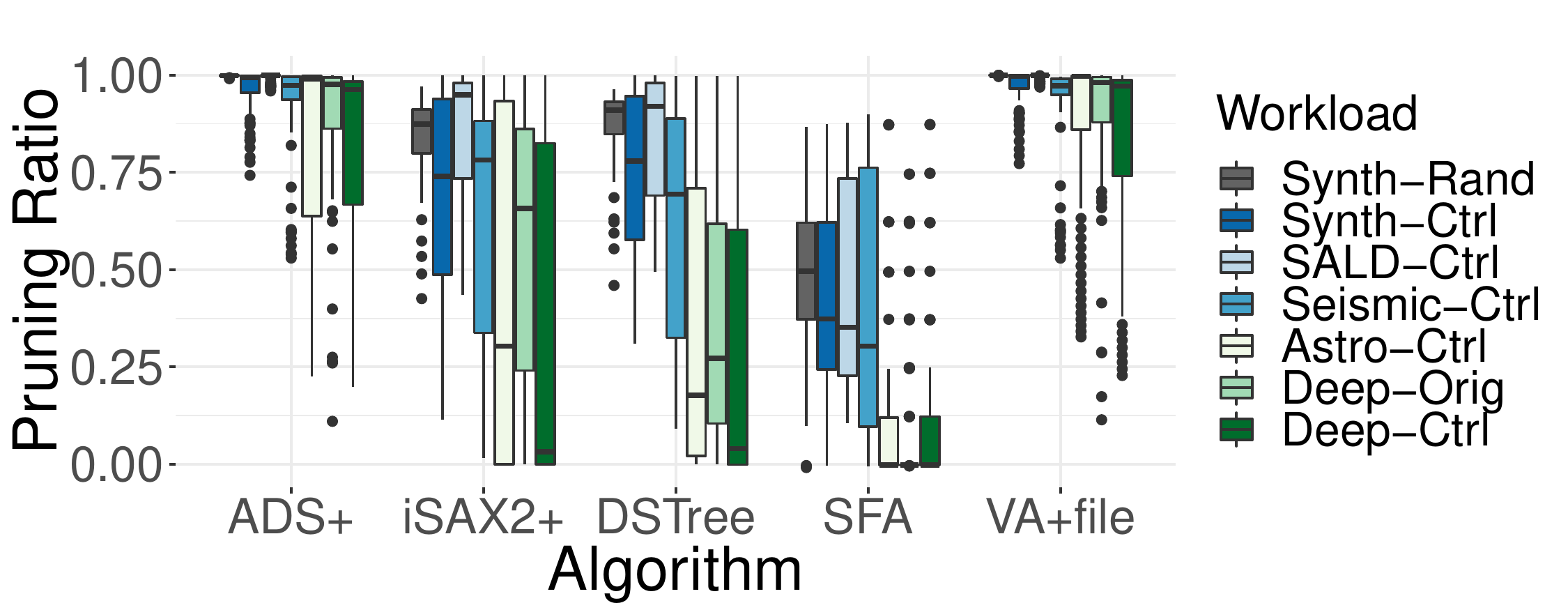}
    \caption{Pruning ratio \\
    	     (Dataset Size= 100GB, Workload = 100 Queries)}
	\label{fig:exact:data::pruning}
\end{figure}

\noindent\textbf{Tightness of the Lower Bound.}
Figure~\ref{fig:exact:tlb:combined} shows the TLB (defined in Section~\ref{subsec:framework}) of each method for increasing data series lengths. We observe that the TLBs of ADS+ and VA+file increase rapidly with increasing lengths, then stabilize when they reach a value close to 1. This explains why the performance of both methods improves with longer series. We also note that VA+file has a slightly tighter lower bound than ADS+, thanks to its non-uniform discretization scheme, which helps explain why VA+file incurs less random I/O than ADS+, and thus performs better. The TLB of the SFA trie is low compared to the other methods, although we used the tight lower bounding distance of SFA (which uses the DFT MBRs). We believe this is due to the optimal alphabet size of 8, which is rather small compared to the default alphabet size of 256 for the SAX-based methods. As for iSAX2+ and DSTree, the main difference in the TLB is that it becomes virtually constant as the length increases.

\noindent\textbf{Pruning Ratio.}
We measure the pruning ratio (higher is better) for all indexes across datasets and data series lengths. For the $Synth$-$Rand$ workload on synthetic datasets, we varied the size from 25GB to 1TB and the length from 128 to 16384. We observed that the pruning ratio remained stable for each method and that overall 
ADS+ and VA+file have the best pruning ratio, followed by DSTree, iSAX2+ and SFA. We also ran experiments with a real workload ($Deep$-$Orig$), a controlled workload on the 100GB synthetic dataset ($Synth$-$Ctrl$), and controlled workloads on the real datasets ($Astro$-$Ctrl$, $Deep$-$Ctrl$, $SALD$-$Ctrl$ and $Seismic$-$Ctrl$). In the controlled workloads, we extract series from the dataset and add noise.
Figure~\ref{fig:exact:data::pruning} summarizes these results. For lack of space, we only report the pruning ratio for the real datasets (all of size 100GB) and the 100GB synthetic dataset. 
The pruning ratio for $Synth$-$Rand$ is the highest for all methods. We observe that the $Synth$-$Ctrl$ workload is more varied than $Synth$-$Rand$ since it contains harder queries with lower pruning ratios. The trend remains the same with ADS+ and VA+file having the best pruning ratio overall, followed by DSTree, iSAX2+ then SFA. For real dataset workloads, ADS+ and VA+file achieve the best pruning, followed by iSAX2+, DSTree, and then SFA. 
The relatively low pruning ratio for the SFA is most probably due to the large leaf size of 1,000,000. Once a leaf is retrieved, SFA accesses all series in the leaf, which reduces the pruning ratio significantly. 
VA+file has a slightly better pruning ratio than ADS+, because it performs less random and sequential I/O, thanks to its tighter lower bound. 
We note that the pruning ratio alone does not predict the performance of an index. In fact, this ratio provides a good estimate of the number of sequential operations that a method will perform, but it should be considered along with other measures like the number of random disk I/Os. \newline

\ifJournal
\noindent\textbf{Accuracy of Approximate Query Answering.}
We measure the effective error (smaller is better) of approximate query search of all indexes when varying dataset size and data series length. Recall that the effective error measures the accuracy of the approximate answer that exact methods calculate in their filtering step (refer to Section~\ref{subsec:framework}). 
Figures~\ref{fig:exact:synthetic:datasize:error:combined} and~\ref{fig:exact:synthetic:length:error:combined} show that for synthetic datasets, neither the dataset size nor the data series length affects the quality of the approximate answer, except for ADS+ whose effective error slightly decreases with increasing lengths. All methods have comparable accuracy, except for ADS+, which has a higher effective error because of the smaller query answering leaf size, as it sacrifices accuracy for speed.
The accuracy varies across real datasets, as can be observed in Figure~\ref{fig:exact:real:error:combined}. In particular, approximate query answering with indexes is more accurate on the {\color{red}SALD-Ctrl and the Deep-Orig} queries. 
 \fi
\noindent{\textbf{Scalability/Search Efficiency with Real Datasets.}} 
In Table~\ref{tab:summary}, we report the name of the best method for each scenario. 
In addition to the four scenarios discussed earlier, we also consider two new scenarios: the average time of the 20 easiest queries (Easy-20) and the average time of the 20 hardest queries (Hard-20) of the corresponding workload. A query is considered easy, or hard, depending on its pruning ratio (computed as the average across all techniques)~\cite{johannesjoural2018}. 

It is important to note that while queries are categorized as easy and hard, easy queries on one dataset may be harder than easy queries on another dataset, as the average pruning ratio for each dataset differs. 
This is because some datasets can be summarized more effectively than others.
We averaged the results over 20 hard queries and 20 easy queries. In-memory datasets are labeled \emph{small} and the others \emph{large}. 

We observe that UCR-Suite wins in exact query answering and on hard queries for the Astro/Deep1B scenarios. This is due to the very low pruning ratio for these workloads. 
DSTree is fast on easy queries and exact query answering on the SALD/Seimic scenarios. ADS+ always wins in indexing on HDD, but is sometimes surpassed by iSAX2+/SFA on SSD.
Similarly to synthetic datasets, the methods behave differently on real datasets when experiments are ran on the SSD platform. VA+file and iSAX2+ have a superior performance overall. DSTree also performs well, while UCR-Suite wins only on hard queries on the Deep1B dataset.

\ifJournal

\begin{figure*}[tb]
	\captionsetup{justification=centering}
	\captionsetup[subfigure]{justification=centering}
	\begin{subfigure}{0.48\textwidth}
		\centering
		 \includegraphics[height={\setHeight},width={\linewidth}]{dual_effeps_accuracy}
		 \caption{
		 	{\color{blue}{\bf Dual Methods Actual Approximation Relative Error $\epsilon_{eff}$ for the 100M dataset\\
		 			{\color{red} x-axis:} Increasing values of $k$ ($k$-NN) \\
		 			{\color{red} y-axis:} Average Actual Epsilon $\epsilon_{eff}$ (range TBD) \\
		 			{\color{red} z-axis:} Dual Method \\
		 			{\color{red} curves:} The actual approximation error averaged for the query workload
		 		}}
		 	}
		\label{fig:dual:effeps:accuracy}
	\end{subfigure}
	\hspace*{\fill} 
	\begin{subfigure}{0.48\textwidth}
		\centering
		\includegraphics[height={\setHeight},width={\linewidth}]{dual_recall_accuracy}
		 \caption{
		 	{\color{blue}{\bf Dual Methods Recall (How many of the real NN did we select true selected knn divided by total true k nn.)\\
		 			{\color{red} x-axis:} Increasing values of $k$ ($k$-NN) \\
		 			{\color{red} y-axis:} Recall \\
		 			{\color{red} z-axis:} Dual Method \\
		 			{\color{red} curves:} The average recall for the query workload  for increasing k values
		 		}}
		 	}
		\label{fig:dual:recall:accuracy}
	\end{subfigure}
\caption{{\color{blue}{\bf  This is relevant for the journal version\\
Dual Methods Accuracy. Add standard deviation of error and max error to see how error varies for each query. Add comparison for range query. Also add extra time on top of approximate to get exact answer.
		}}
	}
	\label{fig:dual:accuracy}
\end{figure*}

\fi


\ifJournal
\begin{figure}[t]
	\captionsetup{justification=centering}
	\captionsetup[subfigure]{justification=centering}
	\begin{subfigure}\columnwidth}
	\centering
	\hspace*{0.5cm}
	\includegraphics[scale=0.5]{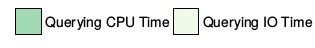}\\
\end{subfigure}

	\begin{subfigure}{0.25\columnwidth}
		\centering
		\includegraphics[width=\columnwidth]{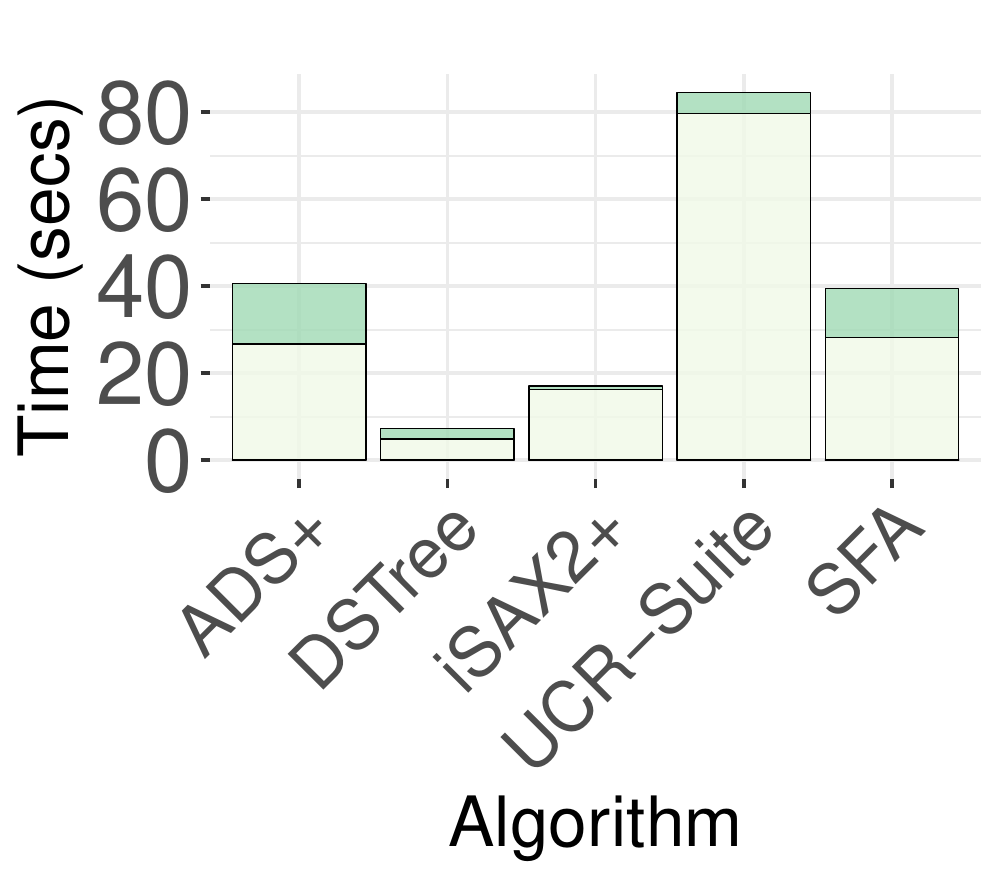}
		\caption{Synthetic Easy}
		\label{fig:exact:real:time:idxproc:cache:synthetic:easy}
	\end{subfigure}
	\begin{subfigure}{0.25\columnwidth}
		\centering
		\includegraphics[width=\columnwidth]{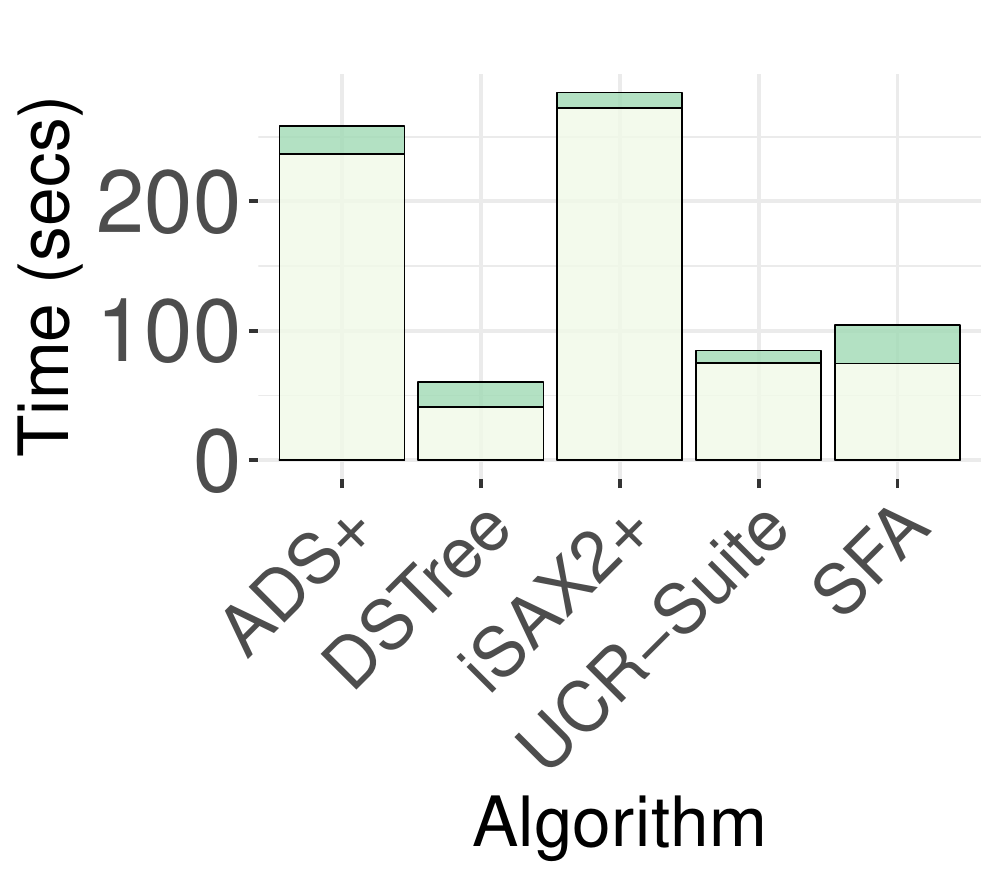}
		\caption{Synthetic Hard}
		\label{fig:exact:real:time:idxproc:cache:synthetic:hard}
	\end{subfigure}
		\begin{subfigure}{0.25\columnwidth}
			\centering
			\includegraphics[width=\columnwidth]{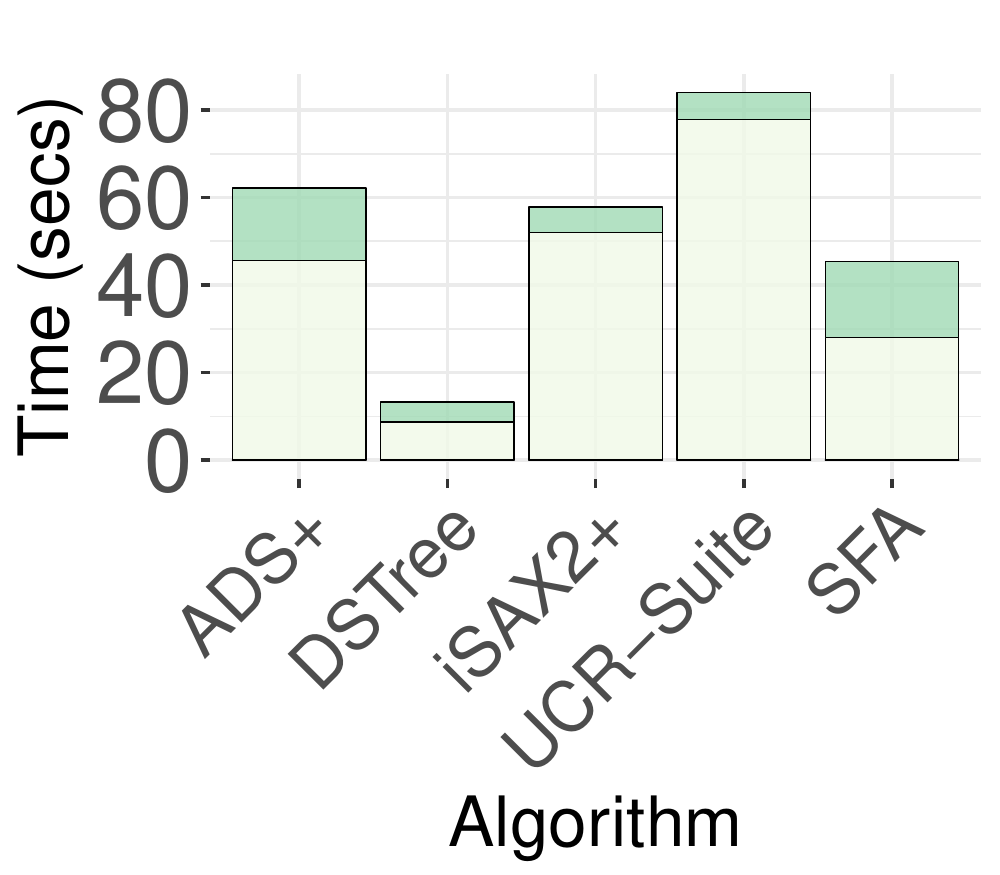}
			\caption{Synthetic Random}
			\label{fig:exact:real:time:idxproc:cache:synthetic:easy}
		\end{subfigure}
		
	\caption{Scalability with Increasing Hardness\\
		(HDD, Synthetic Datasets, Average Query Time)\\}
	\label{fig:exact:real:time:idxproc:cache:hardness}
}
\end{figure}

\begin{figure}[t]
	\captionsetup[subfigure]{justification=centering}
	\begin{subfigure}\columnwidth}
	\centering
	\includegraphics[scale=0.5]{{bar_chart_partial_legend}}\\
	\end{subfigure}	
		\begin{subfigure}{0.25\columnwidth}
			\centering
			\includegraphics[width=\columnwidth]{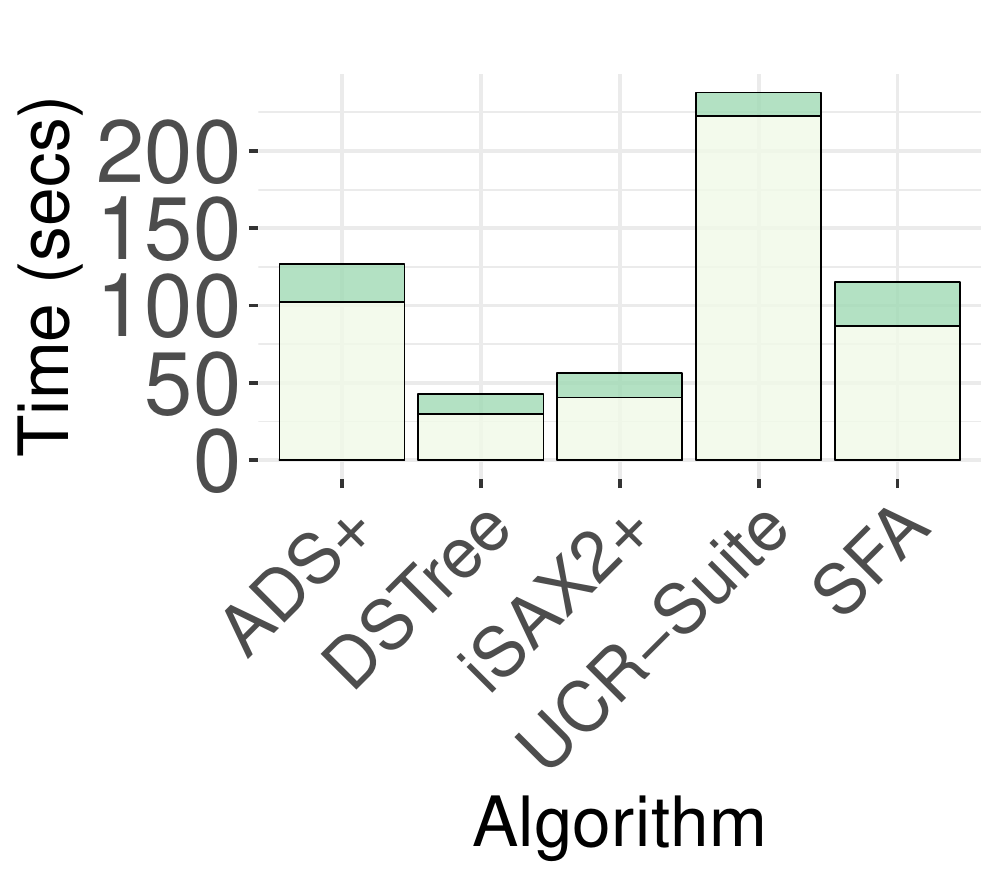}
			\caption{Synthetic Random}
			\label{fig:exact:real:time:idxproc:cache:synthetic:easy}
		\end{subfigure}
			\begin{subfigure}{0.25\columnwidth}
				\centering
				\includegraphics[width=\columnwidth]{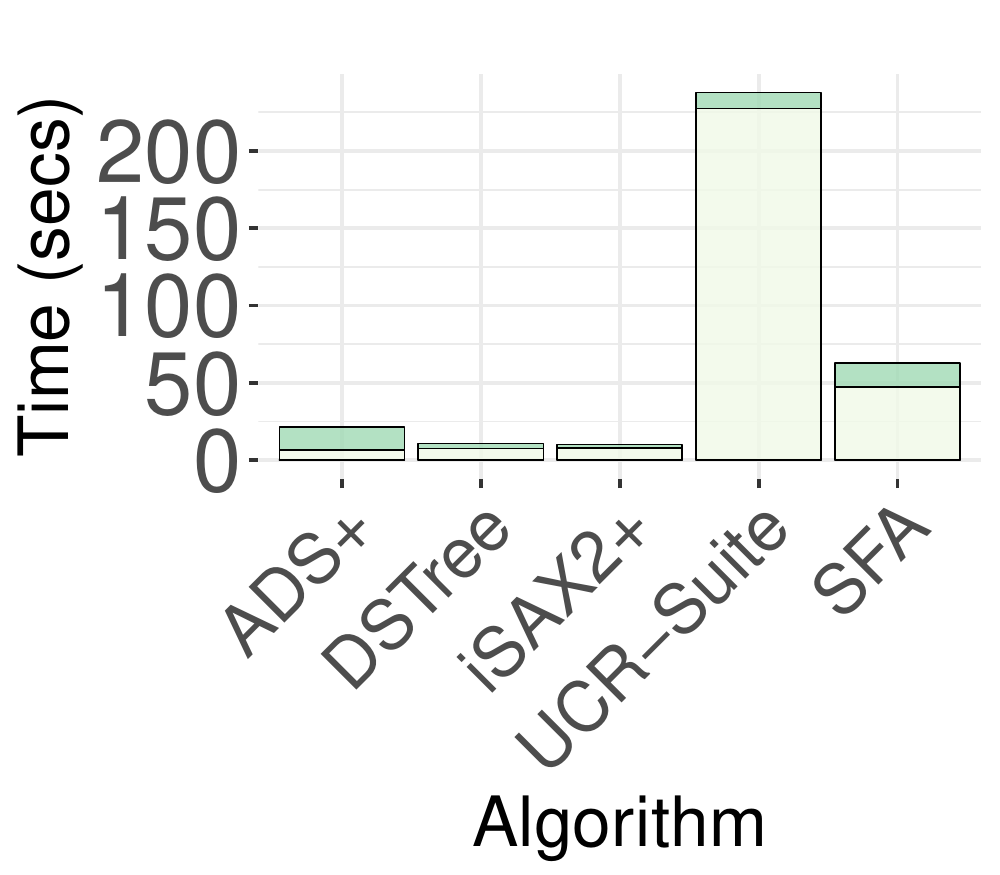}
				\caption{Synthetic Easy}
				\label{fig:exact:real:time:idxproc:cache:synthetic:easy}
			\end{subfigure}
			\begin{subfigure}{0.25\columnwidth}
				\centering
				\includegraphics[width=\columnwidth]{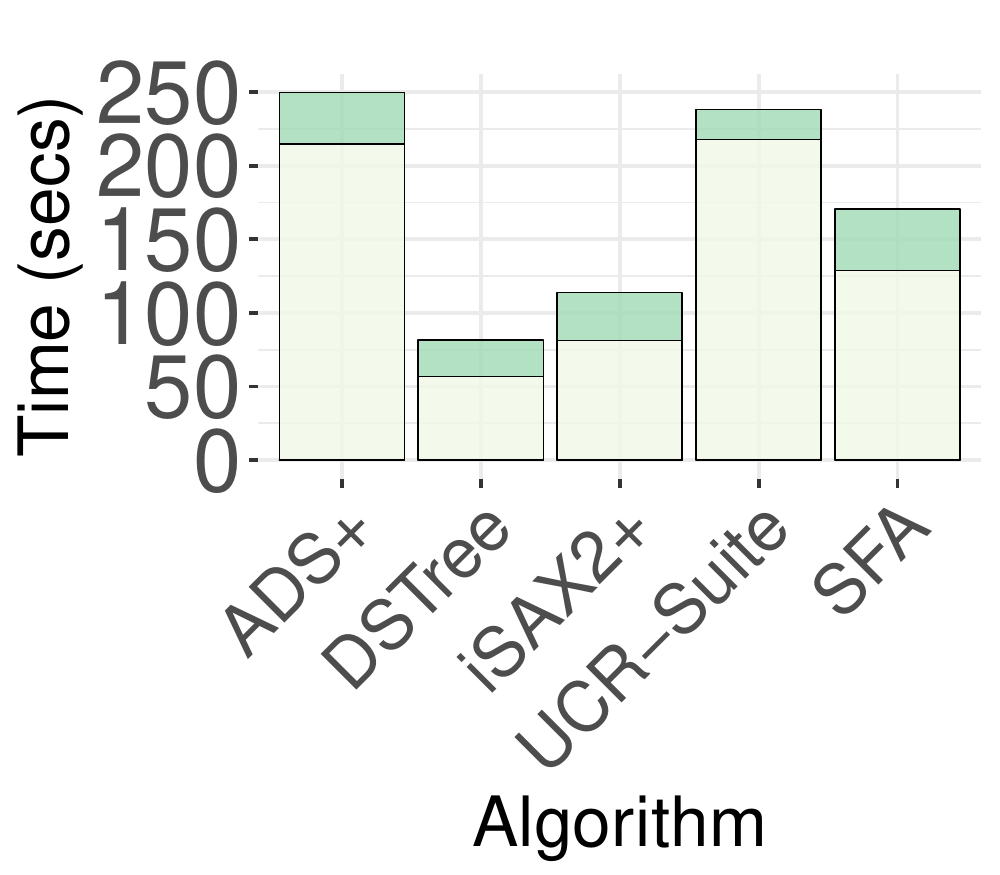}
				\caption{Synthetic Hard}
				\label{fig:exact:real:time:idxproc:cache:synthetic:hard}
			\end{subfigure}
			
	\caption{Scalability with Increasing Hardness\\
		(SSD, Synthetic Datasets, Average Query Time)\\}
	\label{fig:exact:real:time:idxproc:cache:hardness}
}
\end{figure}
\fi
\ifJournal
\begin{figure*}[tb]
	\captionsetup{justification=centering}
	\captionsetup[subfigure]{justification=centering}		
	\begin{subfigure}\textwidth}
		\centering
		\hspace*{5.5cm}
		\includegraphics[scale=0.5]{{bar_chart_partial_legend}}
    \end{subfigure}

	\begin{subfigure}{0.16\textwidth}
		\centering
		\includegraphics[width=\textwidth]{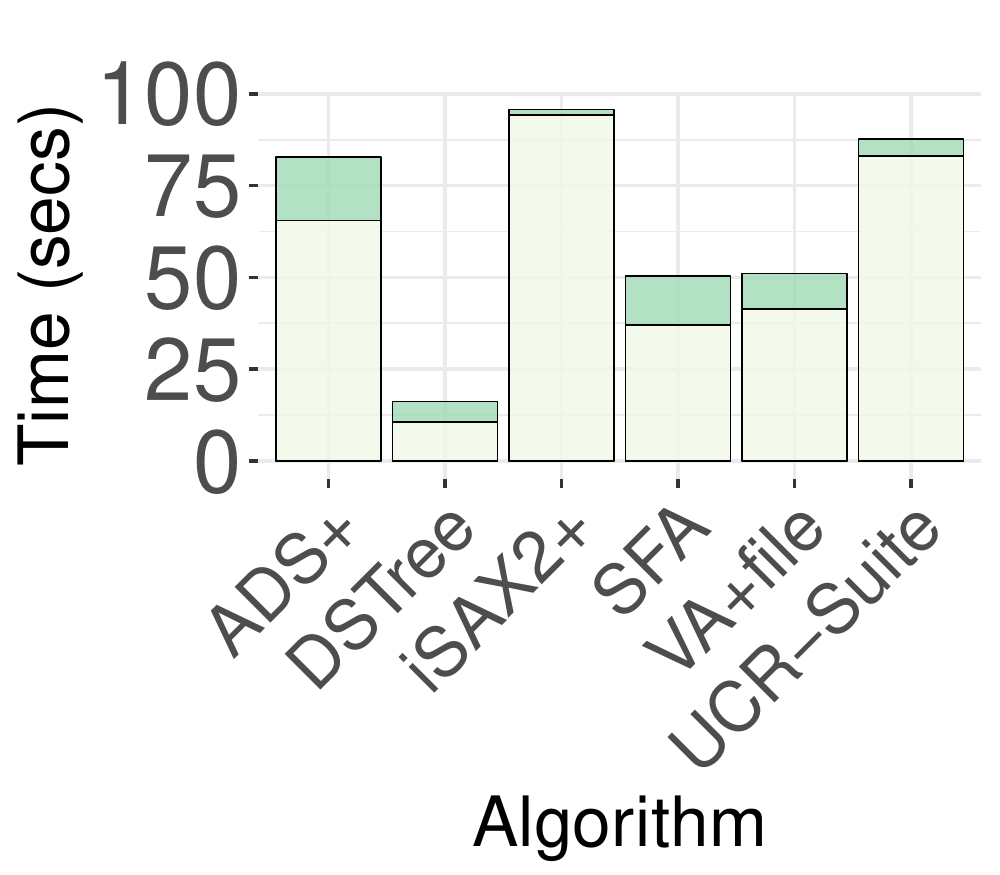}
		\caption{Seismic \\ 
				\hspace*{0.5cm}
				Easy}
		\label{fig:exact:real:time:idxproc:cache:seismic:easy}
	\end{subfigure}
	\begin{subfigure}{0.16\textwidth}
		\centering
		\includegraphics[width=\textwidth]{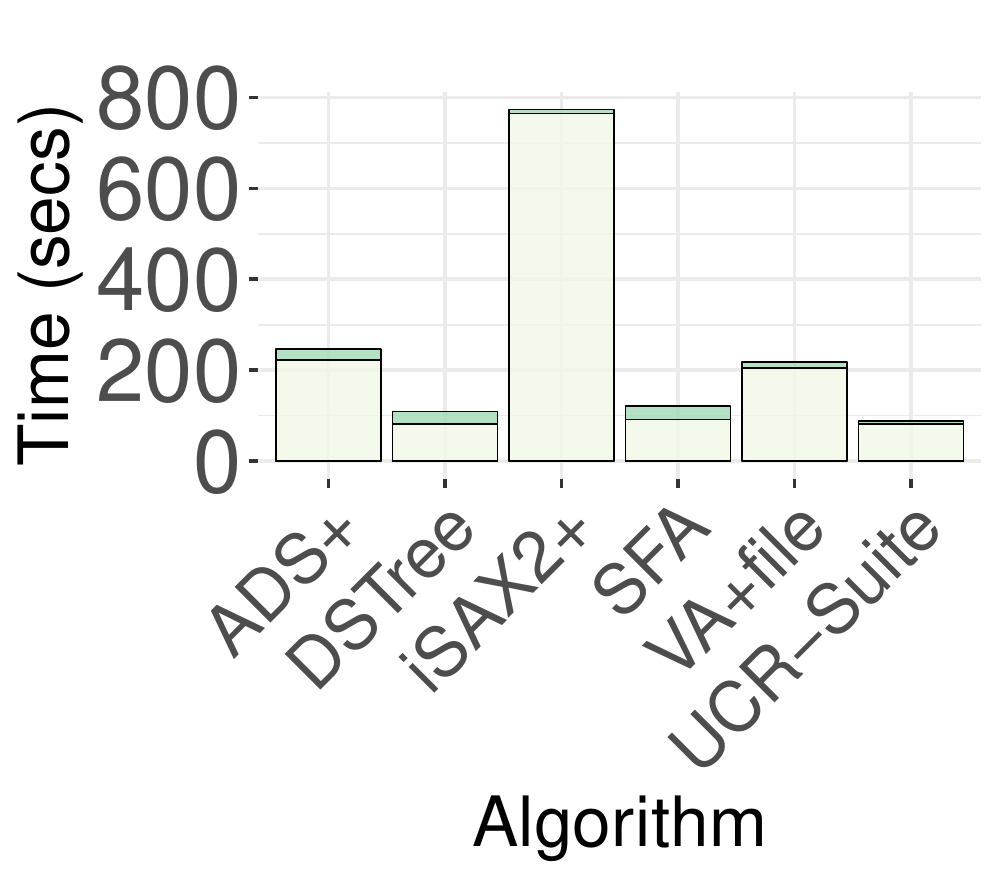}
		\caption{Seismic \\ 
				\hspace*{0.5cm}
				Hard}
		\label{fig:exact:real:time:idxproc:cache:seismic:hard}
	\end{subfigure}
	\begin{subfigure}{0.16\textwidth}
		\centering
		\includegraphics[width=\textwidth]{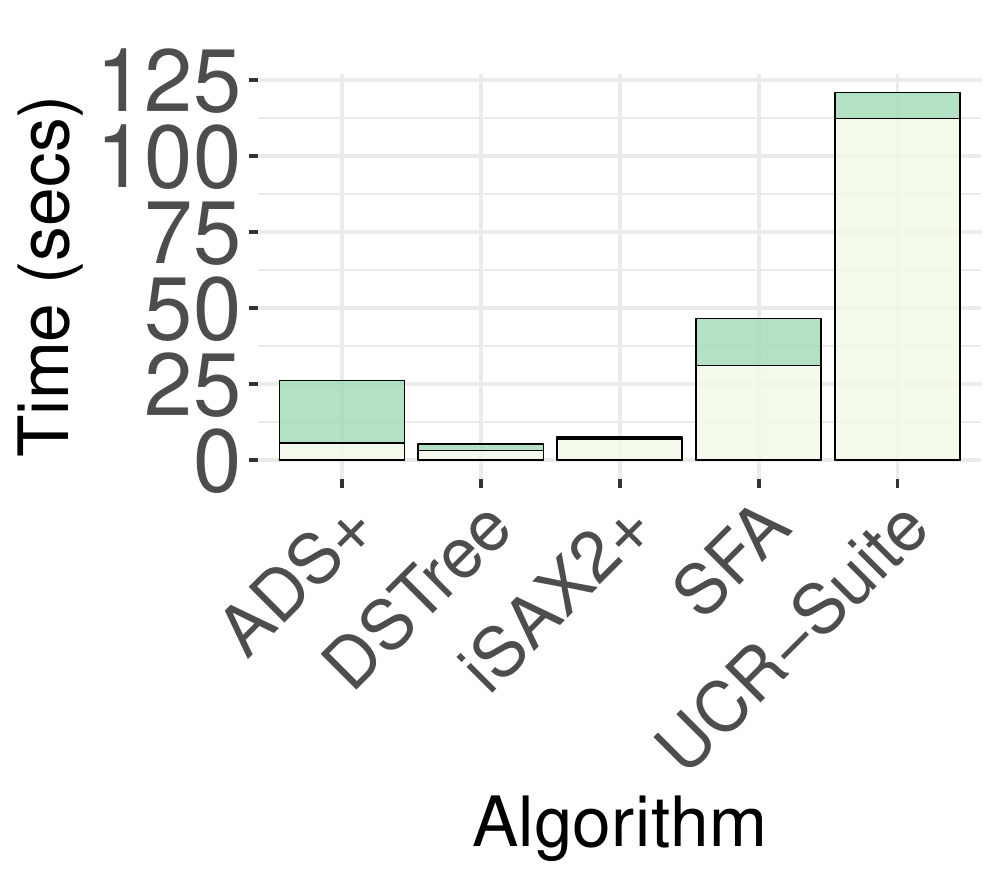}
		\caption{EEG \\ 
				\hspace*{0.5cm}
				Easy}
		\label{fig:exact:real:time:idxproc:cache:eeg:easy}
	\end{subfigure}
	\begin{subfigure}{0.16\textwidth}
		\centering
		\includegraphics[width=\textwidth]{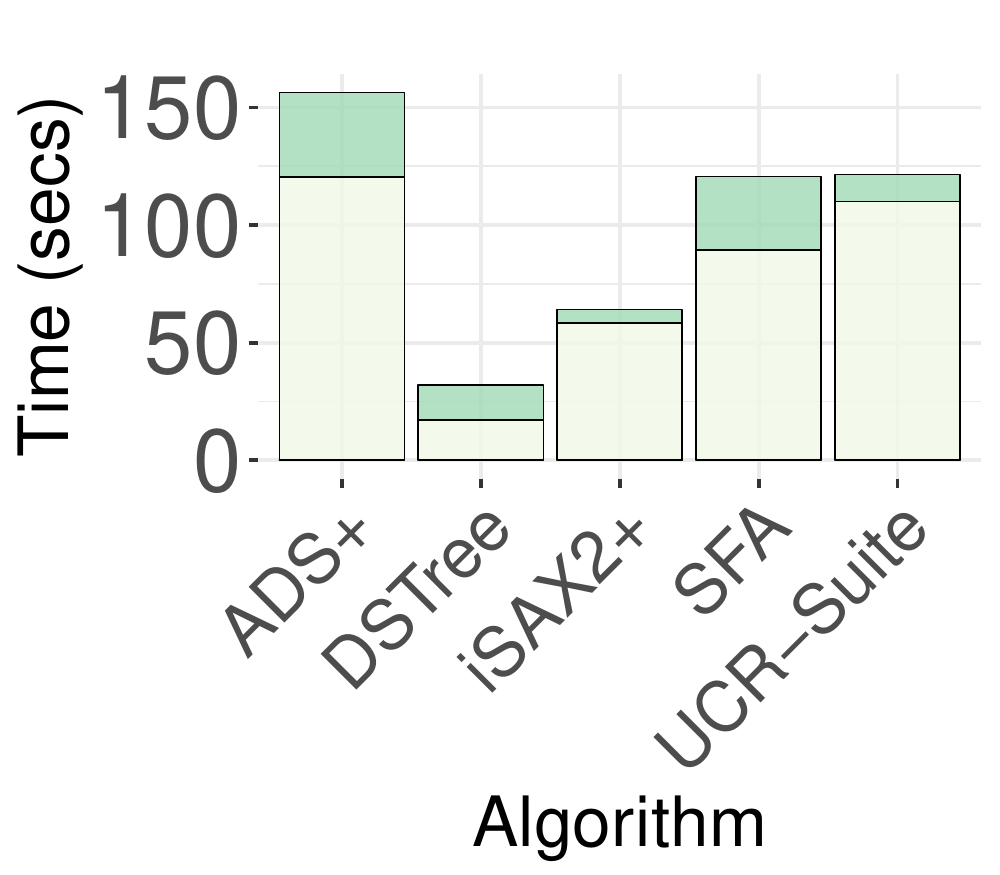}
		\caption{EEG \\ 
				\hspace*{0.5cm}
				Hard}
		\label{fig:exact:real:time:idxproc:cache:eeg:hard}
	\end{subfigure}
	\begin{subfigure}{0.16\textwidth}
		\centering
		\includegraphics[width=\textwidth]{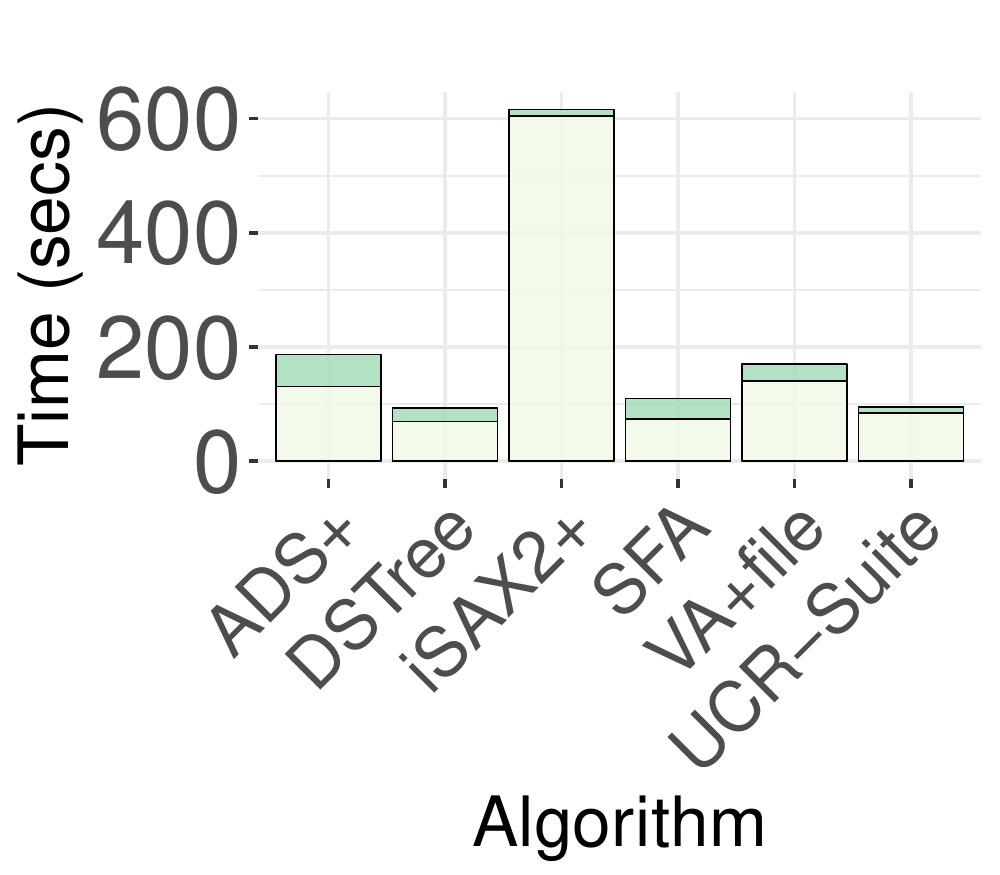}
		\caption{Deep1B \\ 
			\hspace*{0.5cm}
			Easy}
		\label{fig:exact:real:time:idxproc:cache:deep1b:easy}
	\end{subfigure}
	\begin{subfigure}{0.16\textwidth}
		\centering
		\includegraphics[width=\textwidth]{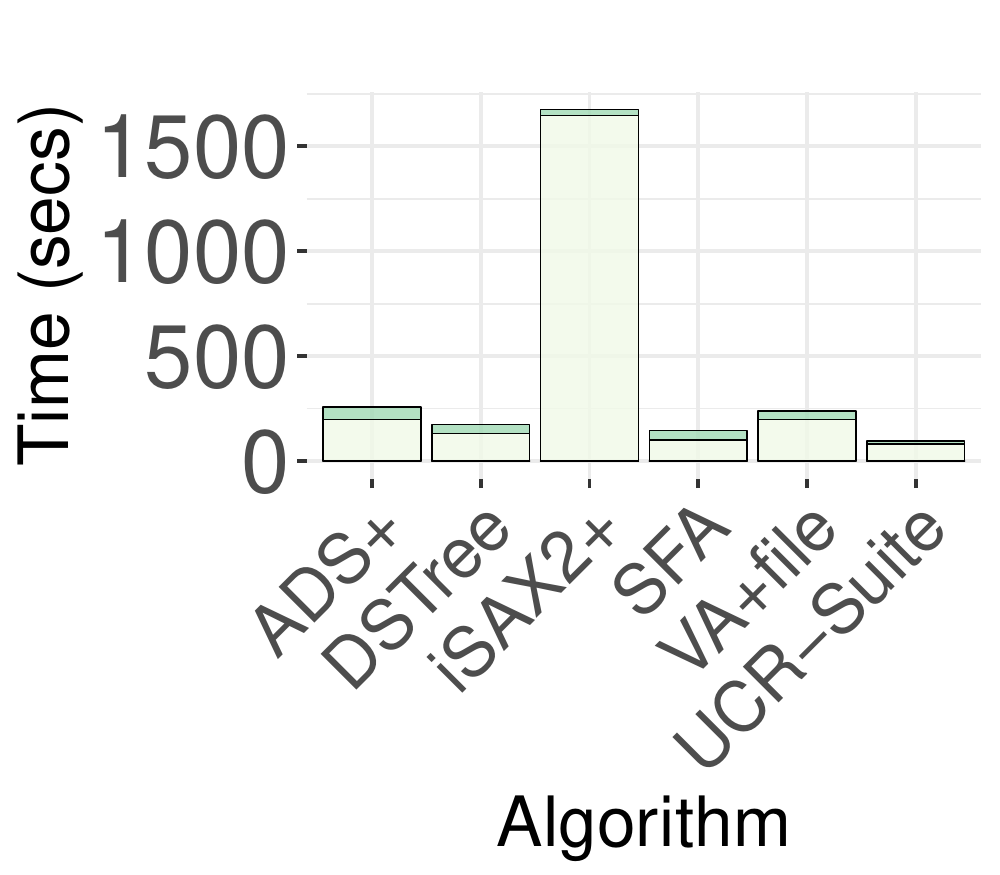}
		\caption{Deep1B \\ 
			\hspace*{0.5cm}
			Hard}
		\label{fig:exact:real:time:idxproc:cache:deep1b:hard}
	\end{subfigure}

	\caption{Average Query Time for Easy and Hard Queries on Real Datasets \\ (HDD)\\}
	\label{fig:exact:real:time:idxproc:cache:hardness}
}
\end{figure*}

\begin{figure*}[tb]
	\captionsetup{justification=centering}
	\captionsetup[subfigure]{justification=centering}		
	\begin{subfigure}\textwidth}
	\centering
	\hspace*{5.5cm}
	\includegraphics[scale=0.5]{{bar_chart_partial_legend}}
\end{subfigure}

\begin{subfigure}{0.16\textwidth}
	\centering
	\includegraphics[width=\textwidth]{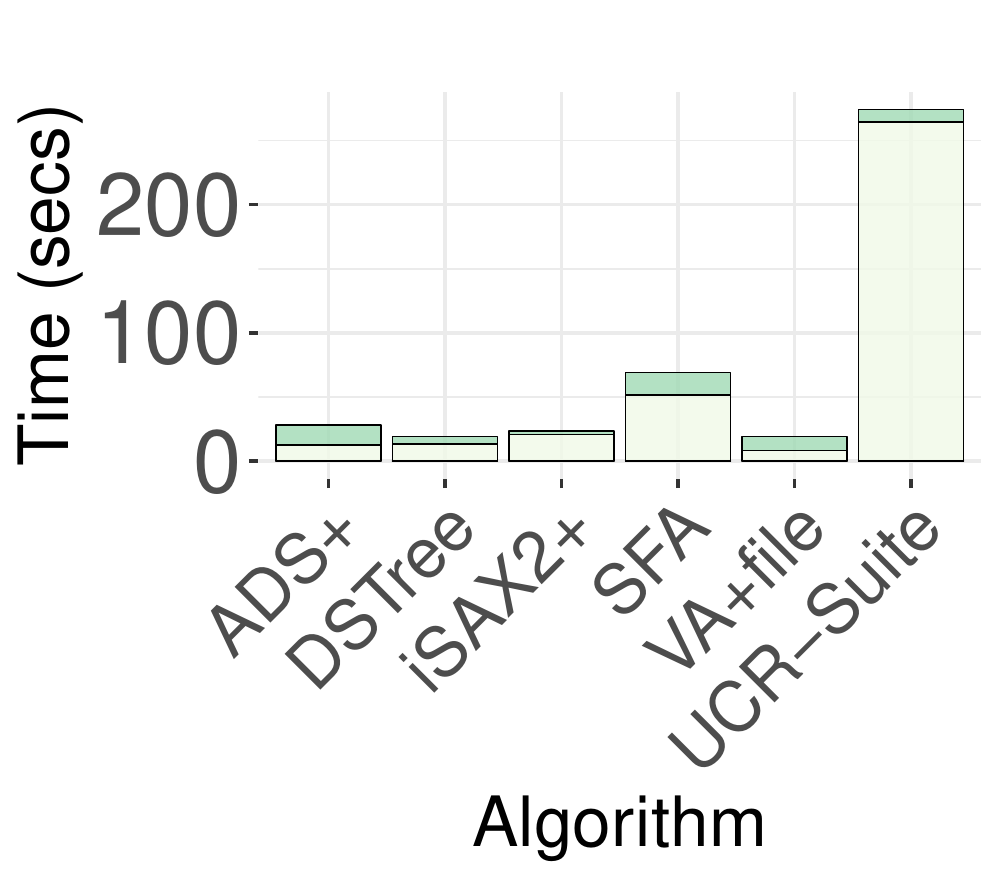}
	\caption{Seismic \\ 
		\hspace*{0.5cm}
		Easy}
	\label{fig:exact:real:time:idxproc:cache:seismic:easy:nefeli}
\end{subfigure}
\begin{subfigure}{0.16\textwidth}
	\centering
	\includegraphics[width=\textwidth]{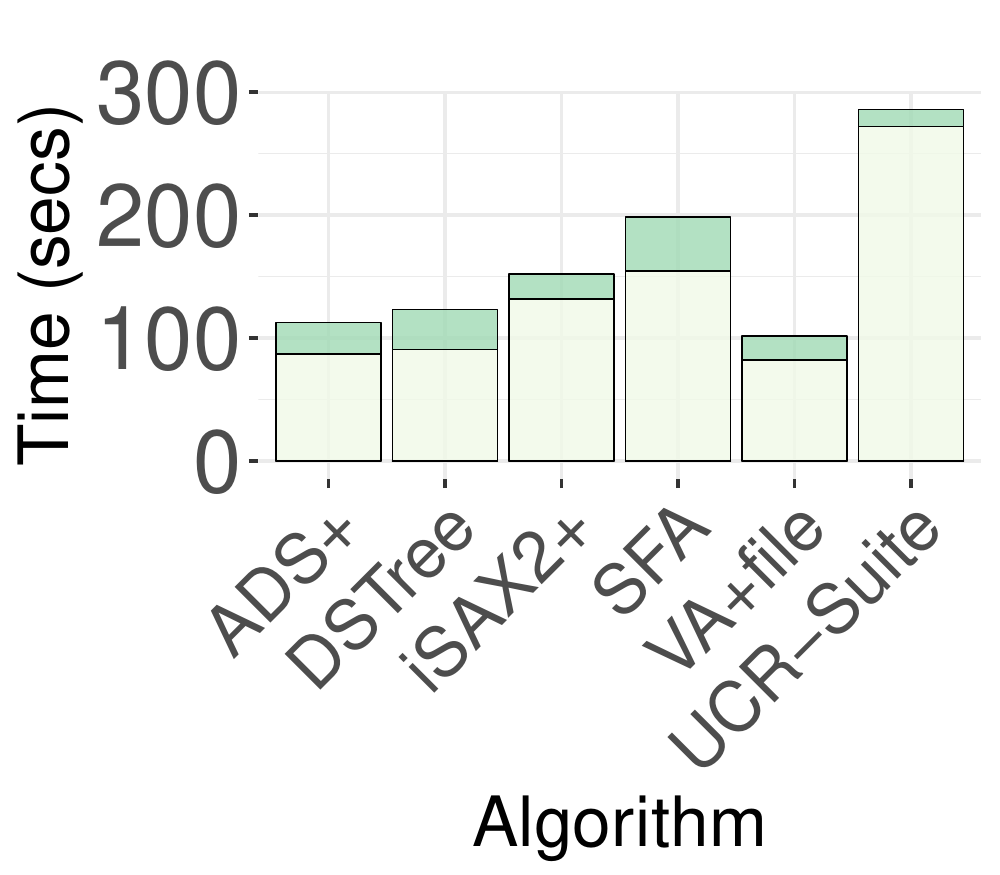}
	\caption{Seismic \\ 
		\hspace*{0.5cm}
		Hard}
	\label{fig:exact:real:time:idxproc:cache:seismic:hard:nefeli}
\end{subfigure}
\begin{subfigure}{0.16\textwidth}
	\centering
	\includegraphics[width=\textwidth]{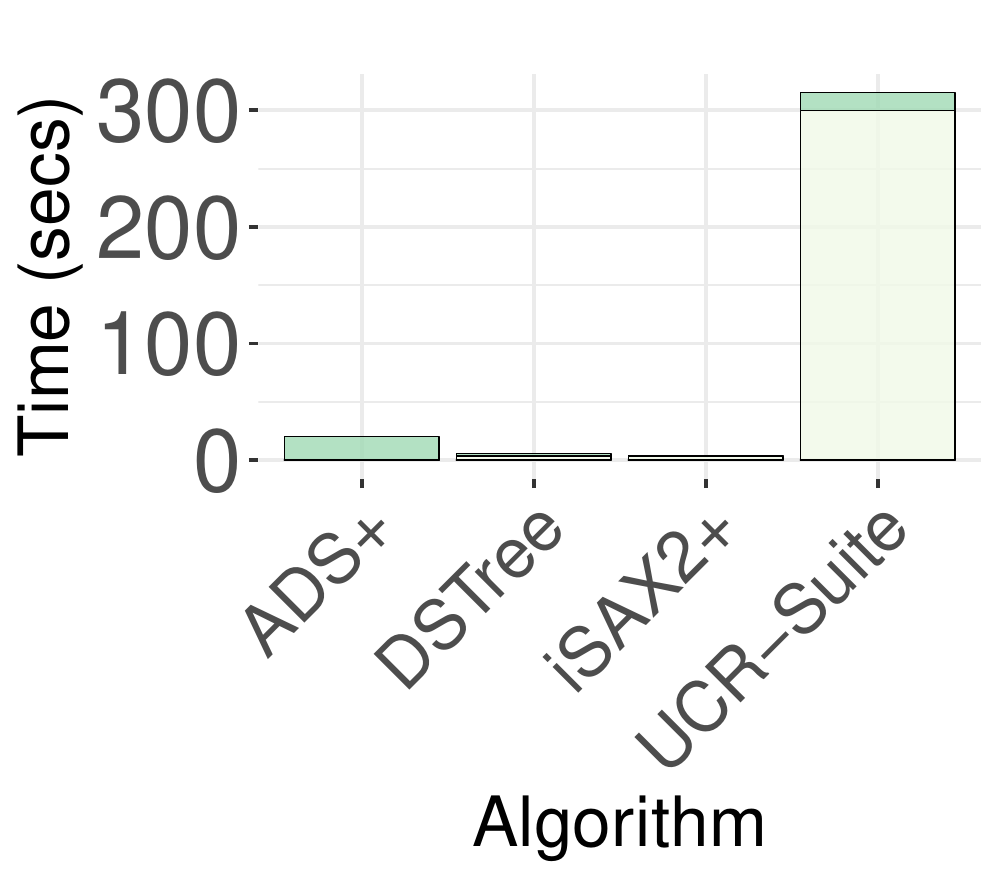}
	\caption{EEG \\ 
		\hspace*{0.5cm}
		Easy}
	\label{fig:exact:real:time:idxproc:cache:eeg:easy:nefeli}
\end{subfigure}
\begin{subfigure}{0.16\textwidth}
	\centering
	\includegraphics[width=\textwidth]{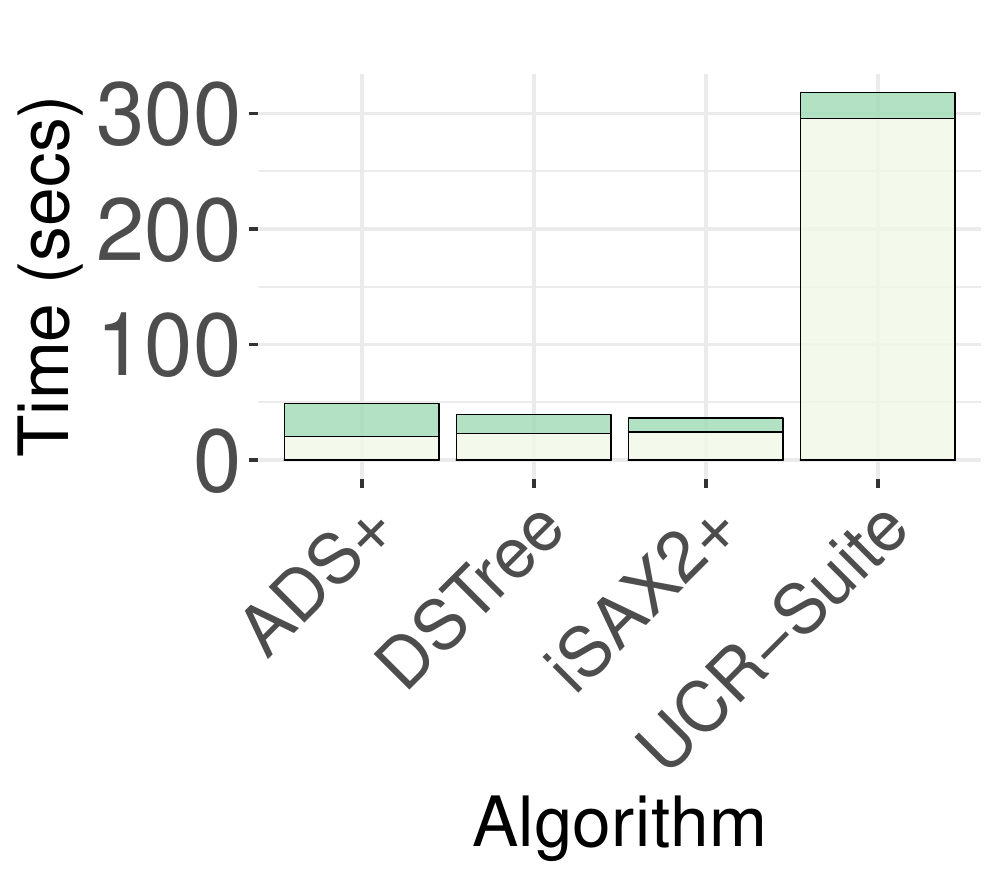}
	\caption{EEG \\ 
		\hspace*{0.5cm}
		Hard}
	\label{fig:exact:real:time:idxproc:cache:eeg:hard:nefeli}
\end{subfigure}
\begin{subfigure}{0.16\textwidth}
	\centering
	\includegraphics[width=\textwidth]{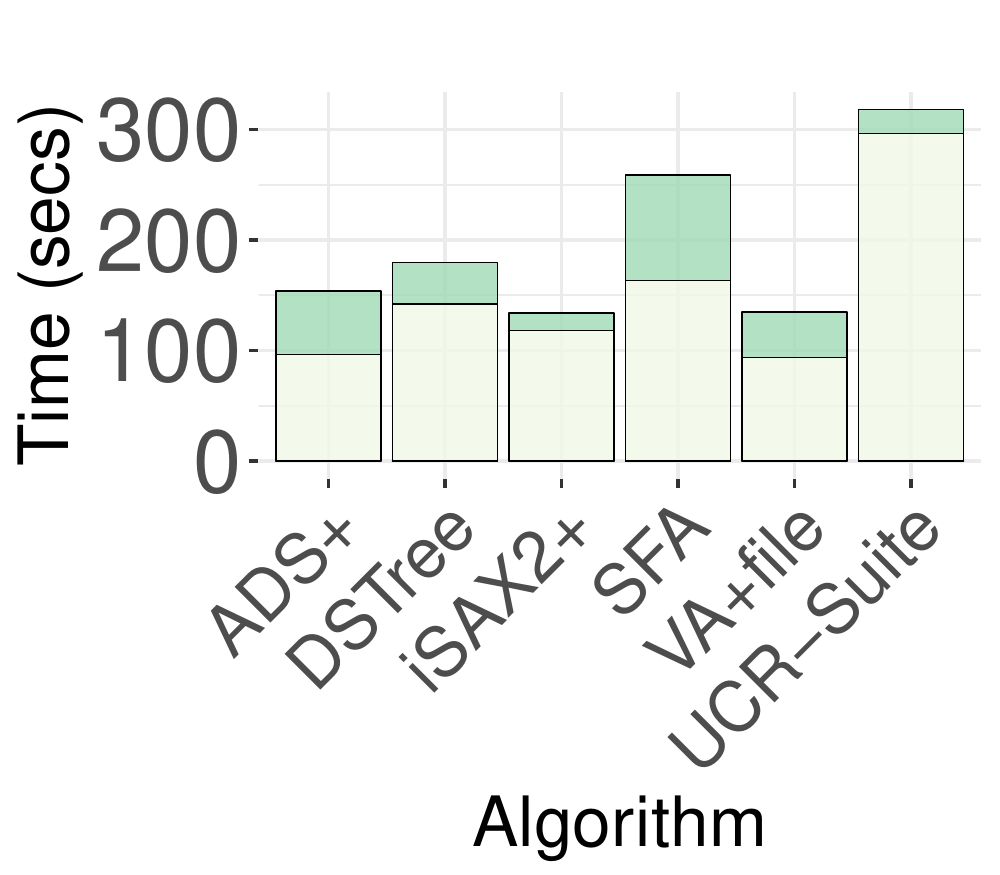}
	\caption{Deep1B \\ 
		\hspace*{0.5cm}
		Easy}
	\label{fig:exact:real:time:idxproc:cache:deep1b:easy:nefeli}
\end{subfigure}
\begin{subfigure}{0.16\textwidth}
	\centering
	\includegraphics[width=\textwidth]{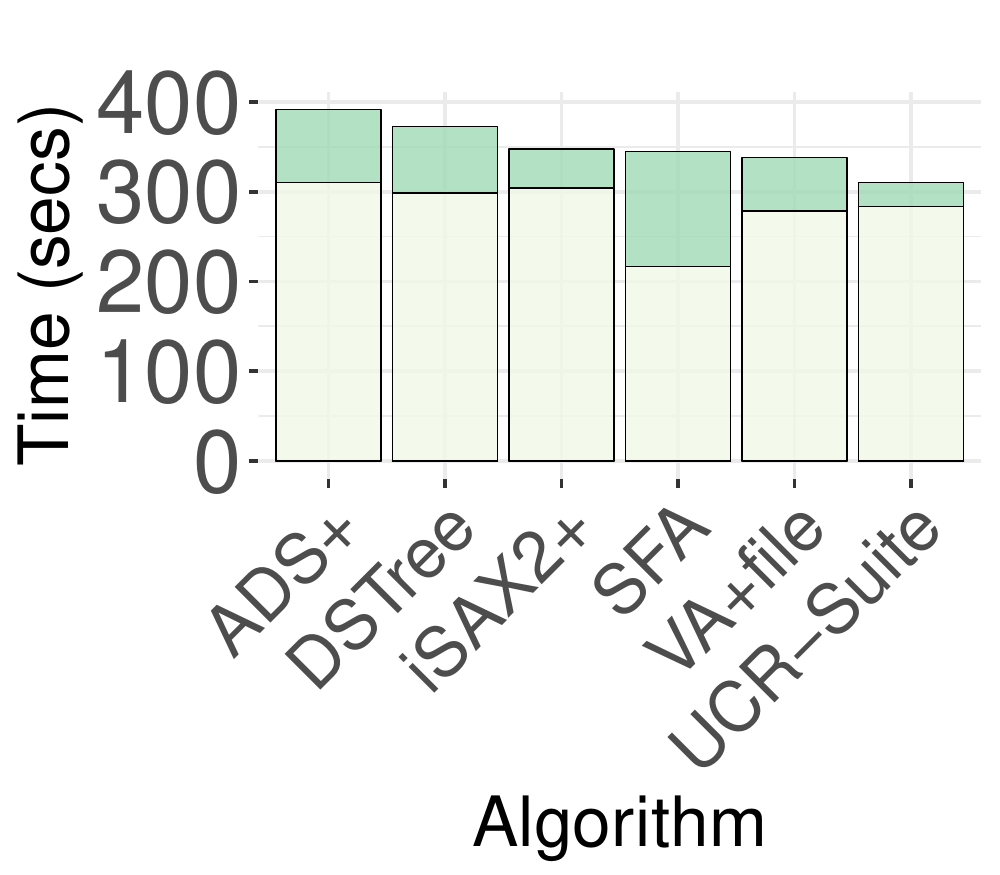}
	\caption{Deep1B \\ 
		\hspace*{0.5cm}
		Hard}
	\label{fig:exact:real:time:idxproc:cache:deep1b:hard:nefeli}
\end{subfigure}

	\caption{Average Query Time for Easy and Hard Queries on Real Datasets \\ (SSD)\\}
\label{fig:exact:real:time:idxproc:cache:hardness:nefeli}
}
\end{figure*}

\fi

\ifFuture

{\color{blue}{\bf \\
	SCOPE: This is relevant for $\epsilon$-Approximate and probabilistic methods \\}
{\it \subsubsection{Approximate, $\epsilon$-Approximate and probabilistic methods}
For each $\epsilon$-approximate and probabilistic method, we test different parameters to find the optimal settings (accuracy vs. performance vs. memory overhead vs. NN hit).
For each probabilistc method, we vary $\delta$ and $\epsilon$ and measure the $\epsilon_{eff}$, the fraction of miss, time performance and memory overhead. The fraction of miss is the fraction of instances for  which  the  method  fails  to  return  the  actual nearest  neighbor \cite{journal/acm/Arya1998}.
The values of $\delta$ and $\epsilon$ that give the best tradeoff are used in subsequent experiments (Figure \ref{fig:prob:eps:delta:datasize}). In the case of $\epsilon$-approximate methods, we run the same experiment with different values of $\epsilon$ (Figure \ref{fig:approx:eps:delta:datasize}).\\
We choose the best parameters from Figures \ref{fig:approx:eps:delta:datasize} and \ref{fig:prob:eps:delta:datasize} to compare probabilistic methods, approximate and $\epsilon$-approximate methods together \ref{fig:approxprob}.
}

\subsubsection{Exact, approximate, $\epsilon$-Approximate and probabilistic methods:}

\begin{itemize}
		\item Do we pick the best exact methods only?
		\item Other than fraction of miss, is there any other way to compare the rank of the approximate result?
		\item Relative error for 1NN, kNN and $r$-range queries
\end{itemize}

Relative Approximatio Error rate for KNN as defined in \cite{conf/ICDE/Abbadi2001}

}
\fi

%

\ifJournal
\begin{figure*}[H]
	\captionsetup{justification=centering}
	\captionsetup[subfigure]{justification=centering}
	\begin{subfigure}{0.31\textwidth}
		\centering
		\includegraphics[scale=0.3]{exact_buffersize_time_idxproc_cache_ads+}
		\caption{ADS+}
		\label{fig:exact:buffersize:disk:idxproc:ADS+}
	\end{subfigure}
	\hspace*{\fill} 
	\begin{subfigure}{0.31\textwidth}
		\centering
		\includegraphics[scale=0.3]{exact_buffersize_time_idxproc_cache_dstree}
		\caption{DSTree}
		\label{fig:exact:buffersize:disk:idxproc:dstree}
	\end{subfigure}
	\hspace*{\fill} 
	\begin{subfigure}{0.31\textwidth}
		\centering
		\includegraphics[scale=0.3]{exact_buffersize_time_idxproc_cache_isax2+}
		\caption{iSAX2+}
		\label{fig:exact:buffersize:disk:idxproc:iSAX2+}
	\end{subfigure}
	\begin{subfigure}{0.31\textwidth}
		\centering
		\includegraphics[scale=0.3]{exact_buffersize_time_idxproc_cache_sfa}
		\caption{SFA trie}
		\label{fig:exact:buffersize:disk:idxproc:sfa}
	\end{subfigure}
	\hspace*{\fill}
	\begin{subfigure}{0.31\textwidth}
		\centering
		\includegraphics[scale=0.3]{exact_buffersize_time_idxproc_cache_m-tree}
		\caption{M-tree}
		\label{fig:exact:buffersize:disk:idxproc:mtree}
	\end{subfigure}
	\hspace*{\fill}
	\begin{subfigure}{0.31\textwidth}
		\centering
		\includegraphics[scale=0.3]{exact_buffersize_time_idxproc_cache_r-tree}
		\caption{R*-tree}
		\label{fig:exact:buffersize:disk:idxproc:rstree}
	\end{subfigure}
	\caption{
		{\color{blue}{\bf SCOPE: This is relevant for the journal version:}
			{\it Parameterization: Buffer Size \\
				(Dataset Size = 100 GB, Data Series Length = 256, Leaf Size = Optimal For Each Method, 100 Exact Queries)
			}}
		}
		\label{fig:exact:buffersize:time:idxproc}
	\end{figure*}

\fi


\ifFuture

\begin{figure*}[tb]
	\captionsetup{justification=centering}
	\captionsetup[subfigure]{justification=centering}
\begin{subfigure}{0.22\textwidth}
        \centering
 \includegraphics[height={\setHeight},width={\linewidth}]{approx_eps_effeps}
  		 \caption{
  		 	{\color{blue}{\bf Approximation Error\\
  		 			{\color{red} x-axis:} Epsilon (0 to 10) \\
  		 			{\color{red} y-axis:} Average Actual Epsilon $\epsilon_{eff}$ (range TBD) \\
  		 			{\color{red} curves:} The average actual approximation error
  		 		}}
  		 	}
\label{fig:approx:eps:epseff:datasize}
\end{subfigure}
\hspace*{\fill} 
\begin{subfigure}{0.22\textwidth}
        \centering
        \includegraphics[height={\setHeight},width={\linewidth}]{dual_recall_accuracy}
  		 \caption{
  		 	{\color{blue}{\bf Recall\\
        	{\color{red} x-axis:} Epsilon (0 to 10) \\
        	{\color{red} y-axis:} Average Recall (0 to 1) \\
        	{\color{red} curves:} The average recall for the query workload for 1NN queries
  		 		}}
  		 	}
        \label{fig:approx:eps:recall:datasize}
\end{subfigure}
\hspace*{\fill} 
\begin{subfigure}{0.22\textwidth}
        \centering
 \includegraphics[height={\setHeight},width={\linewidth}]{approx_eps_footprint}
    		 \caption{
    		 	{\color{blue}{\bf Memory Usage in MB\\
 {\color{red} x-axis:} Epsilon (0 to 10) \\
 {\color{red} y-axis:} Memory Used in MB (range TBD) \\
 {\color{red} curves:} The memory used in MB for different data sets
    		 		}}
    		 	}
\label{fig:approx:eps:footprint:datasize}
\end{subfigure}
\hspace*{\fill} 
\begin{subfigure}{0.22\textwidth}
        \centering
 \includegraphics[height={\setHeight},width={\linewidth}]{approx_eps_time}
\caption{
{\color{blue}{\bf Time Performance\\
 {\color{red} x-axis:} Epsilon (0 to 10) \\
 {\color{red} y-axis:} Total Time in Hours (range TBD) \\
 {\color{red} curves:} The total time to answer the query workload }}
}
 \label{fig:approx:eps:time:datasize}
\end{subfigure}
\caption{{\color{blue}{\bf  This is relevant for $\epsilon$-approximate methods\\
 Accuracy and Performance Tradeoffs for $\epsilon$-approximate Methods
}}
}
   \label{fig:approx:eps:delta:datasize}
\end{figure*}

\begin{figure*}[tb]
	\captionsetup{justification=centering}
	\captionsetup[subfigure]{justification=centering}
	\begin{subfigure}{0.22\textwidth}
		\centering
		\includegraphics[height={\setHeight},width={\linewidth}]{prob_eps_effeps_del}
		\caption{{\color{blue}{\bf Approximation Error\\
					{\color{red} x-axis:} Epsilon (0 to 10) \\
					{\color{red} y-axis:} Average Actual Epsilon $\epsilon_{eff}$ (range TBD) \\
					{\color{red} z-axis:} Tolerance $\delta$ (0 to 1) \\
					{\color{red} curves:} The average actual approximation error
				}}
			}
		\label{fig:prob:eps:epseff:delta:datasize}
	\end{subfigure}
	\hspace*{\fill} 
	\begin{subfigure}{0.22\textwidth}
		\centering
		\includegraphics[height={\setHeight},width={\linewidth}]{dual_recall_accuracy}
		\caption{{\color{blue}{\bf Recall\\
					{\color{red} x-axis:} Epsilon (0 to 10) \\
					{\color{red} y-axis:} Average Recall (0 to 1) \\
					{\color{red} z-axis:} Tolerance $\delta$ (0 to 1) \\
					{\color{red} curves:} The average recall 1NN queries (other k-NN?)
					}}
				}
		\label{fig:prob:eps:recall:delta:datasize}
	\end{subfigure}
	\hspace*{\fill} 
	\begin{subfigure}{0.22\textwidth}
		\centering
		\includegraphics[height={\setHeight},width={\linewidth}]{prob_eps_footprint_del}

		\caption{{\color{blue}{\bf Memory Usage in MB\\
					{\color{red} x-axis:} Epsilon (0 to 10) \\
					{\color{red} y-axis:} Memory Used in MB (range TBD) \\
					{\color{red} z-axis:} Tolerance $\delta$ (0 to 1) \\
					{\color{red} curves:} The memory used in MB for different data sets
					}}
		}
		\label{fig:prob:eps:footprint:delta:datasize}
	\end{subfigure}
	\hspace*{\fill} 
	\begin{subfigure}{0.22\textwidth}
		\centering
		\includegraphics[height={\setHeight},width={\linewidth}]{prob_eps_time_del}

		\caption{{\color{blue}{\bf Time Performance\\
			{\color{red} x-axis:} Epsilon (0 to 10) \\
			{\color{red} y-axis:} Total Time in Hours (range TBD) \\
			{\color{red} z-axis:} Tolerance $\delta$ (0 to 1) \\
			{\color{red} curves:} The total time to answer the query workload			}}
		}
		\label{fig:prob:eps:time:delta:datasize}
	\end{subfigure}
	\caption{{\color{blue}{\bf  This is relevant for probabilistic and $\epsilon$-approximate methods\\
	Accuracy and Performance Tradeoffs for probabilistic Methods.
	Check ICDE00 for more details.
	}}
	}
	\label{fig:prob:eps:delta:datasize}
\end{figure*}

\begin{figure*}[tb]
  \captionsetup{justification=centering}
 \includegraphics[height={\setHeight},width={\linewidth}]{prob_eps_effeps_datasize_accuracy}

  		\caption{{\color{blue}{\bf This is relevant for probabilistic and $\epsilon$-approximate methods \\
  					Comparison of probabilistic, $\epsilon$-approximate and approximate methods.\\
  					{\color{red} x-axis:} TBD (range TBD)\\
  					{\color{red} y-axis:} TDB (range TBD) \\
  					{\color{red} z-axis:} TBD (range TBD) \\
  					{\color{red} curves:} TBD
  				}}
  			}

 \label{fig:approxprob}
\end{figure*}

\begin{figure*}[tb]
	\captionsetup{justification=centering}
	\includegraphics[height={\setHeight},width={\linewidth}]{prob_eps_effeps_datasize_accuracy}
	\caption{{\color{blue}{\bf This is relevant for probabilistic and $\epsilon$-approximate methods \\
				Comparison of probabilistic, $\epsilon$-approximate and exact methods. Three figures:
				Fig 1: Extra time on top of prob, Fig 22d ADS+ \\
	  			Fig 2: Extra mem on top of prob, Fig 22d ADS+ \\
	  			Fig 1: Extra time on top of prob, Fig 22d ADS+ \\
	  			The base case is probabilistic with good accuracy.
	  			{\color{red} x-axis:} TBD (range TBD)\\
	  			{\color{red} y-axis:} TDB (range TBD) \\
	  			{\color{red} z-axis:} TBD (range TBD) \\
	  			{\color{red} curves:} TBD
	  			}}
	  	}

	\label{fig:exactapproxprob}
\end{figure*}
\fi



\section{Discussion}

\label{sec:discussion}

In the data series literature, competing similarity search methods have never been compared together under a unified experimental scheme.
The objective of this experimental evaluation is to consolidate previous work on data series whole-matching similarity search and prepare a solid ground for further developments in the field.

We undertook a challenging and laborious task, where we re-implemented from scratch four algorithms: iSAX2+, SFA trie, DSTree, and VA+file, and optimized memory management problems (swapping, and out-of-memory errors) in R*-tree, M-tree, and Stepwise.
Choosing C/C++ provided considerable performance gains, but also required low-level memory management optimizations.
We believe the effort involved was well worth it since the results of our experimental evaluation emphatically demonstrate the importance of the experimental setup on the relative performance of the various methods.
To further facilitate research in the field we publicize our source code and experimental results~\cite{url/DSSeval}.
This section summarizes the lessons learned in this study.

\noindent{\bf Unexpected Results.}
For some of the algorithms our experimental evaluation revealed some unexpected results.

(1) \emph{The Stepwise method performed lower than our expectations.}
This was both due to the fact that our baseline sequential scan was fully optimized for early abandoning and computation optimization, but most importantly because of \emph{a different experimental setup}.
The original implementation of Stepwise performed batched query answering.
In our case we compared all methods on single query at a time workload scenario.
This demonstrates the importance of \emph{the experimental setup and workload type}.

(2) \emph{The VA+file method performed extremely well.}
Although an older method, VA+file is among the best performers overall. Our optimized implementation, which is much faster than the original 
version, helped unleash the best of this method; this demonstrates the importance of \emph{the implementation framework}.}

(3) \emph{For exact queries on out-of-memory data on the HDD machine, ADS+ is underperforming.}
The reason is that ADS+ performs multiple skips while pruning at a per series level and is thus significantly affected by the hard disk's latency.
In the original study~\cite{journal/vldb/Zoumpatianos2016}, ADS+ was run on a machine with 60\% of the hard disk throughput of the one used in the current work.
The HDD setup with the 6 RAID0 disks gave a significant advantage on methods that perform sequential scans on larger blocks of data and less skips. On the SSD machine, however, the trend is reversed, and ADS+ becomes one of the best contenders overall.
These observations demonstrate the importance of \emph{the hardware setup}. 

(4) \emph{The optimal parameters of most algorithms were different than the ones presented in their respective papers.} 
This is because some methods were not tuned before: the iSAX2+, DSTree and SFA papers have no tuning experiments. We tuned each for varying leaf and buffer sizes (for brevity, we only report results for leaf parametrization in Figure~\ref{fig:exact:leafsize:time:idxproc} (for buffer tuning experiments, see~\cite{url/DSSeval}). For SFA, we also tuned the sample size used to identify the breakpoints, binning method (equi-depth vs. equi-width), and number of symbols for the SFA discretization. 
Another reason is that we studied in more detail methods that were partially tuned (e.g., ADS+ was tuned only for varying leaf size; we also varied buffer size and found that assigning most of RAM to buffering hurts performance).
These findings further demonstrate the need for \emph{careful parameter-tuning}.

(5) \emph{The quality of the summarization, as measured by TLB and pruning, is not necessarily correlated to time performance.} 
An early experimental study~\cite{journal/dmkd/Keogh2003} claimed that the tightness of the lower bound can be used alone to evaluate the efficiency of indexing techniques. While summarization quality is an important factor on its own, we demonstrate that it cannot alone predict the time performance of an index, even in the absence of data and implementation biases. For example, ADS+ achieves very high pruning and TLB, yet, in terms of time, it is outperformed by other methods in some scenarios. 
It is of crucial importance to consider summarization quality 
alongside 
the properties of the index structure and the hardware platform.

\noindent{\bf Speed-up Opportunities.}
Through our analysis, we identified multiple factors that affect the performance of examined methods. In turn, these factors reveal opportunities and point to directions for performance improvements. 

(1) \emph{Stepwise} offers many such avenues.
Its storage scheme could be optimized to reduce the number of random I/O during query answering, and its query answering algorithm would benefit a lot from parallelization and modern hardware optimizations (i.e., through multi-core and SIMD parallelism), as 50\%-98\% of total time is CPU. 

(2) \emph{DSTree} is very fast at query answering, but rather slow at index building.
Nevertheless, a large percentage of this time (85-90\%) is CPU. 
Therefore, also the indexing performance of DSTree can be improved by exploiting modern hardware. 
Moreover, bulk loading during indexing, and buffering during querying, would also make it even faster.

(3) A similar observation holds for \emph{VA+file} \emph{MASS}. Even though MASS is not designed for whole-matching data series similarity search, its performance can be significantly enhanced with parallelism and modern hardware exploitation, since 90\% of its execution time is CPU cost. Similarly, the indexing cost of VA+file 
can be further improved.

(4) 
Finally, we obtained a better understanding of the \emph{ADS+} algorithm. 
Apart from being very fast in index building, our results showed that it also has a leading performance for whole-matching similarity search for \emph{long} data series.
We also discovered that the main bottleneck for 
ADS+ are the multiple skips performed during query answering.
Its effects could be masked by controlling the size of the data segments skipped (i.e., skipping/reading large continuous blocks), and through asynchronous I/O.
Moreover, because of its very good pruning power (that leads to an increased number of skips), we expect ADS+ to work well whenever random access is cheap, e.g., with SSDs and main-memory systems.

\noindent{\bf Data-adaptive Partitioning.}
While the SFA trie and iSAX-based index building algorithms are much faster than the DSTree index building algorithm, their performance during query answering is much worse than that of DSTree.
DSTree spends more time during indexing, intelligently adapting its leaf summarizations when split operations are performed.
This leads to better data clustering and as a result faster query execution.
On the contrary, both iSAX and SFA have fixed maximum resolutions, and iSAX indexes can only perform splits on predefined split-points.
Even though iSAX summarizations at full resolution offer excellent pruning power (see ADS+ in Figure~\ref{fig:exact:data::pruning}), grouping them using fixed split-points in an iSAX-based index does not allow for effective clustering (see Figure \ref{fig:exact:datasize:fill:combined}).
This is both an advantage (indexing 
is extremely fast), but also a drawback as it does not allow clustering to adapt to the dataset distribution.

\ifJournal
\noindent{\bf Approximate Query Answering.} Approximate answers provide a viable alternative when exact solutions are not required, with many methods having a small effective error and great performance (Figures \ref{fig:exact:synthetic:datasize:error:combined}-\ref{fig:exact:real:error:combined}). 
A detailed evaluation of approximate methods is part of our future work.
\fi

\noindent{\bf Access-Path Selection.}
Finally, our results demonstrate that the pruning ratio, along with the ability of an index to cluster together similar data series in large contiguous blocks of data, is crucial for its performance. 
Moreover, our results confirm the intuitive result that the smaller the pruning ratio, the higher the probability that a sequential scan will perform better than an index, as can be observed for the hard queries in Table~\ref{tab:summary}. 
This is because it will avoid costly random accesses patterns on a large part of the dataset. 
However, the decision between a scan or an index, and more specifically, the choice of an index, is not trivial, but is based on a combination of factors: (a) the effectiveness of the summarization used by the index (which can be estimated by the pruning ratio); 
(b) the ability of the index to cluster together similar data series (which determines the access pattern);
and (c) the hardware characteristics (which dictate the data access latencies).
This context gives rise to interesting optimization problems, which have never before been studied in the domain of data series similarity search.

\begin{figure}[!htb]
	\captionsetup{justification=centering}
	\includegraphics[width =\columnwidth]{{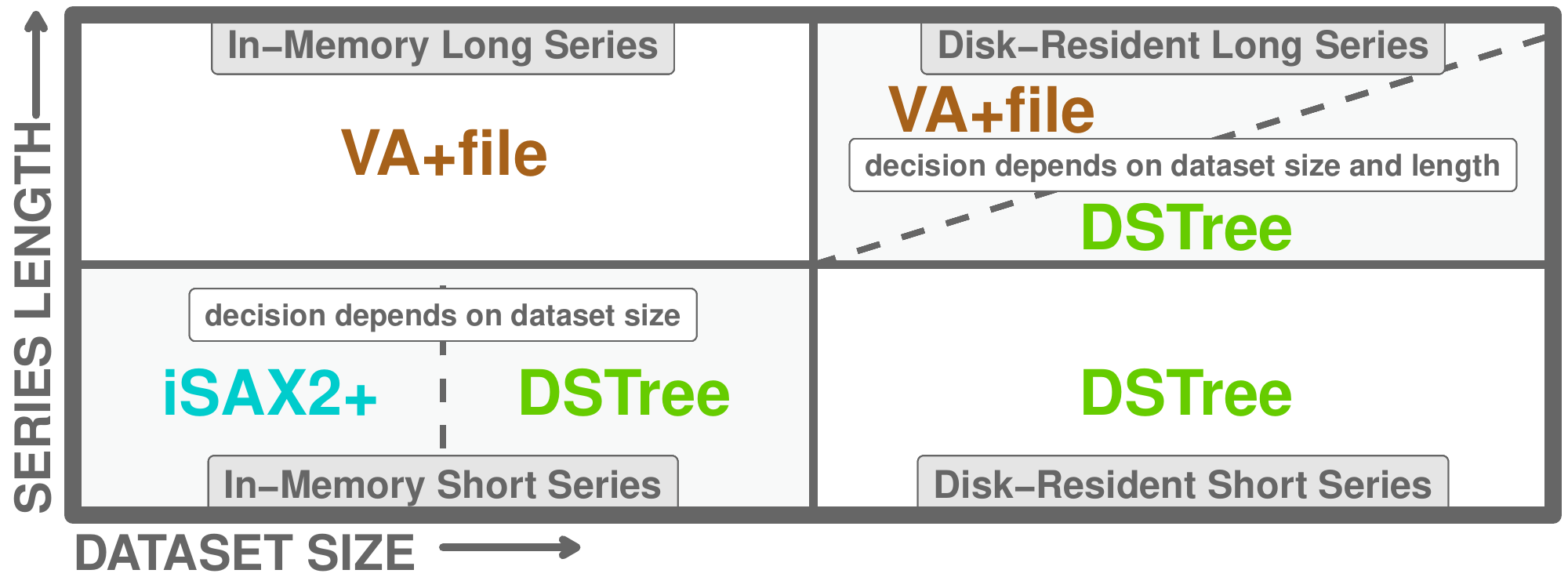}}
	\caption{Recommendations  \\
		(Indexing and answering 10K queries on HDD)		
		}
	\label{fig:recommendations}
\end{figure}

\noindent{\bf Recommendations.}
Figure~\ref{fig:recommendations} presents a decision matrix that reports the best approach to use for problems with different data series characteristics, given a specific hardware setup (i.e., HDD) and query workload (i.e., Indexing + 10K synthetic queries).
In general though, choosing the best approach to answer a similarity query on massive data series is an optimization problem, and needs to be studied in depth. 

\section{Conclusions and Future Work}
\label{sec:conclusions}

In this work, we unified and formally defined the terminology used for the different flavors of data series similarity search problems, and we designed and executed a thorough experimental comparison of several relevant techniques from the literature, which had never before been compared at equal footing to one another. 
Our results paint a clear picture of the strengths and weaknesses of the various approaches, and indicate promising research directions.
Part of our future work is the experimental comparison of approximate methods,  $r$-range queries and sub-sequence matching.

\noindent{\bf Acknowledgments.}
We sincerely thank all authors for generously sharing their code, and M. Linardi for his implementation of MASS~\cite{journal/dmkd/Yeh2017}. Work partially supported by EU project NESTOR (Marie Curie \#748945).

\balance

\bibliographystyle{abbrv}
\def\thebibliography#1{
  \section*{References}
 \normalsize                  
  \list
    {[\arabic{enumi}]}
    {\settowidth\labelwidth{[#1]}
     \leftmargin\labelwidth
     \parsep 1pt                
     \itemsep 0.6pt               
     \advance\leftmargin\labelsep
     \usecounter{enumi}
    }
  \def\newblock{\hskip .11em plus .33em minus .07em}
  \sloppy\clubpenalty10000\widowpenalty10000
  \sfcode`\.=1000\relax
}
\bibliography{ref}  

\end{document}